\documentclass[aps,prd,reprint,superscriptaddress,nofootinbib,showpacs,preprintnumbers]{revtex4-1}
\pdfoutput=1

\usepackage{graphicx} 
\usepackage{atlasphysics}
\usepackage{subfigure}
\usepackage{longtable}
\usepackage{multirow}
\usepackage{textcomp}
\usepackage{units}
\usepackage{lineno}
\usepackage{rotating}
\usepackage{amssymb}
\usepackage{amsmath}
\usepackage{array} 
\usepackage[colorlinks=true,linkcolor=blue,citecolor=blue]{hyperref}

\DeclareGraphicsExtensions{.pdf}

\newcommand{\lumieN}{\ensuremath{36.2\;\mathrm{pb}^{-1}}}
\newcommand{\lumimuN}{\ensuremath{32.6\;\mathrm{pb}^{-1}}}
\newcommand{\dlumi}{\ensuremath{3.4}}

\newcommand{\sigWfid}{ \ensuremath{5.127 \pm 0.011  \pm 0.061  \pm 0.174  \pm 0.005  }}
\newcommand{\sigWplusfid}{ \ensuremath{3.110 \pm 0.008  \pm 0.036  \pm 0.106  \pm 0.004  }}
\newcommand{\sigWminusfid}{ \ensuremath{2.017 \pm 0.007  \pm 0.028  \pm 0.069  \pm 0.002  }}
\newcommand{\sigZfid}{ \ensuremath{0.479 \pm 0.003  \pm 0.005  \pm 0.016  \pm 0.001  }}

\newcommand{\sigW}{ \ensuremath{10.207 \pm 0.021  \pm 0.121  \pm 0.347  \pm 0.164  }}
\newcommand{\sigWplus}{ \ensuremath{6.048 \pm 0.016  \pm 0.072  \pm 0.206  \pm 0.096  }}
\newcommand{\sigWminus}{ \ensuremath{4.160 \pm 0.014  \pm 0.057  \pm 0.141  \pm 0.083  }}
\newcommand{\sigZ}{ \ensuremath{0.937 \pm 0.006  \pm 0.009  \pm 0.032  \pm 0.016  }}

\newcommand{\rWZfid}{ \ensuremath{10.703 \pm 0.078  \pm 0.110  \pm 0.008  }}
\newcommand{\rWplusZfid}{ \ensuremath{6.493 \pm 0.049  \pm 0.064  \pm 0.005  }}
\newcommand{\rWminusZfid}{ \ensuremath{4.210 \pm 0.033  \pm 0.049  \pm 0.003  }}
\newcommand{\rWplusWminusfid}{ \ensuremath{1.542 \pm 0.007  \pm 0.012  \pm 0.001  }}

\newcommand{\rWZ}{ \ensuremath{10.893 \pm 0.079  \pm 0.110  \pm 0.116  }}
\newcommand{\rWplusZ}{ \ensuremath{6.454 \pm 0.048  \pm 0.065  \pm 0.072  }}
\newcommand{\rWminusZ}{ \ensuremath{4.439 \pm 0.034  \pm 0.050  \pm 0.049  }}
\newcommand{\rWplusWminus}{ \ensuremath{1.454 \pm 0.006  \pm 0.012  \pm 0.022  }}

\newcommand{\sigfidZmu}{\ensuremath{0.456 \pm 0.004  \pm 0.004  \pm 0.015  }}

\newcommand{\sigZmu}{\ensuremath{0.935 \pm 0.009  \pm 0.009  \pm 0.032  \pm 0.019  }}

\newcommand{\sigfidWmu}{\ensuremath{4.949 \pm 0.015  \pm 0.081  \pm 0.168  }}

\newcommand{\sigWmu}{\ensuremath{10.210 \pm 0.030  \pm 0.166  \pm 0.347  \pm 0.153  }}

\newcommand{\sigfidWmuplus}{\ensuremath{3.002 \pm 0.011  \pm 0.050  \pm 0.102  }}

\newcommand{\sigWmuplus}{\ensuremath{6.062 \pm 0.023  \pm 0.101  \pm 0.206  \pm 0.099  }}

\newcommand{\sigfidWmuminus}{\ensuremath{1.948 \pm 0.009  \pm 0.034  \pm 0.066  }}

\newcommand{\sigWmuminus}{\ensuremath{4.145 \pm 0.020  \pm 0.072  \pm 0.141  \pm 0.086  }}

\newcommand{\sigfidZe}{\ensuremath{0.426 \pm 0.004  \pm 0.012  \pm 0.014  }}

\newcommand{\sigZe}{\ensuremath{0.952 \pm 0.010  \pm 0.026  \pm 0.032  \pm 0.019  }}

\newcommand{\sigfidWe}{\ensuremath{4.791 \pm 0.014  \pm 0.089  \pm 0.163  }}

\newcommand{\sigWe}{\ensuremath{10.255 \pm 0.031  \pm 0.190  \pm 0.349  \pm 0.156  }}

\newcommand{\sigfidWeplus}{\ensuremath{2.898 \pm 0.011  \pm 0.052  \pm 0.099  }}

\newcommand{\sigWeplus}{\ensuremath{6.063 \pm 0.023  \pm 0.108  \pm 0.206  \pm 0.104  }}

\newcommand{\sigfidWeminus}{\ensuremath{1.893 \pm 0.009  \pm 0.038  \pm 0.064  }}

\newcommand{\sigWeminus}{\ensuremath{4.191 \pm 0.020  \pm 0.085  \pm 0.142  \pm 0.084  }}

\newcommand{\Pythia}{\textsc{Pythia}}
\newcommand{\Powheg}{\textsc{PowHeg}}
\newcommand{\Mcatnlo}{\textsc{Mc@Nlo}}
\newcommand{\Herwig}{\textsc{Herwig}}
\newcommand{\Photos}{\textsc{Photos}}

\newcommand{\Tab}{Tab.}
\newcommand{\Tabs}{Tabs.}
\newcommand{\TTab}{Table}

\newcommand{\Sec}{Sec.}

\newcommand{\Fig}{Fig.}
\newcommand{\Figs}{Figs.}
\newcommand{\FFig}{Figure}
\newcommand{\FFigs}{Figures}

\newcommand{\Zg}{\ensuremath{Z/\gamma^*}}

\newcommand{\CZ}{{\ensuremath{C_Z}}}

\newcommand{\CW}{{\ensuremath{C_W}}}
\newcommand{\AWZ}{{\ensuremath{A_{W/Z}}}}
\newcommand{\CWZ}{{\ensuremath{C_{W/Z}}}}

\newcommand{\muonRecoSF}{\ensuremath{0.993 \pm 0.002\,\mathrm{(sta)} \pm 0.002\,\mathrm{(sys)}}}
\newcommand{\muonIsoSF}{\ensuremath{0.9995 \pm 0.0006\,\mathrm{(sta)} \pm 0.0013\,\mathrm{(sys)} }}
\newcommand{\muonTriggerSF}{\ensuremath{1.020 \pm 0.003\,\mathrm{(sta)} \pm 0.002\,\mathrm{(sys)}}}

\newcommand{\NWmuCands}{\ensuremath{139748}}
\newcommand{\NWmuplusCands}{\ensuremath{84514}}
\newcommand{\NWmuminusCands}{\ensuremath{55234}}

\newcommand{\nZmumuWmu}{\ensuremath{3.3\%}}
\newcommand{\nWtaunuWmu}{\ensuremath{2.8\%}}
\newcommand{\nZtautauWmu}{\ensuremath{0.1\%}}
\newcommand{\nttbarWmu}{\ensuremath{0.4\%}}

\newcommand{\nDiBosonWmu}{\ensuremath{0.1\%}}

\newcommand{\nQCDWmuPlus}{\ensuremath{1.7\%}}

\newcommand{\nQCDWmuMinus}{\ensuremath{2.8\%}}

\newcommand{\nEWKWmuPlus}{\ensuremath{6.1\%}}
\newcommand{\nEWKWmuMinus}{\ensuremath{7.6\%}}

\newcommand{\nWmuBkg}{\ensuremath{12300 \pm 1100}}
\newcommand{\nWmuplusBkg}{\ensuremath{6600 \pm 600}}
\newcommand{\nWmuminusBkg}{\ensuremath{5700 \pm 600}}

\newcommand{\nZtautauZmu}{\ensuremath{0.07\%}}
\newcommand{\nttbarZmu}{\ensuremath{0.1\%}}
\newcommand{\nDiBosonZmu}{\ensuremath{0.2\%}}

\newcommand{\nQCDZmu}{\ensuremath{0.4\%}}

\newcommand{\nQCDZmuStatRel}{\ensuremath{40\%}}

\newcommand{\nQCDZmuSystRel}{\ensuremath{56\%}}

\newcommand{\NZmuCands}{\ensuremath{11709}}
\newcommand{\nZmuBkg}{\ensuremath{86 \pm 32}}

\begin{document}

\preprint{CERN-PH-EP-2011-143}
\preprint{Submitted to Phys. Rev. D}

\title{Measurement of the inclusive $W^{\pm}$ and \Zg cross sections \\
in the $e$ and $\mu$ decay channels 
in $pp$ collisions at $\sqrt{s}=7$\,TeV \\with the ATLAS detector}

\author{The ATLAS Collaboration}
\thanks{Full author list given at the end of the article.}

\noaffiliation

\date{\today}

\begin{abstract}
The production cross sections of the inclusive Drell-Yan processes 
$\ensuremath{W^\pm} \rightarrow \ell \nu$ and
$\Zg \to \ell \ell$ ($\ell=e,\mu$) are measured 
in proton-proton collisions at $\sqrt{s}=7$~TeV with the ATLAS detector.
The cross sections are reported integrated over a fiducial
kinematic range, extrapolated to the full range and also
evaluated differentially as a function of the $W$ decay lepton
pseudorapidity and the $Z$ boson rapidity, respectively.
Based on 
an integrated luminosity of about $35$\,pb$^{-1}$ collected in 2010,
the precision of these measurements reaches a few per cent. 
The integrated and the differential $W^{\pm}$ and $\Zg$ 
cross sections in the $e$ and $\mu$ channels are combined, and
compared with perturbative QCD calculations,
based on a number of different parton distribution 
sets available at NNLO.
\end{abstract}


\maketitle

\renewcommand{\arraystretch}{1.35}

\section{Introduction}
\label{sec:intro}

The inclusive Drell-Yan~\cite{Drell:1970wh} production cross sections of $W$ and $Z$ bosons  
have been
an important testing ground for Quantum Chromodynamics (QCD). Theoretical
calculations of this process extend to next-to-leading order 
(NLO)~\cite{KubarAndre:1978uy,Altarelli:1979ub,Kubar:1980zv} and next-to-next-to
leading order (NNLO) ~\cite{Rijken:1994sh,Hamberg:1990np,vanNeerven:1991gh,Harlander:2002wh,Anastasiou:2003ds} 
perturbation theory. Crucial ingredients of
the resulting QCD cross section calculations
are the parameterisations of the momentum distribution
functions of partons in the proton (PDFs). These have been determined recently
in a variety of phenomenological analyses to NLO QCD
by the CTEQ~\cite{Nadolsky:2008zw,Lai:2010vv} 
group and to NNLO by the  MSTW~\cite{Martin:2009iq}, 
ABKM~\cite{ABKM09, Alekhin:2010iu},
HERAPDF~\cite{HERA:2009wt,Radescu:2011cn}, JR~\cite{JimenezDelgado:2008hf}  
and NNPDF~\cite{Ball:2011mu,Ball:2011uy} groups.

The present measurement determines the cross sections times leptonic
branching ratios,
$\sigma_{W^{\pm}} \cdot$ BR($W \to \ell \nu$) and
$\sigma_{Z/\gamma^*} \cdot$ BR($Z/\gamma^* \to \ell\ell$),
of inclusive $W$ and $Z$ production 
for electron and muon final states, where $\ell=e,~\mu$.
Compared to the initial measurement by the ATLAS Collaboration~\cite{Aad:2010yt},
the data set is enlarged by one hundred and the
luminosity uncertainty significantly reduced~\cite{lumiConf}
from $11$\,\% to \dlumi\,\%.
The CMS Collaboration has updated their initial measurement
of total $W$ and $Z$ cross sections~\cite{Khachatryan:2010xn} to include
data corresponding to an integrated luminosity similar to that used
here~\cite{:2011nx}. Similar measurements have been performed at the
$p\bar{p}$ collider TeVatron by the CDF and D0
collaborations~\cite{Abulencia:2005ix, Abbott:1999tt}.

The presented cross section values are integrated over the 
fiducial region of the analysis and also extrapolated to the
full kinematic range.
The data are also reported differentially, as functions of the
lepton pseudorapidity~\footnote{ATLAS uses a right-handed coordinate system 
with its origin at the nominal interaction point (IP) in the centre of the detector 
and the $z$-axis along the beam pipe. The $x$-axis points from the IP to the centre 
of the LHC ring, and the $y$ axis points upward. Cylindrical coordinates $(r,\phi)$ 
are used in the transverse plane, $\phi$ being the azimuthal angle around the beam 
pipe. The pseudorapidity is defined in terms of the polar angle $\theta$ as 
$\eta=-\ln\tan(\theta/2)$. 
Distances are measured as $\Delta R = \sqrt{\Delta \eta^2 + \Delta \phi^2}$.}, 
$\eta_l$, for the $W^{\pm}$ cross sections,
and of the boson rapidity, $y_Z$, for the $Z/\gamma^*$ cross section.
For the ``$\Zg$'' case, which will subsequently often be denoted simply as ``$Z$'',
all values refer to the dilepton mass window from $66$ to $116$\,GeV.
The $Z$ cross section measurement in the electron channel is
significantly extended by the inclusion of the forward
detector region, which allows the upper limit of the
pseudorapidity range for one of the electrons to be increased 
from $2.47$~\cite{Aad:2010yt} to $4.9$.

The electron and muon $W^{\pm}$ and $Z$ cross sections are combined
 to form a single joint measurement taking into account the systematic error
correlations between the various data  sets.
This also leads to an update of the initial differential measurement of the
$W$ charge asymmetry published by  ATLAS~\cite{Aad:2011yn}.
Normalised cross sections as function of the $Z$ boson rapidity and
$W$ boson and lepton charge asymmetry measurements have been performed also by the
CMS~\cite{Chatrchyan:2011wt, Chatrchyan:2011jz} and the CDF and D0
collaborations~\cite{%
Aaltonen:2010zza, 
Abazov:2007jy,  
Aaltonen:2009ta, 
Acosta:2005ud, 
Abazov:2007pm, 
Abazov:2008qv}. 

The combined $W^{\pm}$ and $Z$ cross sections, integrated
and differential, are compared with QCD predictions based on recent 
determinations of the parton distribution functions of the proton. 
In view of the per cent level precision of the measurements,
such comparisons are restricted to PDFs obtained to NNLO.

A brief overview of the ATLAS detector, trigger and simulation and the analysis 
procedure are presented in Sec.~\ref{sec:datasimul}. 
The acceptance corrections and their uncertainties are discussed in Sec.~\ref{sec:accext},
while Sec.~\ref{sec:seleffbkg} presents the selection, the efficiencies and the backgrounds for both 
electron and muon channels.
The cross section results are first given, in Sec.~\ref{sec:crosseclep}, 
separately for each lepton flavour.
In Sec.~\ref{sec:combicross} the $e$ and $\mu$ data sets are combined 
and the results are compared to theoretical predictions.
The paper is concluded with a brief summary of the results.

\section{Data and Simulation} 
\label{sec:datasimul}

\subsection{ATLAS Detector}
\label{sec:detector}

The ATLAS detector~\cite{DetectorPaper:2008}  comprises a
superconducting solenoid surrounding the inner detector (ID) and a
large superconducting toroid magnet system enclosing the calorimeters.
The ID system is immersed in a $2$\,T axial magnetic 
field and provides  tracking information for charged particles in a pseudorapidity
range matched by the precision measurements of the electromagnetic 
calorimeter. The silicon pixel and strip (SCT) tracking detectors 
cover the pseudorapidity range $|\eta|< 2.5$.
The Transition Radiation Tracker (TRT), which surrounds the silicon detectors, 
enables tracking  up to $|\eta| = 2.0$ and contributes to
electron identification.

The liquid argon (LAr) electromagnetic (EM) calorimeter
is divided into one barrel ($|\eta| < 1.475$) and 
two end-cap components ($1.375 < |\eta| < 3.2$, EMEC). It uses an accordion geometry to 
ensure fast and uniform response and fine segmentation
for optimum reconstruction and identification of electrons and photons.
The  hadronic scintillator tile calorimeter
consists of a barrel covering the region $|\eta| < 1.0$, 
and two extended barrels in the range $0.8 < |\eta| < 1.7$.  The LAr Hadronic 
End-cap Calorimeter (HEC) 
($1.5<|\eta|<3.2$) is located behind the 
end-cap electromagnetic calorimeter. The Forward Calorimeter (FCal) covers the
range $3.2< |\eta| < 4.9$ and also
uses LAr as the active material.

The muon spectrometer (MS) is based on three large superconducting
toroids with coils arranged in an eight-fold symmetry around
the calorimeters, covering a range of $|\eta|<2.7$.
Over most of the $\eta$ range, precision measurements
of the track coordinates in the principal bending direction of the magnetic field are
 provided by Monitored Drift Tubes (MDTs). At large pseudorapidities ($2.0 < |\eta| < 2.7$), 
Cathode Strip Chambers (CSCs) with higher granularity are used in the innermost station.
The muon trigger detectors consist of  Resistive Plate Chambers (RPCs) in the barrel ($|\eta|<1.05$) and 
Thin Gap Chambers (TGCs) in the end-cap regions ($1.05 < |\eta| < 2.4$), with a 
small overlap in the $|\eta| \simeq $1.05 region.

The ATLAS detector has a three-level trigger system consisting of Level-1 (L1), Level-2 (L2)
and the Event Filter (EF). The L1 trigger rate at design luminosity is approximately 75\,kHz.
The L2 and EF triggers reduce the event rate to approximately 200~Hz before data transfer to
mass storage.

\subsection{Triggers}
\label{sec:data}

The analysis uses data taken in the year 2010 with proton beam
energies of $3.5$\,TeV. For the electron channels
the luminosity is \lumieN. 
For the muon channels the luminosity is smaller, \lumimuN,
as a fraction of the available data, 
where the muon trigger conditions varied too rapidly, 
is not included

Electrons are triggered in the pseudorapidity range $|\eta_e|<$ 2.5,
where the electromagnetic calorimeter is finely segmented. A single
electron trigger with thresholds in transverse energy of $10$\,GeV at
L1 and $15$\,GeV at the higher trigger levels is used for the main
analysis. Compact electromagnetic energy depositions triggered at
L1 are used as the seed for the higher level trigger algorithms, which
are designed for identifying electrons based on calorimeter and fast
track reconstruction.

The electron trigger efficiency is determined 
from $W \rightarrow e\nu$ and $Z \rightarrow ee$ events 
as the fraction of triggered electrons
with respect to the offline reconstructed signal~\cite{Aad:2011mk}.
The  efficiency is found
to be close to $100$\,\%, being constant in 
both the transverse energy $E_T$ and the pseudorapidity $\eta_e$, with  
a small reduction by about $2$\,\% towards the limits of the fiducial
region ($E_T = 20$\,GeV and $|\eta_e| = 2.5$, see Sec.~\ref{sec:sigdef}). 
A systematic uncertainty of $0.4$\,\% is assigned to the efficiency determination.

The muon trigger is based at L1 on a coincidence of layers of RPCs in the 
barrel region and TGCs in the end caps.
The parameters of muon candidate tracks are then derived 
by fast reconstruction algorithms in both inner detector and muon spectrometer.
Events are triggered with a single muon trigger with an EF threshold of transverse momentum $p_T$ = 13\,GeV.

The muon trigger efficiency is determined from a 
study of $Z \rightarrow \mu \mu$ events.
The average efficiency is measured to be $85.1$\,\% with a total uncertainty of $0.3$\,\%.
The lower efficiency of the muon trigger system is due to the reduced geometrical
acceptance in the barrel region.

\subsection{Simulation}
\label{sec:simulation}

The properties of both signal and background
processes, including acceptances and efficiencies,
are modelled 
using the \Mcatnlo~\cite{mcatnlo},
\Powheg~\cite{Nason:2004rx,Frixione:2007vw,Alioli:2010xd,Alioli:2008gx},
\Pythia~\cite{pythia} and \Herwig~\cite{herwig} Monte Carlo (MC) programs. 
All generators are interfaced to \Photos~\cite{Golonka:2005pn} to simulate the effect of final state QED radiation.
The response of the ATLAS detector to the 
generated particles is modelled using GEANT4~\cite{:2010wqa, geant4}. 
The CTEQ 6.6 PDF set~\cite{Nadolsky:2008zw} is used for the \Mcatnlo\ and \Powheg\ samples. 
For the \Pythia\ and \Herwig\ samples the MRST LO$^*$~\cite{mrst} parton distribution 
functions are used. MC parameters describing the properties of minimum bias events and the
underlying event are tuned to the first ATLAS measurements~\cite{mbtune}.
Furthermore, the simulated events are reweighted so that the resulting transverse momentum distributions 
of the $W$ and $Z$ bosons match the data~\cite{Aad:2011gj, Collaboration:2011fp}.

The effect of multiple $pp$ interactions per bunch crossing (``pile-up'') 
is modelled by overlaying simulated minimum bias events over the 
original hard-scattering event. MC events are then reweighted so that 
the reconstructed vertex distribution agrees with the data.

The Monte Carlo simulation is also corrected with respect to the data
in the lepton reconstruction and identification efficiencies 
as well as in the energy (momentum) scale and resolution.

\TTab~\ref{tab:samples} summarises the information
on the simulated event samples used for the measurement, including the cross sections
used for normalisation. 
The $W$ and $Z$ samples are
normalised to the NNLO cross sections from the FEWZ
program~\cite{Gavin:2010az, Aad:2010yt}.
The uncertainties on those cross sections arise from the choice of
PDF, from factorisation and renormalisation scale dependence
and from the $\alpha_s$ uncertainty. An uncertainty of $(+7,-10)$\,\% is taken
for the $t\bar{t}$ cross section \cite{Moch:2008ai, Langenfeld:2009tc,Aad:2011yb}.

\subsection{Analysis Procedure}
\label{sec:sigdef}

The integrated and differential $W$ and $Z$ production cross sections
are measured in the fiducial volume of the ATLAS detector using the equation
\begin{equation} 
 \sigma_{\rm fid} = \frac{N - B}{\CWZ \cdot L_{\rm int}} \,,
\label{eq:WZxsecfid} 
\end{equation} 
\noindent where
$N$ is the number of candidate events observed in data,
$B$  the number of background events, determined
using data and simulation, and $L_{\rm int}$  the integrated luminosity 
corresponding to the run selections and trigger employed. 
The correction by the efficiency
factor $\CWZ$ determines the cross sections
$\sigma_{\rm fid}$ within the fiducial regions of the measurement.
These regions are defined as

\noindent\begin{tabular}{ll}
  $W \to e\nu$\,:\, & $p_{T,e} > 20\ \GeV\,,\; |\eta_{e}| < 2.47\,,$ \\
             & \mbox{excluding}\; $1.37<|\eta_{e}| < 1.52\,,$\\
             & $p_{T,\nu} > 25\ \GeV\,,\; m_T > 40\ \GeV\,;$ \\
  $W \to \mu\nu$\,:\, & $p_{T,\mu} > 20\ \GeV\,,\; |\eta_{\mu}| < 2.4\,,$\\
               & $p_{T,\nu} > 25\ \GeV\,,\; m_T > 40\ \GeV\,;$ \\
  $Z \to ee$\,:\, & $p_{T,e} > 20\ \GeV\,,\; \mbox{both}\; |\eta_{e}| < 2.47\,,$\\
              & \mbox{excluding}\; $1.37<|\eta_{e}| < 1.52\,,$\\
              & $66 < m_{ee}< 116\,\mathrm{GeV}\,;$\\
  \mbox{Forward} \; $Z \to ee$\,:\, & $p_{T,e} > 20\ \GeV\,,\; \mbox{one}\; |\eta_{e}| < 2.47\,,$\\
                               & \mbox{excluding}\; $1.37<|\eta_{e}| < 1.52\,,$\\
                               & \mbox{other}\; $2.5<|\eta_{e}| < 4.9\,,$\\
                               & $66 < m_{ee}< 116\,\mathrm{GeV}\,;$\\
  $Z \to \mu\mu$\,:\, & $p_{T,\mu} > 20\ \GeV\,,\; \mbox{both}\; |\eta_{\mu}| < 2.4\,,$ \\
                 & $66 < m_{\mu\mu}< 116\,\mathrm{GeV}\,.$ \\
\end{tabular}

For the $W$ channels the transverse mass, $m_T$, is defined as $m_T =
\sqrt{2 p_{T,\ell} p_{T,\nu} \cdot (1 -
  \cos{\Delta\phi_{\ell, \nu}})}$, where $\Delta\phi_{\ell, \nu}$ is the azimuthal
separation between the directions of the charged lepton and the neutrino.

The main analysis, used to determine the integrated cross sections,
is performed for the $W$ and $Z$ electron and muon decay channels for leptons in the
central region of the detector of $|\eta_e| < 2.47$ and  $|\eta_\mu| <2.4$, respectively. 
A complementary analysis of the $Z \rightarrow ee$ channel is used in addition
to measure the differential cross section at larger rapidity. Here the 
allowed pseudorapidity range 
is chosen from  $|\eta_e| = 2.5$ to $4.9$ for one of the electrons.

The differential cross sections are measured, as a function of the absolute values
of the $W$ decay lepton pseudorapidity and $Z$ boson rapidity, in bins with boundaries at

\noindent\begin{tabular}{ll}
   $\eta_\ell\,=\,$ & $[\,0.00\,,\,0.21\,,\,0.42\,,\,0.63\,,\,0.84\,,\,1.05\,,\,1.37\,,\,1.52\,,$\\
                   & $\,1.74\,,\,1.95\,,\,2.18\,,\,2.47\,(e)\,\,\mathrm{or}\,\,2.40\,(\mu)\,]\,;$ \\
   $y_Z\,=\,$      & $[\,0.0\,,\,0.4\,,\,0.8\,,\,1.2\,,\,1.6\,,\,2.0\,,\,2.4\,,\,2.8\,,\,3.6\,]\,,$\\ 
\end{tabular}
\noindent where the notation for absolute $\eta$ and $y$ is omitted.

\begin{table*}[ht]
  \begin{center}
    \begin{tabular}{lrrr}
      \hline
      \hline
      \raisebox{-0.4ex}{Physics process}         & Generator & $\sigma \cdot$ BR [nb] &  \\
      \hline
      $W^+ \to \ell^+\nu$ ($\ell=e,\mu$)              & \Mcatnlo  & 6.16$\pm$0.31     & NNLO \\
      $W^- \to \ell^-\bar{\nu}$ ($\ell=e,\mu$)              & \Mcatnlo  & 4.30$\pm$0.21     & NNLO \\
      $\Zg \to \ell\ell$ ~~($m_{\ell\ell}>60$~GeV, $\ell=e,\mu$)  & \Mcatnlo  & 0.99$\pm$0.05      & NNLO\\
      \hline
      $W \rightarrow \tau \nu$                    & \Pythia  & 10.46$\pm$0.52     & NNLO \\
      $\Zg \to \tau \tau$ ~~($m_{\tau\tau}>60$~GeV) & \Pythia  & 0.99$\pm$0.05      & NNLO \\
      $t\bar{t}$                                  & \Mcatnlo &
      0.165$^{+0.011}_{-0.016}$      & ~$\approx\,$NNLO \\
      $WW$                                        & \Herwig  & 0.045$\pm$0.003    & NLO \\
      $WZ$                                        & \Herwig  & 0.0185$\pm$0.0009  & NLO \\
      $ZZ$                                        & \Herwig  & 0.0060$\pm$0.0003  & NLO \\
      \hline
      Dijet ($e$ channel, $\hat{p}_{\mathrm{T}}>15$~GeV) 
                                                  & \Pythia  & 1.2 $\times 10^6$  & LO \\

      Dijet ($\mu$ channel, $\hat{p}_{\mathrm{T}}> 8$~GeV)        
                                                  & \Pythia  & 10.6 $\times 10^6$  & LO \\
      $b\overline{b}$ ($\mu$ channel, $\hat{p}_{\mathrm{T}}>18$~GeV, $p_{\rm{T}}(\mu) >$ 15 GeV) 
                                                  & \Pythia  & $73.9$             & LO \\
      $c\overline{c}$ ($\mu$ channel, $\hat{p}_{\mathrm{T}}>18$~GeV,
      $p_{\rm{T}}(\mu) >$ 15 GeV)                   &  \Pythia & 28.4               & LO \\
      \hline                 
      \hline
    \end{tabular}
    \caption{\it Signal and background Monte Carlo samples as well as
      the generators used in the simulation. For each sample the
      production cross section, multiplied by the relevant branching
      ratios (BR), to which the samples are normalised, is given. The
      electroweak $W$ and $Z$ cross sections are calculated at
      NNLO in QCD, $t\bar{t}$ at approximate NNLO and dibosons at NLO in QCD. The inclusive
      jet and heavy quark cross sections are given at leading order
      (LO). These samples are generated with requirements on the
      transverse momentum of the partons involved in the
      hard-scattering process, $\hat{p}_{\mathrm{T}}$. No systematic
      uncertainties are assigned for the jet and heavy-quark cross sections, 
      since methods are used to extract their normalisation and their
      systematic uncertainties from data (see text).
    }
  \label{tab:samples}
  \end{center}
\end{table*}

The combined efficiency factor $\CWZ$~is calculated from simulation
and corrected for differences in reconstruction, identification
and trigger efficiencies between data and simulation 
(see Sec.~\ref{sec:seleffbkg}). 
Where possible, efficiencies in data and MC are derived from $Z \rightarrow \ell\ell$
and, in the case of the electron channel, $W \to e\nu$
events~\cite{Aad:2011mk,muEffConf}. The efficiency estimation is
performed by triggering and selecting such events with good purity
using only one of the two leptons in the $Z \rightarrow \ell\ell$ case
and a significant missing transverse energy in the $W \to e\nu$ case, 
a procedure often referred to as ``tagging''.
Then the other very loosely identified lepton
can be used as a probe to estimate various efficiencies after
appropriate background subtraction. The method is therefore often
referred to as the ``tag-and-probe'' method.

The total integrated cross sections are measured using the equation
\begin{equation} 
 \sigma_{\rm tot} = \sigma_{W/Z} \times BR(W/Z \to \ell\nu/\ell\ell) = \frac{\sigma_{\rm fid}}{\AWZ}\,, 
\label{eq:WZxsectot} 
\end{equation} 
where the acceptance $\AWZ$ is used to extrapolate the
cross section measured in the fiducial volume, $\sigma_{\rm fid}$, to the full kinematic region.
The acceptance is derived from MC, and the uncertainties on the simulation modeling 
and on parton distribution functions constitute an additional uncertainty 
on the total cross section measurement.
The total and fiducial
cross sections are corrected for QED radiation effects in the final
state.

The correction factors \CWZ\  and \AWZ\  are obtained as follows
\begin{equation}
  \CWZ = \frac{N_{\rm MC, rec}}{N_{\rm MC, gen, cut}}
  \:\:\:\:\:\:\mbox{and}\:\:\:\:\:\: \AWZ = \frac{N_{\rm MC, gen, cut}}{N_{\rm MC, gen, all}}\,,
\end{equation}
where $N_{\rm MC, rec}$ are sums of weights of events after simulation, reconstruction and
selection, $N_{\rm MC, gen, cut}$ are taken at generator level after
fiducial cuts and $N_{\rm MC, gen, all}$ are the sum of weights of all
generated MC events (for the \Zg\ channels within $66 < m_{\ell\ell} < 116$\,GeV).

For the measurement of charge-separated $W^\pm$ cross sections, the
\CW\ factor is suitably modified to incorporate a correction for event
migration between the $W^+$ and $W^-$ samples as
\begin{equation}
  C_{W+} = \frac{N_{\rm MC, rec+}}{N_{\rm MC, gen+, cut}}
  \:\:\:\:\:\:\mbox{and}\:\:\:\:\:\: C_{W-} = \frac{N_{\rm MC, rec-}}{N_{\rm MC, gen-, cut}}\,,
\end{equation}
where $N_{\rm MC, rec\pm}$ and $N_{\rm MC, gen\pm, cut}$ are sums of weights of
events reconstructed or generated as $W^\pm$, respectively, without
any further charge selection. For example, $N_{\rm MC, rec+}$ includes a small
component of charge misidentified events generated as $W^-$, while 
$N_{\rm MC, gen+, cut}$ contains only events generated as $W^+$ without
requirements on the reconstructed charge. This charge
misidentification effect is only relevant for the electron channels,
and is negligible in the muon channels.

Electron and muon integrated measurements are combined after extrapolation to the 
full phase space available for $W$ and $Z$ production and decay
and also to a common fiducial region, chosen to minimise the extrapolation needed 
to adjust the electron and muon cross sections to a common basis.
This kinematic region is defined extrapolating both channels to
$|\eta_{\ell}|<2.5$ and interpolating the electron measurement over the
region $1.37<|\eta_{e}| < 1.52$. The differential cross sections are
combined extrapolating all $Z$ measurements to full phase space in
lepton pseudorapidity accessible in $Z$ production and decay and extending the range of the most forward bin
of $W$ measurements to $2.18<|\eta_{\ell}|<2.5$. The experimental selections
on the transverse momenta of the leptons and on the transverse or invariant
mass are retained for the differential cross sections.

\section{Acceptances and Uncertainties}
\label{sec:accext}

The acceptances \AWZ\ are determined using the \Mcatnlo\ Monte Carlo program and the CTEQ 6.6 PDF set. 
The central values and their systematic uncertainties are listed in \Tab~\ref{tab:acc}, separately for $W^+$, $W^-$, 
$W^{\pm}$ and $\Zg$ production. The uncertainties due to the finite statistics of the Monte Carlo samples are negligible.
The systematic uncertainties are obtained by combining four different components:

\begin{itemize}
\item The uncertainties within one PDF set ($\delta A^\mathrm{pdf}_\mathrm{err}$). 
They are derived from the CTEQ 6.6 PDF~\cite{Nadolsky:2008zw} eigenvector error sets at the 90\% C.L. limit. 

\item The uncertainties due to differences between PDF sets ($\delta A^\mathrm{pdf}_\mathrm{sets}$). 
They are estimated as the maximum difference between the CTEQ 6.6,
  ABKM095fl~\cite{ABKM09, Alekhin:2010iu}, HERAPDF
  1.0~~\cite{HERA:2009wt}, MSTW2008~\cite{Martin:2009iq}, CT10,
  CT10W~\cite{Lai:2010vv} and NNPDF2.1~\cite{Ball:2011mu} sets, where 
 samples generated with CTEQ 6.6 are reweighted event by event to other PDFs~\cite{Bourilkov:2006cj}.

\item  The uncertainties due to the modelling of the hard-scattering processes of $W$ and $Z$ production ($\delta A_\mathrm{hs}$). These are derived from comparisons of \Mcatnlo\ and \Powheg\ simulations, using the CTEQ 6.6 PDF set and the parton shower and hadronisation models based on the \Herwig\ simulation.

\item  The uncertainties due to the parton shower and hadronisation description ($\delta A_\mathrm{ps}$). These are derived as the difference in the acceptances calculated with \Powheg\ Monte Carlo, using the CTEQ 6.6 PDF set but different models for parton shower and hadronisation descriptions, namely the \Herwig\ or \Pythia\ programs.
\end{itemize}

In addition, to compute the total cross section ratios (see
\Sec~\ref{sec:ratios}), the correlation coefficients between the full $W$ and
$Z$ acceptance uncertainties are used.
They are 0.80 for $W^{\pm}-Z$, 0.83 for $W^--Z$, 0.78 for $W^+-Z$ and 0.67 for $W^+-W^-$.

\begin{table}
\centering
\begin{tabular}{m{1cm}m{1cm}ccccc}
  \hline
  \hline
   & \multicolumn{1}{c}{$A$} 
   & \: $\delta A^\mathrm{pdf}_\mathrm{err}$  \:
   & \: $\delta A^\mathrm{pdf}_\mathrm{sets}$   \:
   & \: $\delta A_\mathrm{hs}$   \:
   & \: $\delta A_\mathrm{ps}$   \:
   & \: $\delta A_\mathrm{tot}$   \: \\
  \hline
  \multicolumn{7}{c}{Electron channels} \\
  \hline
  $W^{+}$ & 0.478 & 1.0 & 0.7 & 0.9 & 0.8 & 1.7 \\
  $W^{-}$ & 0.452 & 1.5 & 1.1 & 0.2 & 0.8 & 2.0 \\
  $W^{\pm}$
         & 0.467 & 1.0 & 0.5 & 0.6 & 0.8 & 1.5 \\
  $Z$      & 0.447 & 1.7 & 0.6 & 0.2 & 0.7 & 2.0 \\
  \hline
  \multicolumn{7}{c}{Muon channels} \\
  \hline
  $W^{+}$ & 0.495 & 1.0 & 0.8 & 0.6 & 0.8 & 1.6 \\
  $W^{-}$ & 0.470 & 1.5 & 1.1 & 0.3 & 0.8 & 2.1 \\
  $W^{\pm}$
         & 0.485 & 1.0 & 0.5 & 0.4 & 0.8 & 1.5 \\
  $Z$      & 0.487 & 1.8 & 0.6 & 0.2 & 0.7 & 2.0 \\
  \hline
  \hline
\end{tabular}
\caption{\it Acceptance values ($A$) and their relative uncertainties
  ($\delta A$) in percent for $W$ and $Z$ production in electron and muon channels.
  The various components of the uncertainty are defined in the text. The total uncertainty
  ($\delta A_{tot}$) is obtained as the quadratic sum of the four parts.
}
\label{tab:acc}
\end{table}

The corrections, and their uncertainties, to extrapolate the electron 
and the muon measurements from each lepton fiducial region to the common
fiducial region, where they are combined, are calculated
with the same approach as described for the acceptances. The extrapolations 
contribute $\sim$3\% to the \Wmn\ and  $\sim$7\% to the \Wen\ cross sections. 
Similarly, the fiducial measurement of the $Z$ cross section is enhanced by
$\sim$5\% in the muon channel and by $\sim$12\% in the electron channel.
 The uncertainties on these corrections are found to be on the 0.1\,\% level.
The combined fiducial measurements are therefore characterised by 
negligible theoretical uncertainty due to the extrapolation to the unmeasured phase space.

The differential cross sections for the electron and the muon channels are also combined 
after extrapolating each measurement to the common fiducial kinematic region. 
In the case of the $W$ measurements the applied correction is effective only 
in the highest $\eta_\ell$ bin and is about 30\% in the muon channel 
and about 9\% in the electron channel. The extrapolation factors needed to combine 
the $Z$ electron and muon measurements, and their systematic 
uncertainties, are listed in \Tab~\ref{tab:extr}.
The uncertainty is of the order of 0.1\,\% in most of the rapidity
intervals and increases to 1-2\% near the boundary of the measurement
fiducial regions.

\begin{table}
\small
\centering
\begin{tabular}{ccccc}
  \hline
  \hline
     $y_{Z}^{min}$ & $y_{Z}^{max}$         &  \Zmm     &   Central \Zee   & Forward \Zee \\
  \hline
   0.0 & 0.4 &    1.000(0)      &  0.954(1)     &    -      \\
   0.4 & 0.8 &    1.000(0)      &  0.903(1)      &    -      \\
   0.8 & 1.2 & 0.984(1) &  0.855(2)      &    -      \\
   1.2 & 1.6 & 0.849(2)  &  0.746(3)      & 0.103(1)  \\ 
   1.6 & 2.0 & 0.578(5)  &  0.512(4)      & 0.327(3)  \\
   2.0 & 2.4 & 0.207(5)  &  0.273(5)      & 0.590(7)  \\
   2.4 & 2.8 &   -       &    -           & 0.797(1) \\
   2.8 & 3.6 &   -       &    -           & 0.404(4)  \\
  \hline
  \hline
\end{tabular}
\caption{\it Central values and absolute uncertainties (in
  parenthesis) of extrapolation correction factors 
from fiducial regions to full lepton pseudorapidity $\eta$ phase space. The factors are provided in bins 
of $Z$ boson rapidity for \Zmm\ and for central and forward \Zee\ measurements.}
\label{tab:extr}
\end{table}

\section{Event Selection, Efficiencies and Background Determination}
\label{sec:seleffbkg}

\subsection{Electron Channels}
\label{sec:eleeffbgd}

\paragraph{Event Selection:}

Events are required to have at least one primary vertex formed by at
least three tracks. To select $W$ boson events in the electron
channel, exactly one well reconstructed electron is required with $E_T
>$ 20 GeV and $|\eta|<$ 2.47. Electrons in the
transition region between barrel and end-cap calorimeter,
1.37$<|\eta|<$1.52, are excluded, as the reconstruction quality is
significantly reduced compared to the rest of the pseudorapidity
range. The transverse energy is calculated from
calorimeter and tracker information. The electron is required to pass 
``medium'' identification criteria~\cite{Aad:2011mk}. To
reject efficiently the QCD background, the electron track must 
in addition have a hit in the innermost layer of the
tracking system, the ``pixel b-layer''. The additional calorimeter energy
deposited in a cone of size $\Delta R \leq 0.3$ around the electron cluster
is required to be small, where the actual selection is optimised as a
function of electron $\eta$ and $p_T$ to
have a flat 98\% efficiency in the simulation for isolated electrons from the decay
of a $W$ or $Z$ boson. The missing transverse energy, \met, is
determined from all measured and identified physics objects, as well as
remaining energy deposits in the calorimeter and tracking information~\cite{Aad:2011re}. 
It is required to be larger than $25$\,GeV. Further, the transverse mass,
$m_T$, has to be larger than $40$\,GeV.

The selection as described is also used for the $Z$ boson case with the following 
modifications: instead of one, two electrons are required to be reconstructed 
and pass the ``medium'' criteria without the additional  ``pixel b-layer'' and
isolation cuts; their charges have to be opposite,
and their invariant mass has to be within the interval $66$ to $116$\,GeV. 

For the selection of $Z$ events at larger rapidities, a
central electron passing ``tight''~\cite{Aad:2011mk} criteria as
well as the calorimeter isolation requirement described above for the
$W$ channel is required. A second electron candidate with $E_T > 20$\,GeV has to
be reconstructed in the forward region, $2.5 \leq |\eta| \leq 4.9$,
and to pass ``forward loose'' identification
requirements~\cite{Aad:2011mk}. Its transverse energy is determined
from the calorimeter cluster energy and position. As the forward
region is not covered by the tracking system, no charge can be
measured and the electron identification has to rely on calorimeter cluster
shapes only. The invariant mass of the selected pair is required to be
between $66$ and $116$\,GeV.

\paragraph{Calibration and Efficiencies:}

Comprehensive studies of the electron performance
are described in~\cite{Aad:2011mk}.
Energy scale and resolution corrections are determined from data as a function 
of $\eta$ in the central and forward region, by comparing the
measured \Zee\ line shape to the one predicted by the simulation.
For the central region, the linearity and
resolution are in addition cross checked using $J/\psi\rightarrow ee$
and single electron $E/p$ measurements in \Wen\ events.

The electron efficiencies are evaluated in two steps called
reconstruction and identification. The reconstruction step consists of
the loose matching of a good quality track to a high $p_T$ calorimeter
cluster. Identification summarises all the further requirements to
reduce the background contamination.

The electron reconstruction efficiency in the central region is
obtained from the $Z$ tag-and-probe method. The efficiency in data is
found to be slightly higher by $1.3\%$ than in MC, and the simulation is
adjusted accordingly with an absolute systematic uncertainty of
$0.8\%$.

The identification efficiency for electrons from $W$ or $Z$ decay in
the central region is determined using two different
tag-and-probe methods, which are performed on selected $W$ and $Z$
data samples, respectively. The $W$-based determination employs the
significant missing transverse energy in those events to obtain
an unbiased electron sample. The method benefits from larger statistics but
needs more involved procedures for background subtraction, as compared to
the $Z$-related determination. Consistent correction factors to be
applied to the simulation are derived from the two methods as a
function of the electron rapidity. For the ``medium'' identification
criteria, the Monte Carlo efficiency is adjusted by about $-2.5\%$ on
average, with a resulting absolute uncertainty of typically less than
$1$\,\% on this correction. 
The quality of the data to MC agreement in the ``tight''
identification criteria efficiency
is found to depend significantly on electron $\eta$,
and an adjustment by
on average $+2\%$ with an absolute uncertainty of about $1$\,\% is performed. 
The additional requirements on b-layer hits and calorimeter isolation
are found to be very efficient and rather well described in the
simulation, resulting in small adjustments and small systematic 
uncertainties only.

To distinguish $W^+$ from $W^-$ events, the charge
of the decay electron has to be known. The charge misidentification
probability as a function of $\eta$ is determined from a sample of $Z \rightarrow ee $ events
where both electrons are reconstructed with the same sign.
 It depends on the identification criteria and in
general increases at large $|\eta|$. For electrons passing the
``medium'' criteria, about 1\% of all electrons are assigned the wrong
charge, while for ``tight'' electrons this figure is about halved. 
From these measurements, additional uncertainties are
derived from the opposite charge requirement on the $Z$ cross section ($0.6\%$) and
from migration and charge dependent effects on the $W^+$ and $W^-$
cross sections (0.1\%).

In the forward region ($|\eta| > 2.5$), the electron reconstruction
is nearly 100\% efficient and taken from MC. The
identification efficiency is determined using the $Z$ tag-and-probe
method in two forward electron rapidity bins, which
correspond to the inner part of the EMEC ($2.5 < |\eta| < 3.2$) and
the FCal ($3.2 < |\eta| < 4.9$), respectively. 
The simulation overestimates the efficiency by $8.4\%$ and $1.7\%$
in these two bins and is adjusted accordingly, with absolute uncertainties of
$5.8\%$ and $8.8\%$, respectively.

\paragraph{Background Determination:}

The largest electroweak background in the \Wen\ channel is given by
the $W\to \tau \nu$ production, mainly from decays involving true electrons, $\tau \to
e\bar{\nu}_e\nu_\tau$. Relative to the number of all $W^\pm$ candidate
events, this contribution is estimated to be $2.6\%$. The background
from $t\bar{t}$ events is determined to be $0.4\%$ and further contributions on
the $0.1-0.2\%$ level arise from $\ensuremath{Z}\rightarrow \tau \tau$, \Zee\
and diboson production. The sum of electroweak and $t\bar{t}$
backgrounds are found to be $3.7\%$ in the $W^-$ and $3.2\%$ in the
$W^+$ channel of the respective numbers of events.

A further significant source of background in the \Wen\ channel,
termed ``QCD background'', is given by jet production
faking electron plus missing transverse energy final states.
The QCD background is derived from the data using a template fit of 
the \met\ distribution in a control sample selected without 
\met\ requirement and inverting a subset of the electron
identification criteria. The \met\ template for the signal and the
other electroweak and $t\bar{t}$ backgrounds are taken from the
simulation.
The QCD background in the signal region
is determined to be $3.4\%$ and $4.8\%$ for the $W^+$ and $W^-$
channels, respectively. The statistical uncertainty of this fit is
negligible. The background as well as the signal templates are varied
to assess the systematic uncertainty on the fraction of QCD
background. The relative uncertainty is estimated to be $12\%$ for $W^+$ and
$8\%$ for $W^-$, corresponding to a fraction of about $0.5\%$ of the
$W^+$ or $W^-$ candidates. The fit is performed in each bin
of electron pseudorapidity separately to obtain the background for the
differential analysis.

The relative background contributions in the central \Zee\ analysis due
to electroweak processes, \Wen, $Z \rightarrow \tau\tau$ and $W\to
\tau \nu$, and to $t\bar{t}$ production are estimated using the
corresponding MC samples to be $0.3\%$ in total. The fraction of
candidate events due to diboson decays is $0.2\%$.

The QCD background in the central \Zee\ analysis is estimated from
data by fitting the invariant mass distribution using a background
template selected with inverted
electron identification cuts and the signal
template from MC. This procedure yields a fraction of QCD background
of $1.6\%$. The relative systematic uncertainty on this fraction is
dominant and
evaluated to be $40\%$ using different background templates and fit
ranges, as well as an alternative method based on fitting a sample selected 
with looser identification criteria. For the differential analysis, the sum of 
background is determined from the global
fit, and the relative contributions of each bin are taken from the background
template. Differences between templates lead to further relative 25\%
bin-to-bin uncorrelated uncertainties on the QCD background fraction.

In the forward \Zee\ analysis the main electroweak background comes from 
\Wen\ events with an associated jet faking an electron in the forward region. It is estimated to be
$1.9\%$. The QCD background is estimated by fitting the $m_{ee}$
distribution in a similar manner as for the central analysis. Due to
the larger level of background the fit can be performed directly in all
boson rapidity $y_Z$ bins. In total the QCD background is estimated to
be $9.4\%$ with relative statistical and systematic uncertainties of
$8\%$ and $17\%$. Differentially the QCD background fraction varies from $7\%$ to
$20\%$ with typical relative total uncertainties of $20\%$ to $40\%$.

\subsection{Muon Channels}
\label{sec:mueffbgd}

\paragraph{Event Selection:}
Collision events are selected with the same vertex requirement as for
the electron channels. In addition, the vertex with the
highest squared transverse momentum sum of associated tracks is selected as the
primary vertex for further cuts. To reduce fake collision candidates from cosmic-ray or
beam-halo events, the position of the primary vertex along the beam axis is required to be
within 20~cm of the nominal position. The efficiency of this
requirement is larger than 99.9\% in both data and simulation.

Muon track candidates are formed from pairs of stand-alone tracks in the 
inner detector and the muon spectrometer, combined using a chi-square matching procedure \cite{ATLAS-CONF-2010-036}.
$W$ and $Z$ events are selected requiring at least one or two combined track 
muons with $p_T > 20$\,GeV and $|\eta| < 2.4$, respectively. 
The $z$ position of the muon track extrapolated to the
beam line has to match the $z$ coordinate of the primary vertex within
$\pm~1$\,cm. A set of ID hit requirements \cite{muEffConf} is applied to 
select high quality tracks also demanding at least one hit in
the ``pixel b-layer''.

A track-based isolation criterion is defined requiring the sum of transverse 
momenta, $\sum p_T^{ID}$, of ID tracks with $p_T > 1$~GeV within a cone $\Delta R <0.2$ 
around the muon direction, divided by the muon transverse momentum $p_T$, to be less than $0.1$.
When analysed after all other selection cuts, this requirement  
has a high QCD background rejection power, while keeping more than 
$99$\,\% of  the signal events in both the $W$ and $Z$ channels.

\Wmn\ events are further selected requiring the missing transverse energy, 
defined as in the electron analysis, to be larger than $25$\,GeV 
and the transverse mass to be larger than $40$\,GeV.
In the \Zmm\ analysis, the two decay muons are required 
to be of opposite charge, and the invariant mass of the $\mu^+\mu^-$
pair to be within the interval $66$ to $116$\,GeV.

\paragraph{Calibration and Efficiencies:}

Muon transverse momentum resolution corrections are determined comparing data and MC as a function of $\eta$
in barrel and end-cap regions \cite{muResolConf}. 
They are derived by fitting the invariant mass distribution from \Zmm\ events and 
the curvature difference between inner detector and muon spectrometer tracks 
weighted by the muon electric charge in \Zmm\ and \Wmn\ events. 
Muon transverse momentum scale corrections are measured comparing the peak position of
the  \Zmm\ invariant mass distribution between data and MC
and fitting the muon transverse momentum distributions in \Zmm\ events \cite{Aad:2011yn,muResolConf}. 
Scale corrections are well below 1\% in the central pseudorapidity region and they increase to about 1\% in the 
high-$\eta$ regions due to residual misalignment effects in the ID and MS. 

Muon trigger and identification efficiencies are measured in a sample of \Zmm\ events selected with looser requirements on the second muon and with tighter cuts on the invariant mass window and on the angular correlation between the two muons than in the main analysis in order to reduce the contamination from background events \cite{muEffConf}.
The efficiencies are measured using a factorised approach: the efficiency of the combined reconstruction is derived with respect to the ID tracks, and the isolation cut is tested relative to combined tracks; finally the trigger efficiency is measured relative to isolated combined muons. 
The residual background contamination is measured from data, by fitting the invariant mass spectrum 
with a signal template plus a background template describing the shape of multijet events 
 measured from a control sample of non-isolated muons.
The total background contamination, subtracted from the signal sample, is estimated to be 1.0\% in the measurement of the reconstruction efficiency and negligible for other selections.
The data-to-Monte Carlo correction factors are all measured to be very close to $1$, i.e.
 \muonRecoSF\ for the combined reconstruction, \muonIsoSF\ for the isolation and \muonTriggerSF\ for 
the trigger efficiencies. Systematic uncertainties are evaluated by varying
the relevant selection cuts within their resolution and 
the amount of subtracted background within its uncertainty. 
For the ID reconstruction efficiency, no correction has to be applied.

\paragraph{Background Determination:}

The electroweak background in the $\ensuremath{W} \rightarrow \mu \nu$ channel is dominated by the  $\ensuremath{Z}\rightarrow \mu \mu$ 
and the  $\ensuremath{W}\rightarrow \tau \nu$ channels. 
Relative to the number of $W^{\pm}$ candidate events, 
these contributions are determined to be \nZmumuWmu\ and \nWtaunuWmu, respectively.
The contribution from  $\ensuremath{Z} \rightarrow \tau \tau$ decay is \nZtautauWmu\ while the $t\overline{t}$ contribution is estimated to be \nttbarWmu.
Diboson decays contribute \nDiBosonWmu. 
Overall these backgrounds are found to be 
\nEWKWmuPlus\ in the $W^+$ and \nEWKWmuMinus\ in the $W^-$ channel, respectively.

The QCD background in the \Wmn\ channel is primarily composed of heavy-quark decays, with smaller contributions from pion and kaon decays in flight and hadrons faking muons. Given the uncertainty in the dijet cross section
prediction and the difficulty of simulating fake prompt muons, the QCD background is 
derived from data. The number of expected events is 
determined extrapolating from control regions defined by reversing the isolation and missing transverse energy requirements. 
This analysis yields a fraction of background events of \nQCDWmuPlus\ in the $W^+$ 
and of \nQCDWmuMinus\ in the $W^-$ channel respectively. 
The systematic uncertainty is dominated by the uncertainty on the extrapolation of the isolation efficiency for QCD events from the control to the signal sample, which is estimated to be 
about 23\% relative to the number of background events.

The relative background contributions in the \Zmm\ channel
due to $t\overline{t}$ events, $Z \rightarrow \tau\tau$ and diboson decays 
are estimated to be \nttbarZmu, \nZtautauZmu, and \nDiBosonZmu, 
respectively.
The background contaminations from \Wtau\ and \Wmn\ are found
to be negligible.

The QCD background in the \Zmm\ channel is also estimated from data. 
The number of events is measured in control samples, selected using inverted
isolation and $m_{\mu\mu}$ requirements, corrected for the signal and electroweak background contamination, 
and extrapolated to the signal region. The measured fraction of background events is \nQCDZmu.
The systematic uncertainty is evaluated testing a different isolation definition 
for the control region, propagating the uncertainties in the electroweak 
background subtraction and checking the stability of the method against
boundary variations of the control regions.
Additional cross checks of the background estimation are done comparing with
the result of a closure test on simulated events and of an analysis of the invariant 
mass spectrum based on fit templates, derived from the data and the Monte Carlo. 
The relative systematic uncertainty amounts to \nQCDZmuSystRel\,
while the relative statistical
uncertainty is \nQCDZmuStatRel.

Cosmic ray muons overlapping in time with a collision event are another potential source of background. From a study of non-colliding bunches this background contribution is found to be negligible.

\section{Cross Section Measurements}
\label{sec:crosseclep}

\subsection{Electron Cross Sections}
\label{sec:elcross}

\paragraph{Control distributions:}

\begin{figure}[tbp]
  \begin{center}
    \includegraphics[width=0.4\textwidth]{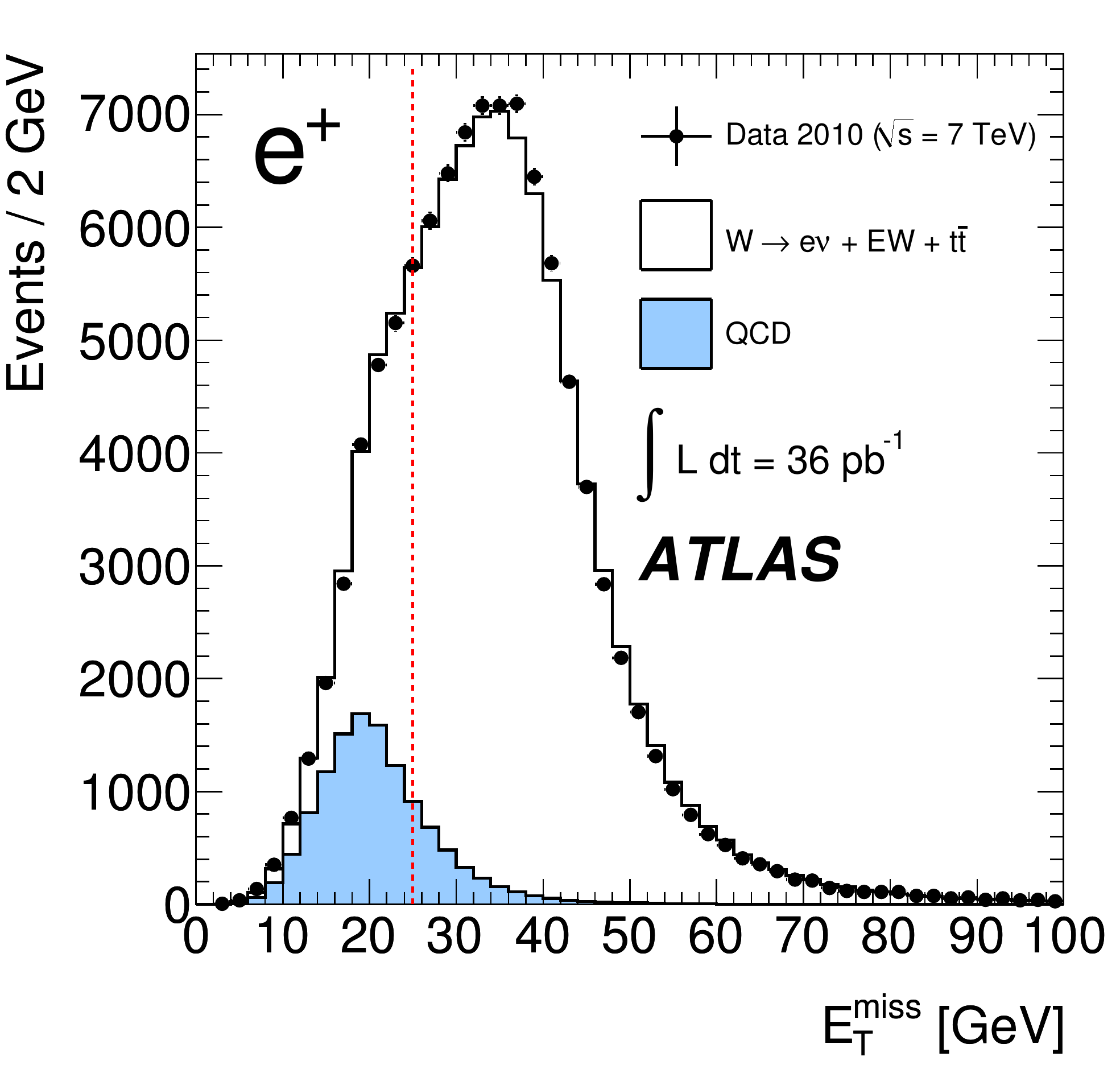}
    \includegraphics[width=0.4\textwidth]{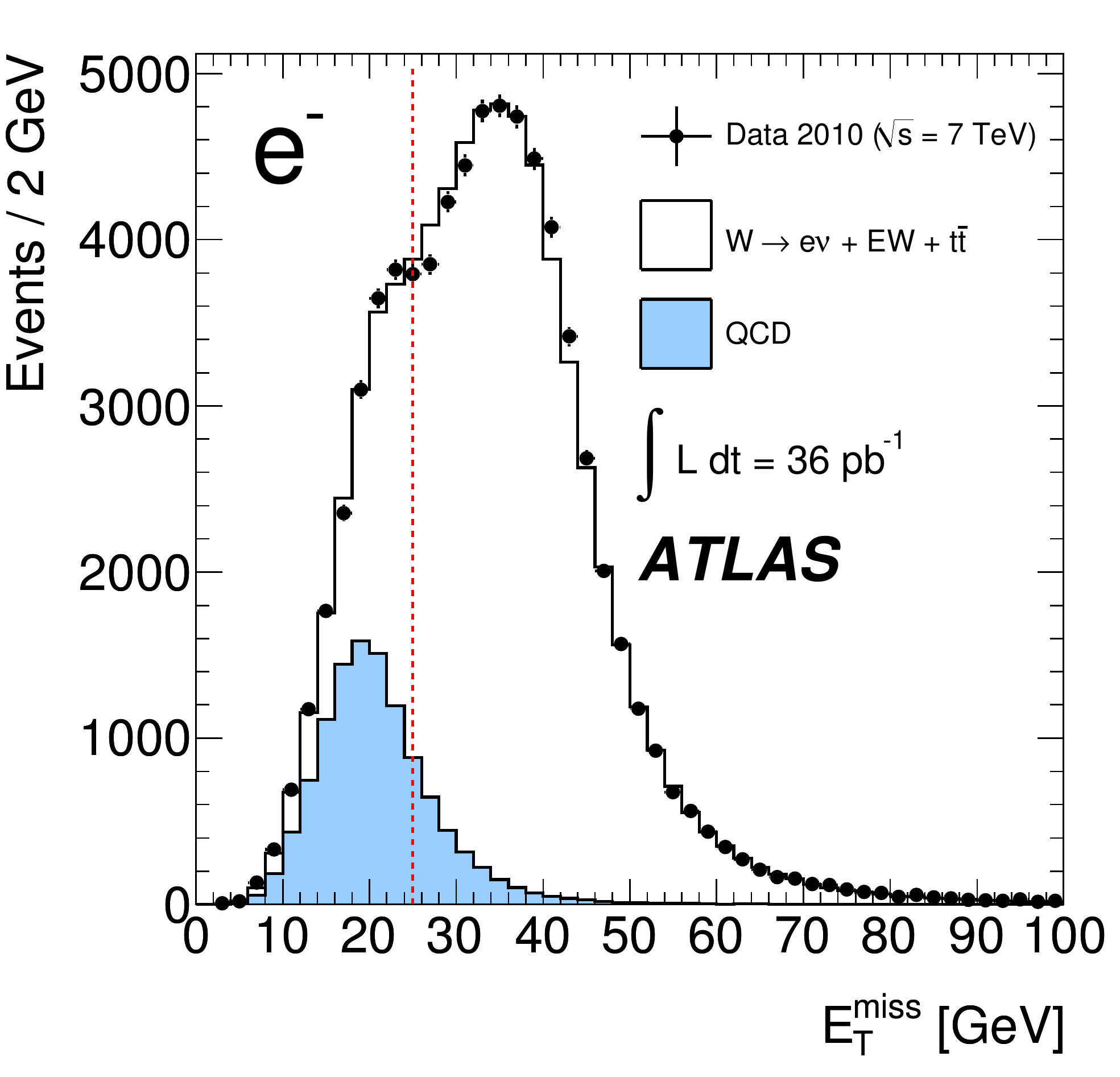}
    \caption{\it Distributions of \met\   
      in the selected \Wen\ candidate events for
      positive (top) and negative (bottom) charge. 
      The QCD background is represented by a background template taken from
      data, see text. 
      The analysis uses the requirement $\met > 25$\,GeV,
      indicated by the red line.}
    \label{fig:WenuFitQCD}
  \end{center}
\end{figure}

\begin{figure*}[ptb]
  \begin{center}
    \includegraphics[width=0.4\textwidth]{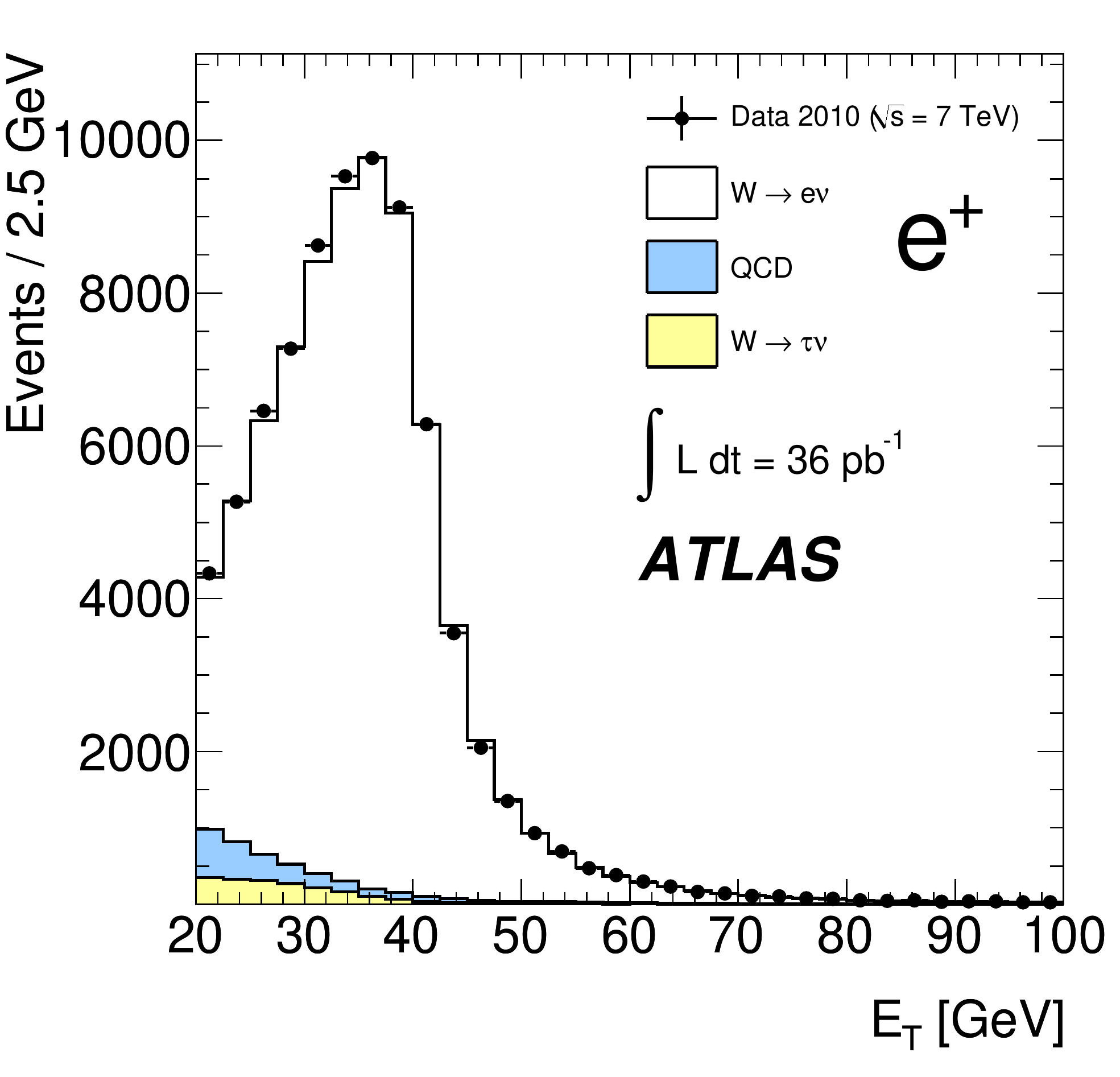}
    \includegraphics[width=0.4\textwidth]{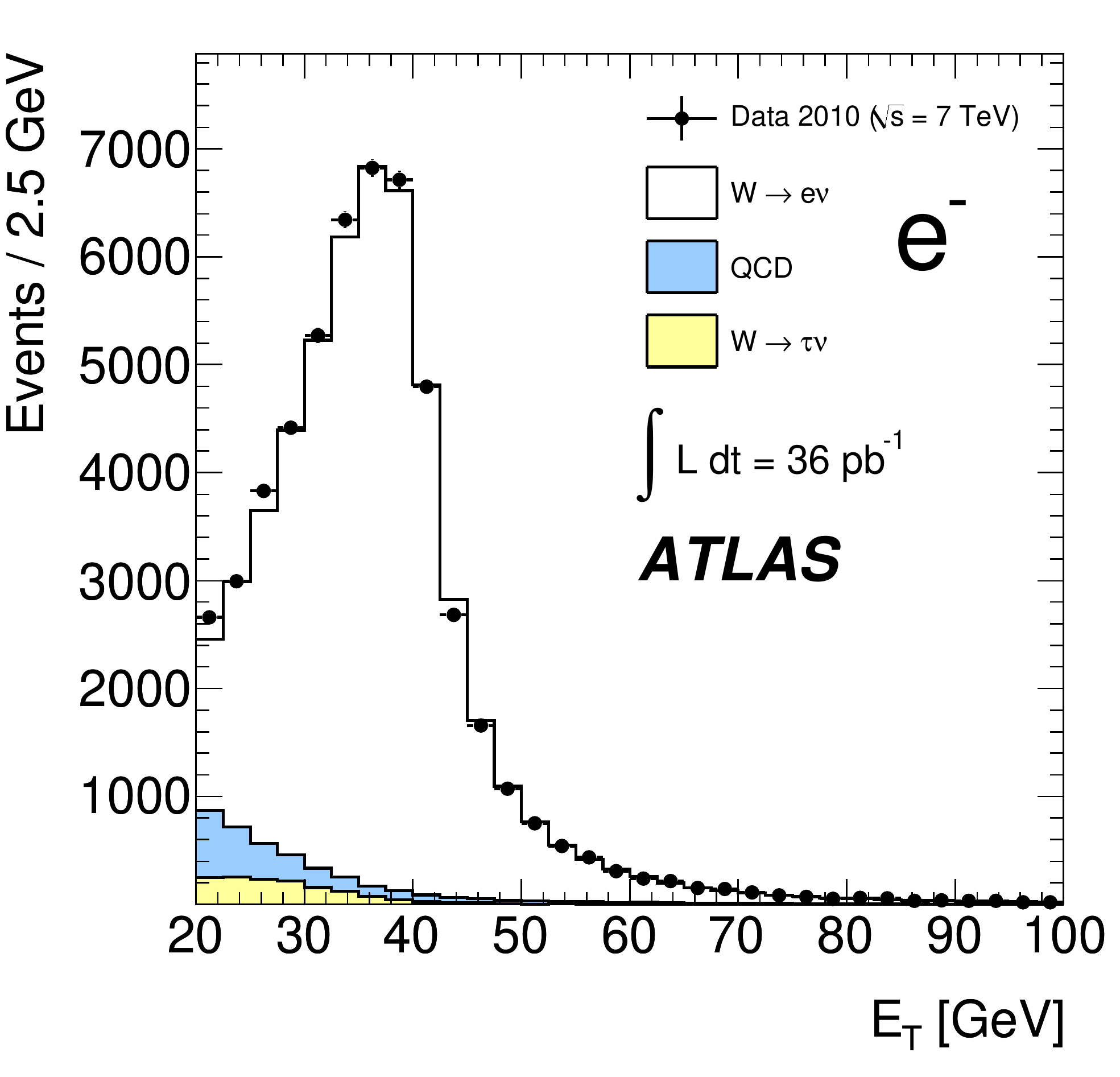}
    \includegraphics[width=0.4\textwidth]{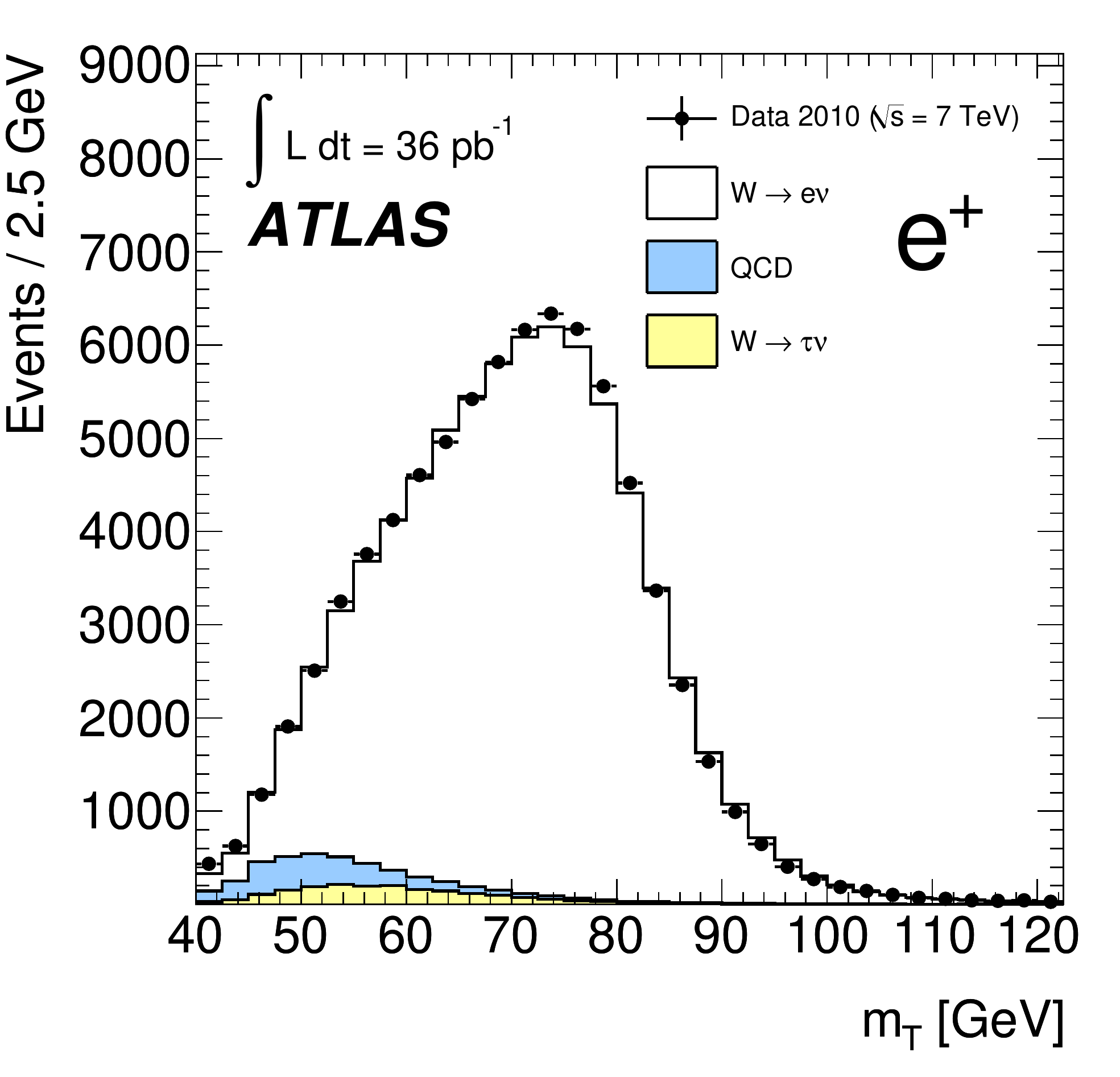}
    \includegraphics[width=0.4\textwidth]{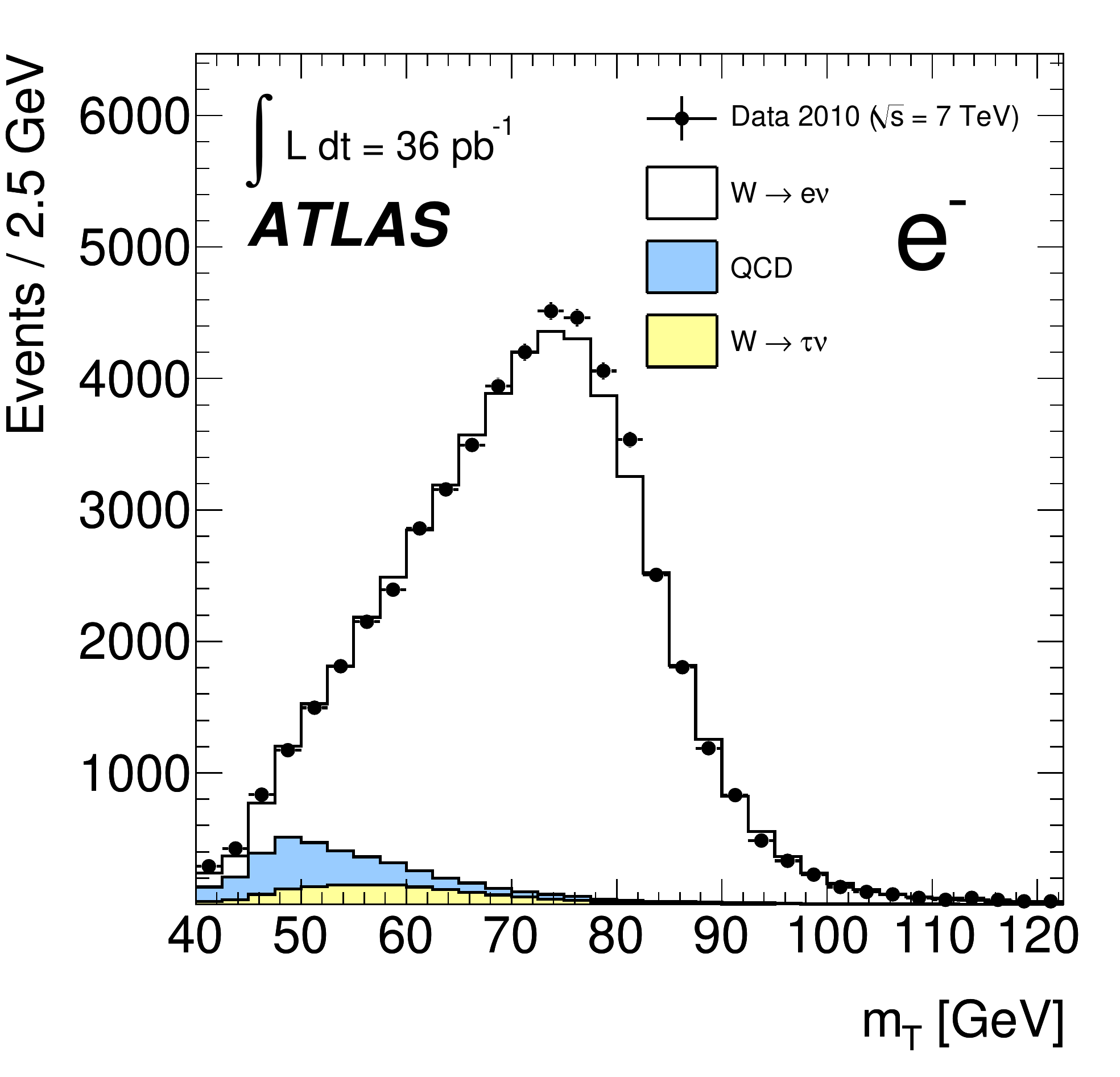}
    \caption{\it Top: Distribution of the electron transverse energy $E_T$
      in the selected \Wen\ candidate  events after all cuts for
      positive (left) and negative (right) charge. Bottom: Transverse mass
      distributions for $W^+$ (left) and $W^-$ (right) candidates.
      The simulation is normalised to the data.       
      The QCD background shapes are taken from
      background control samples (top) or MC simulation with relaxed
      electron identification criteria (bottom) and are normalised to
      the total number of QCD events as described in the text.}
    \label{fig:Wenu:ETMT}
  \end{center}
\end{figure*}

The understanding of the $W$ and $Z$ measurements
can be illustrated by comparing the measured with the
simulated distributions.
A total of $77885$ \Wplus\ and $52856$
\Wminus\ events are selected in the electron channel.
A crucial quantity in the $W$ measurement is the missing transverse
energy \met, for which the distributions for the two charges are shown 
in \Fig~\ref{fig:WenuFitQCD}.
The requirement $\met > 25\,\GeV$ is seen to 
suppress a large fraction of the QCD background.
\FFig~\ref{fig:Wenu:ETMT} shows the distributions of the electron
transverse energy $E_T$
and the transverse mass $m_T$ of the \Wen\ candidates. 
The observed agreement between data and MC is good.

\begin{figure*}[tbp]
  \centering
  \includegraphics[width=0.4\textwidth]{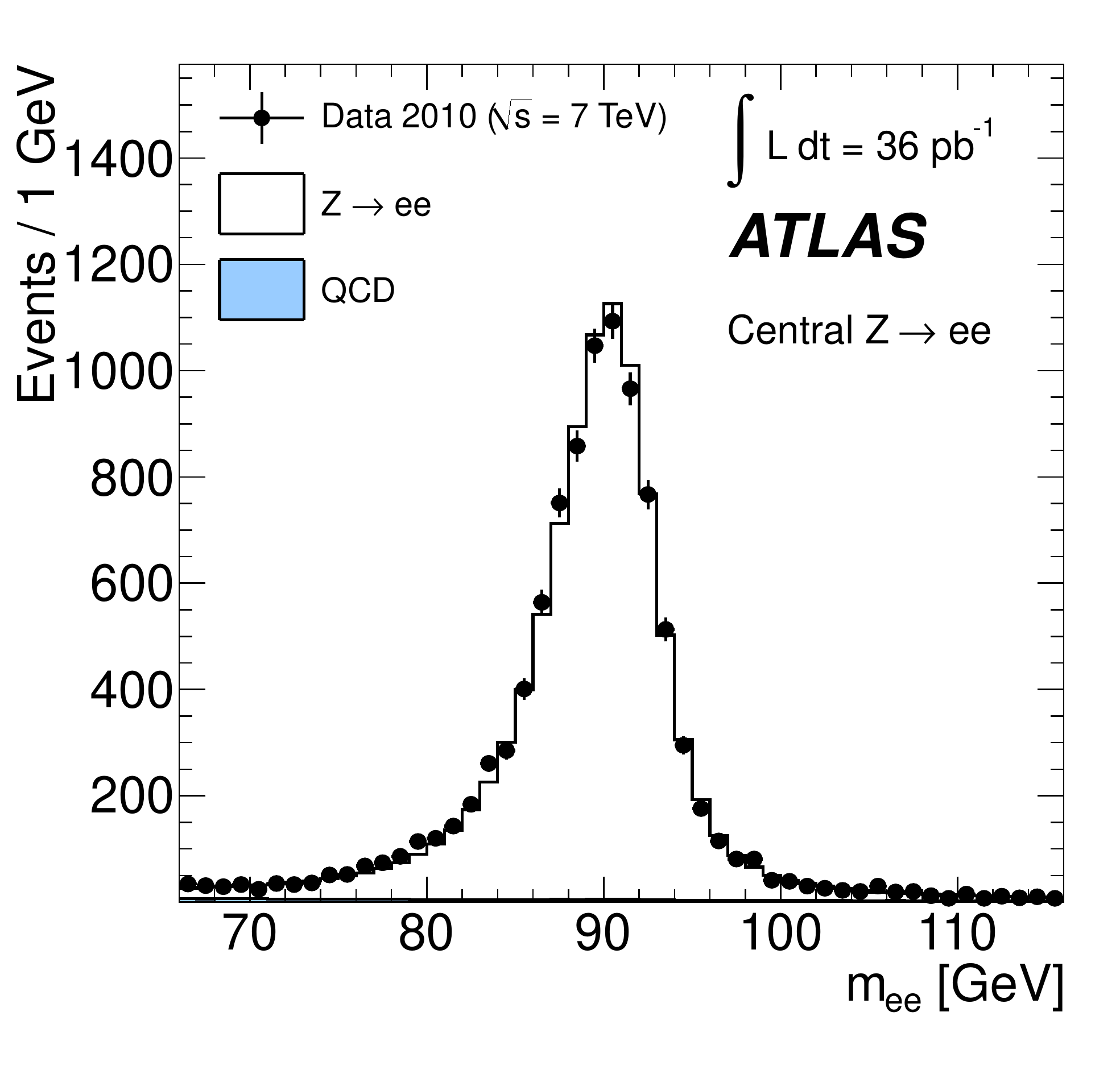}
  \includegraphics[width=0.4\textwidth]{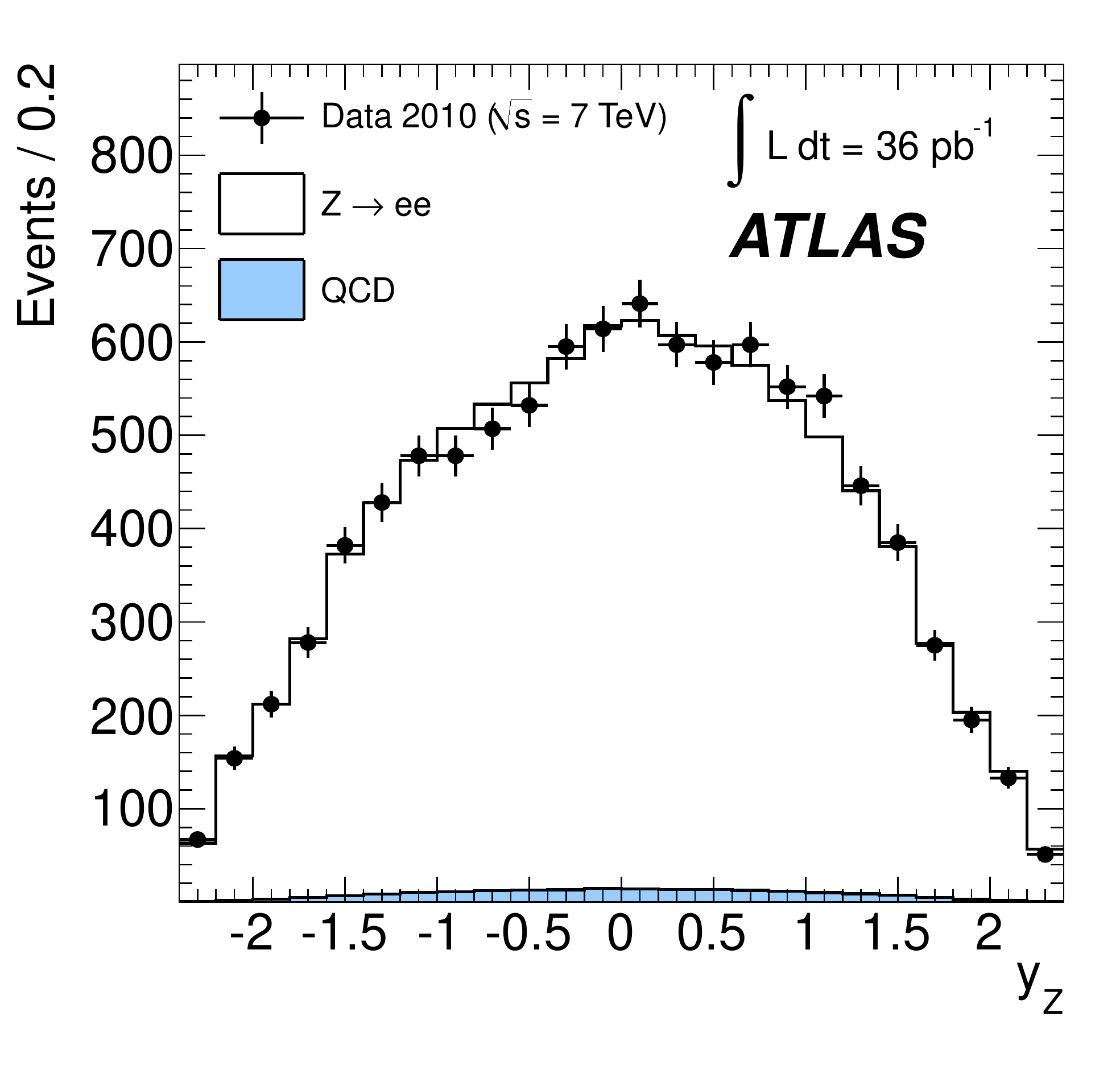}
  \caption{\it Dielectron invariant mass $m_{ee}$ (left) and rapidity
    $y_Z$ distribution (right) for the central \Zee\ analysis. 
    The simulation is normalised to the data.       
    The QCD background shapes are taken from a
    background control sample and normalised to the result of the
    QCD background fit.}
  \label{fig:zee_central}
\end{figure*}

\begin{figure*}[tbp]
  \centering
  \includegraphics[width=0.4\textwidth]{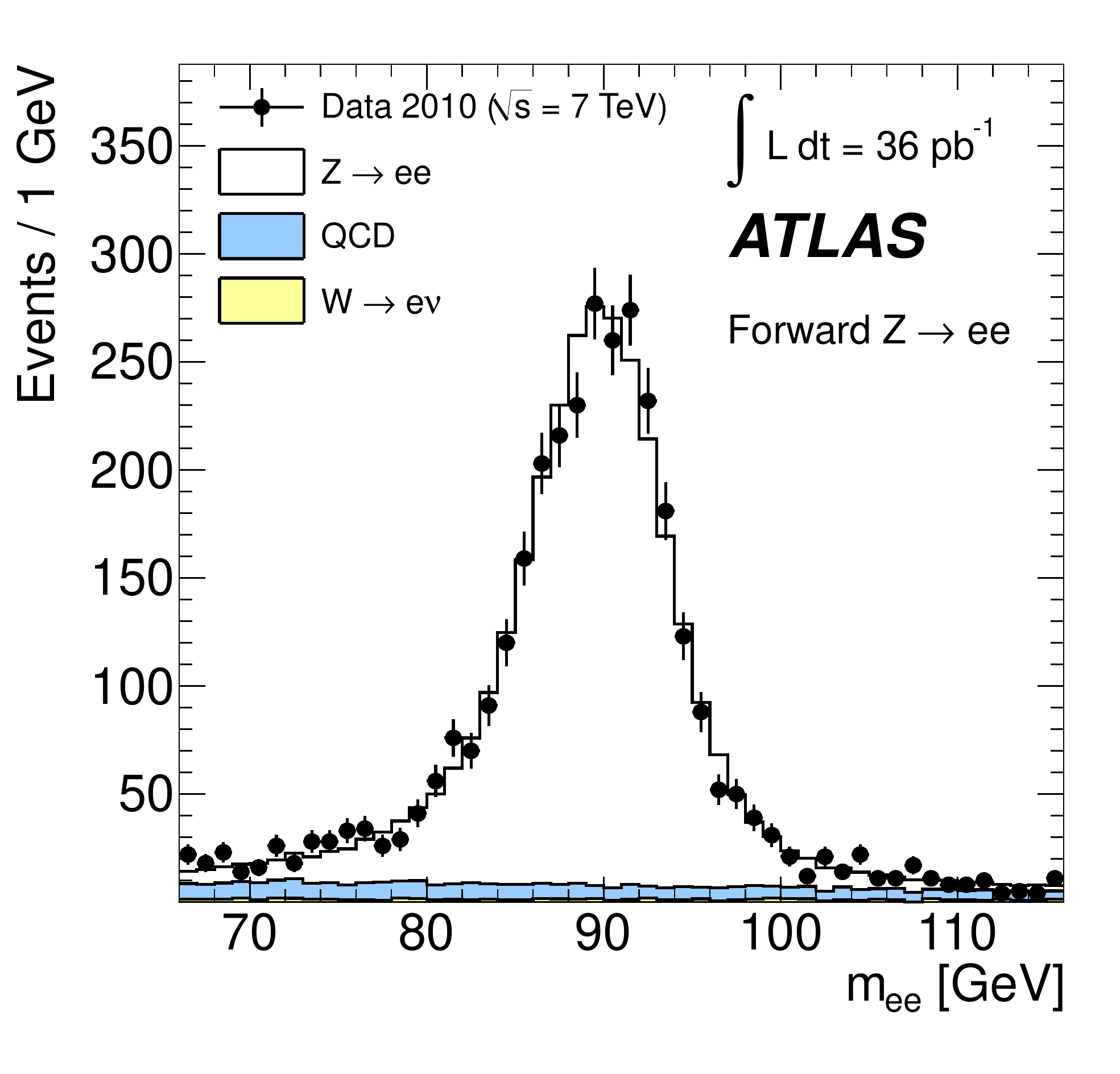}
  \includegraphics[width=0.4\textwidth]{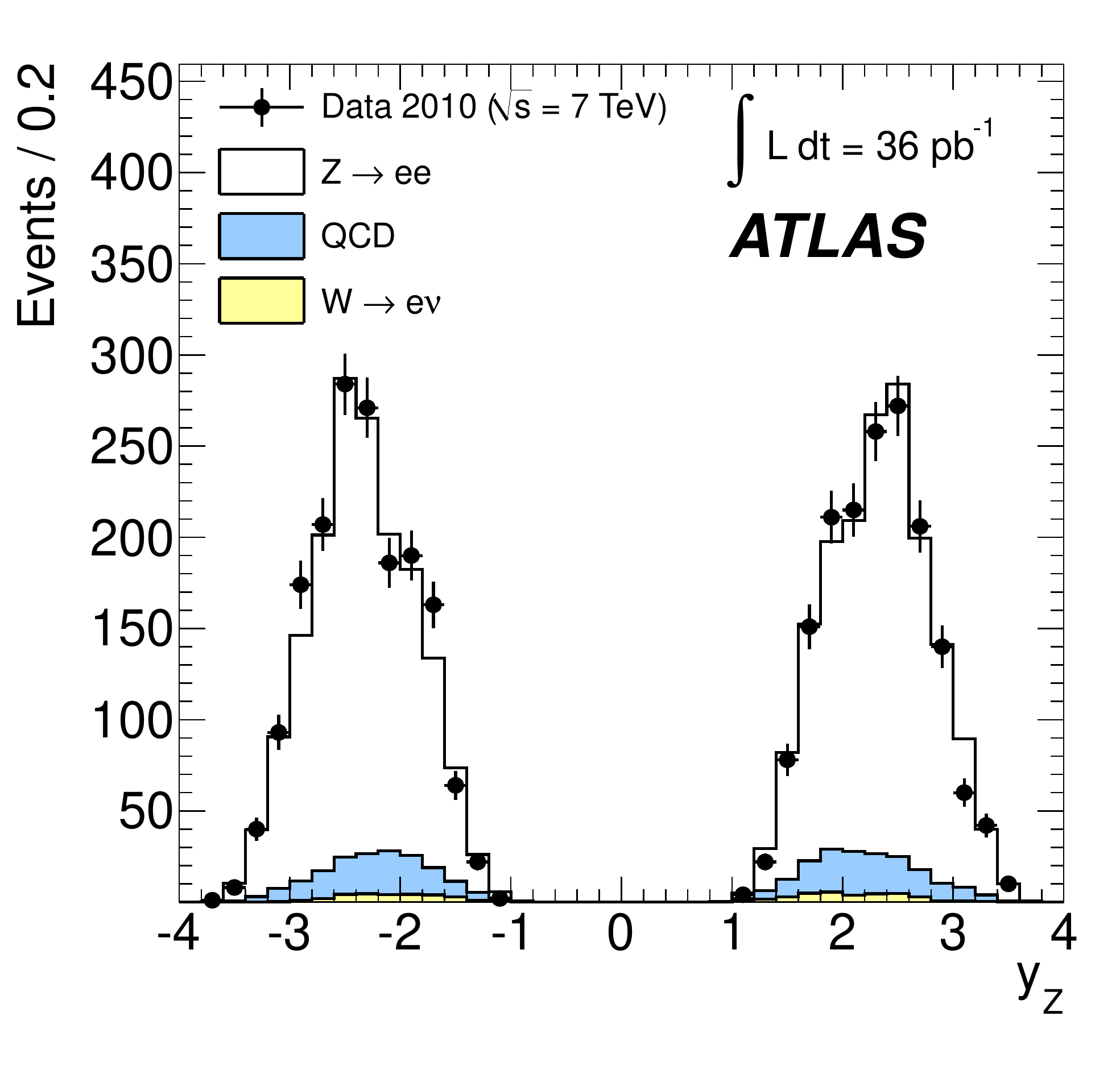}
  \caption{\it Dielectron invariant mass $m_{ee}$ (left) and rapidity
    $y_Z$ distribution (right) for the forward \Zee\ analysis. 
    The simulation is normalised to the data.       
    The QCD background shapes are taken from a
    background control sample and normalised to the result of the
    QCD background fit.}
  \label{fig:zee_forward1}
\end{figure*}

\begin{figure}[htbp]
  \centering
  \includegraphics[width=0.4\textwidth]{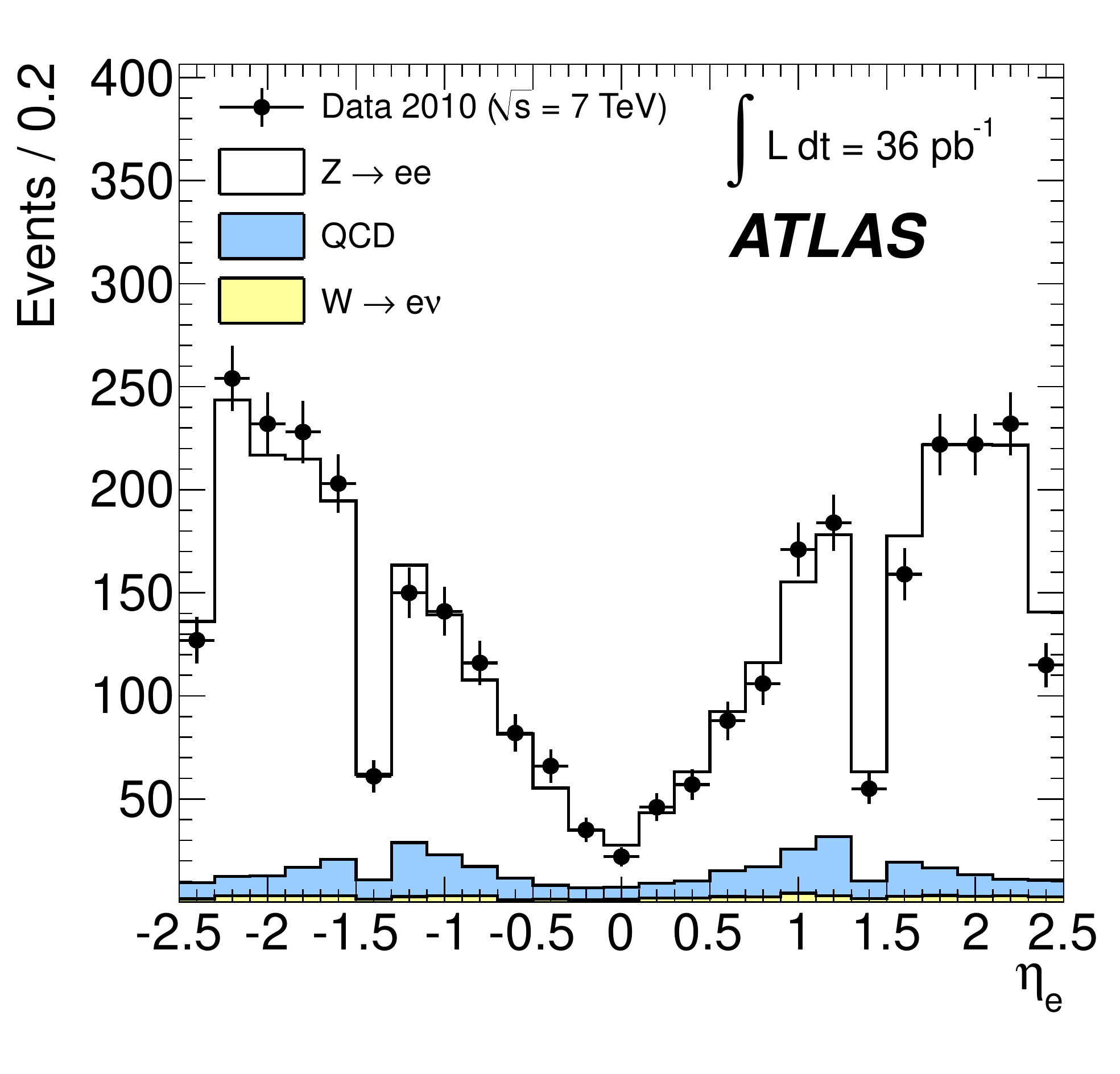}
  \includegraphics[width=0.4\textwidth]{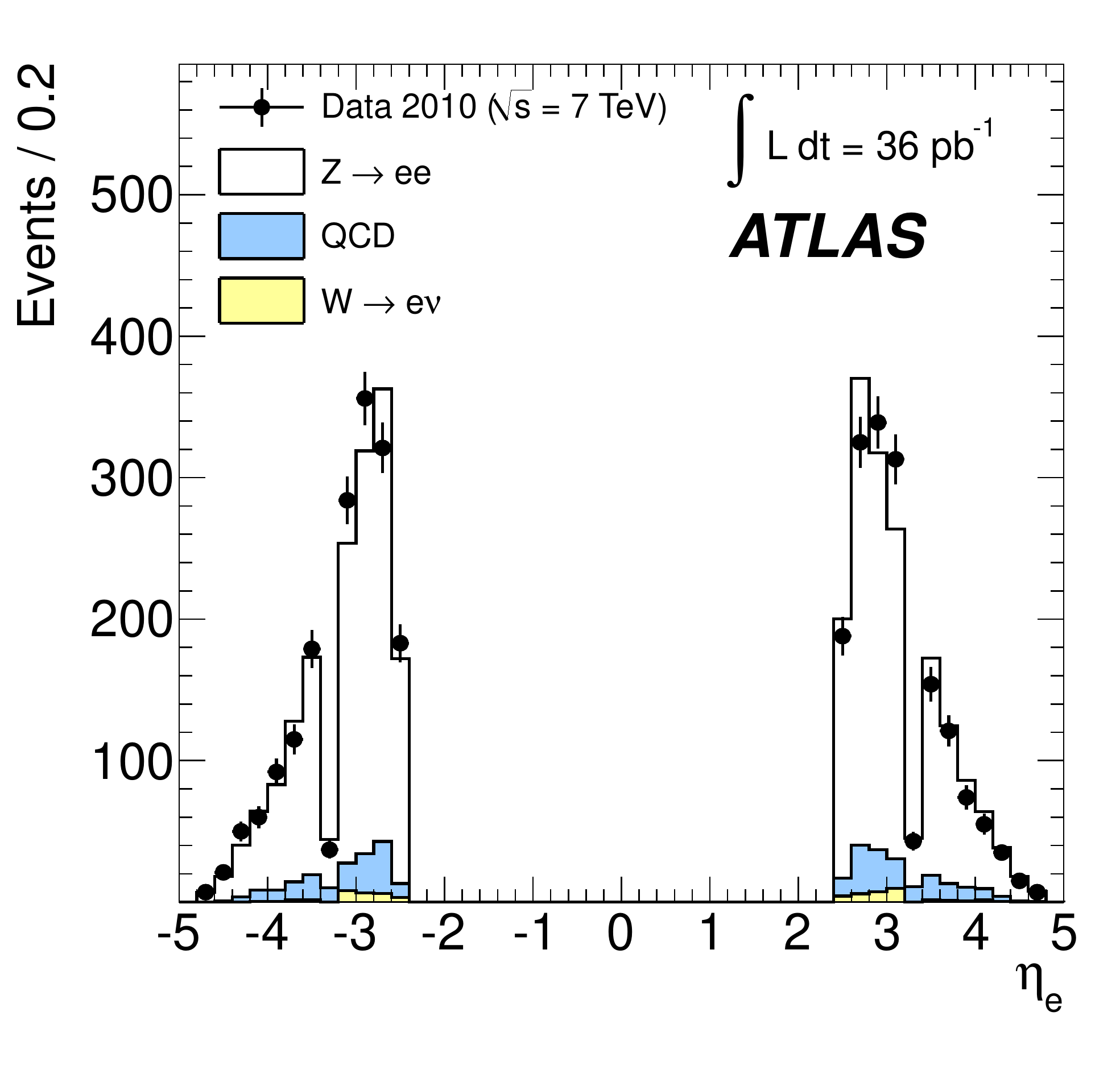}
  \caption{\it Pseudorapidity for the central (top) and the
    forward (bottom) electron in the forward \Zee\ analysis.  
    The simulation is normalised to the data.       
    The QCD background shapes are taken from a
    background control sample and normalised to the result of the
    QCD background fit.}
  \label{fig:zee_forward2}
\end{figure}

A total of $9725$ and $3376$ candidates are selected by the central and
forward \Zee\ analysis, respectively.
The invariant mass and boson rapidity distributions are
compared to the simulation in \Figs~\ref{fig:zee_central} and
\ref{fig:zee_forward1} for the two analyses. The
complementarity in rapidity region covered is easily visible. For the
forward \Zee\ analysis the lepton rapidity distributions
for the two electrons are shown in \Fig~\ref{fig:zee_forward2}. The forward
electron reaches pseudorapidities up to $|\eta| = 4.9$. The agreement between
data and Monte Carlo is good in all cases. 
Due to a small number of non-operational LAr readout channels,
the rapidity distributions show an asymmetry, which is 
well described by the simulation. The overlaps between different
calorimeter parts are visible as regions with significantly lower
efficiency.

\paragraph{Results:}

\begin{table}
  \centering
  \begin{tabular}{lcccc}
    \hline
    \hline
    & N & B & $\CWZ$ & $\AWZ$ \\
    \hline
    $W^+$     & $77885$  & $5130 \pm 350$ &$0.693 \pm 0.012$ & $0.478 \pm 0.008$ \\
    $W^-$     & $52856$  & $4500 \pm 240$ & $0.706 \pm 0.014$ & $0.452 \pm 0.009$ \\
    $W^\pm$    & $130741$ & $9610 \pm 590$ & $0.698 \pm 0.012$ & $0.467 \pm 0.007$\\
    $Z$       & $9725$ & $206 \pm 64$ & $0.618 \pm 0.016$ & $0.447 \pm 0.009$\\
    \hline 
  \end{tabular}
  \caption{\it Number of observed candidates N and expected background
    events B, efficiency and acceptance correction factors for the $W$
    and $Z$ electron channels. Efficiency scale factors used to correct the simulation for
    differences between data and MC are  included in the reported
    \CWZ\ factors. The given uncertainties are the quadratic sum of
    statistical and systematic components.
    The statistical uncertainties on the $\CWZ$ and $\AWZ$ factors are negligible.}
  \label{tab:WZaccElectron}
\end{table}

\TTab~\ref{tab:WZaccElectron} reports the number of candidates, estimated
background events and the \CWZ\ and \AWZ\
correction factors used, where the uncertainties on \AWZ\ are obtained from \Tab~\ref{tab:acc}.
The cross sections for all channels are reported in
\Tab~\ref{tab:xsec_echannels} with fiducial and total
values and the uncertainties due to data statistics, luminosity,
further experimental systematic uncertainties and the acceptance extrapolation in case of
the total cross sections.  

\TTab~\ref{tab:electronsyst} presents the sources of systematic uncertainties in all channels. 
Excluding  the luminosity contribution of
\dlumi\,\%, the $W$ cross sections are measured with an experimental
uncertainty of 1.8\% to 2.1\%, where the main contributions are due to
electron reconstruction and identification as well as missing
transverse energy performance related to the hadronic recoil~\cite{Aad:2011re}.

The $Z$ cross section is measured, apart from the luminosity contribution, 
with an experimental precision of 2.7\%.
This is dominated by the uncertainty on the electron 
reconstruction and identification efficiency.

The theoretical uncertainties on \CWZ\ are evaluated by comparisons of
\Mcatnlo\ and \Powheg\ Monte Carlo simulations and by testing the effect
of different PDF sets, as described in Sec.~\ref{sec:accext} for the acceptances.
The total theoretical uncertainty is found to be 0.6\% for \CW\ and 0.3\% for \CZ.

The theoretical uncertainty on the extrapolation from the fiducial
region to the total phase space for $W$ and $Z$ production is between 1.5\% and 2.0\%,
as mentioned above.

The cross sections measured as a function of the $W$ electron pseudorapidity, for separated charges,
and of the $Z$ rapidity are presented in \Tabs~\ref{tab:shortelwp},
\ref{tab:shortelwm}, \ref{tab:shortzcc} and \ref{tab:shortzcf}.
The statistical, bin-correlated and uncorrelated systematic and total
uncertainties are provided. The overall luminosity uncertainty is not
included.
The statistical uncertainty in each bin is about 1-2\% for the $W$
differential measurements, while the total uncertainty is at the
2.5-3\% level. For the $Z$ rapidity measurement the
statistical uncertainty is about 2\,\% for $|y_Z|<1.6$ and
grows to 3-5\% in the more forward bins. The total uncertainty on the
$Z$ cross sections is 3-4\% in the central region and up to 10\%
in the most forward bins. It is mainly driven by the uncertainties on
the electron reconstruction and identification efficiencies. 

\begin{table}[tbqh]
  \begin{center}
\begin{tabular}{lccccc}
\hline
\hline
  & \multicolumn{5}{c}{\bf $\sigma_W^{\rm fid} \cdot$ BR($W \to e \nu$) \ \ [nb]} \\
\hline
  &  & & \hspace{-0.45cm} sta & \hspace{-0.6cm} sys & \hspace{-0.9cm} lum  \\
\hline
${W^+} $      &   \multicolumn{5}{c}{$~\sigfidWeplus$}  \\
${W^-} $      &   \multicolumn{5}{c}{$~\sigfidWeminus$}  \\
$ W^\pm $         &   \multicolumn{5}{c}{$~\sigfidWe$}       \\
\hline
  & \multicolumn{5}{c}{\bf $\sigma_W^{\rm tot} \cdot$ BR($W \to e \nu$) \ \ [nb]} \\
\hline
  & \hspace{1.26cm} & sta & \hspace{0.48cm} sys & \hspace{0.43cm} lum & \hspace{0.3cm} acc \\
\hline
${W^+} $      &   \multicolumn{5}{c}{$~\sigWeplus$}  \\
${W^-} $      &   \multicolumn{5}{c}{$~\sigWeminus$}  \\
$ W^\pm $     &   \multicolumn{5}{c}{$~\sigWe$}       \\
\hline
\hline
 & \multicolumn{5}{c}{\bf $\sigma_{Z/\gamma^*}^{\rm fid} \cdot$ BR($Z/\gamma^* \to ee$)\ [nb]} \\
\hline
  &  & & \hspace{-0.45cm} sta & \hspace{-0.6cm} sys & \hspace{-0.9cm} lum  \\
\hline
$Z/\gamma^*$      &   \multicolumn{5}{c}{$~\sigfidZe$}   \\
\hline
 & \multicolumn{5}{c}{\bf $\sigma_{Z/\gamma^*}^{\rm tot} \cdot$ BR($Z/\gamma^* \to ee$)\ [nb]} \\
\hline
  & \hspace{1.26cm} & sta & \hspace{0.48cm} sys & \hspace{0.43cm} lum & \hspace{0.3cm} acc \\
\hline
$Z/\gamma^*$      &   \multicolumn{5}{c}{$~\sigZe$}   \\
\hline
\hline
\end{tabular}
\caption{\it Fiducial and total  cross sections times branching ratios
  for $W^+$, $W^-$, $W^{\pm}$ and $\Zg$ production in the electron
 decay channel.
 The electron fiducial regions are defined in Sec.~\ref{sec:sigdef}.  
   The uncertainties denote the statistical (sta), the experimental systematic (sys),
   the luminosity (lum), and the extrapolation (acc) uncertainties.}
\label{tab:xsec_echannels}
\end{center}
\end{table}

\begin{table}
  \begin{center}
    \begin{tabular}{lcccc}
      \hline
      \hline
      & $\delta \sigma_{W^\pm}$  &  $\delta \sigma_{W+}$    &  $\delta \sigma_{W-}$   & $\delta \sigma_Z$  \\
      \hline
      Trigger                        & 0.4   & 0.4   & 0.4  & $<$0.1 \\
      Electron reconstruction        & 0.8  & 0.8  & 0.8 & 1.6\\
      Electron identification        & 0.9  & 0.8  & 1.1 & 1.8\\
      Electron isolation      & 0.3  & 0.3  & 0.3 & ---\\
      Electron energy scale and resolution & 0.5  & 0.5  & 0.5 & 0.2\\
      Non-operational LAr channels           & 0.4   & 0.4   & 0.4 &  0.8\\
      Charge misidentification         & 0.0   & 0.1   & 0.1 &  0.6\\
      QCD background                     & 0.4 & 0.4 & 0.4& 0.7\\
      Electroweak+$t\bar{t}$ background                     & 0.2 & 0.2 & 0.2& $<$0.1\\
      \MET scale and resolution      & 0.8 & 0.7 & 1.0 & ---\\
      Pile-up modeling               & 0.3   & 0.3  & 0.3 & 0.3\\
      Vertex position                & 0.1  & 0.1 & 0.1 & 0.1\\
      \CWZ\ theoretical uncertainty             & 0.6 & 0.6 & 0.6& 0.3\\
      \hline                        
      Total experimental uncertainty  & 1.8   & 1.8  & 2.0 & 2.7\\
      \AWZ\ theoretical uncertainty   & 1.5 &  1.7 &  2.0   & 2.0\\
      \hline
      \hline
      Total excluding luminosity & 2.3 &  2.4 & 2.8 & 3.3\\
      \hline
      Luminosity                & \multicolumn{4}{c}{$\dlumi$}   \\
      \hline
      \hline
    \end{tabular}
    \caption{\it Summary of relative systematic uncertainties
             on the measured integrated cross sections in the electron channels in per cent. 
             The theoretical uncertainty of \AWZ\ applies only to the total cross section.}
    \label{tab:electronsyst}
  \end{center}
\end{table}

\clearpage

\subsection{Muon Cross Sections}
\label{sec:mucross}

\paragraph{Control distributions:}

\begin{figure}[b]
  \begin{center}
    \includegraphics[width=0.4\textwidth]{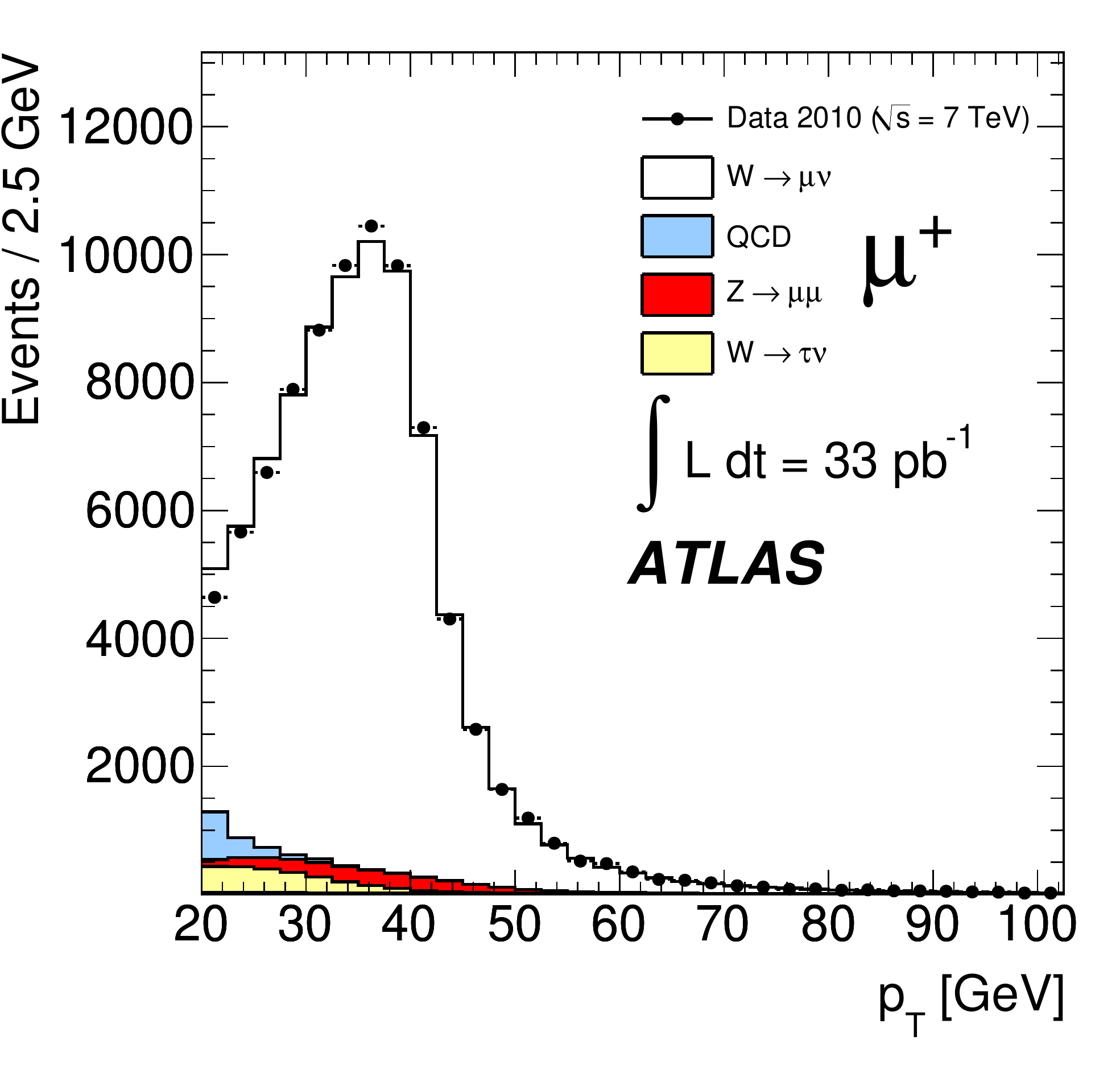}
    \includegraphics[width=0.4\textwidth]{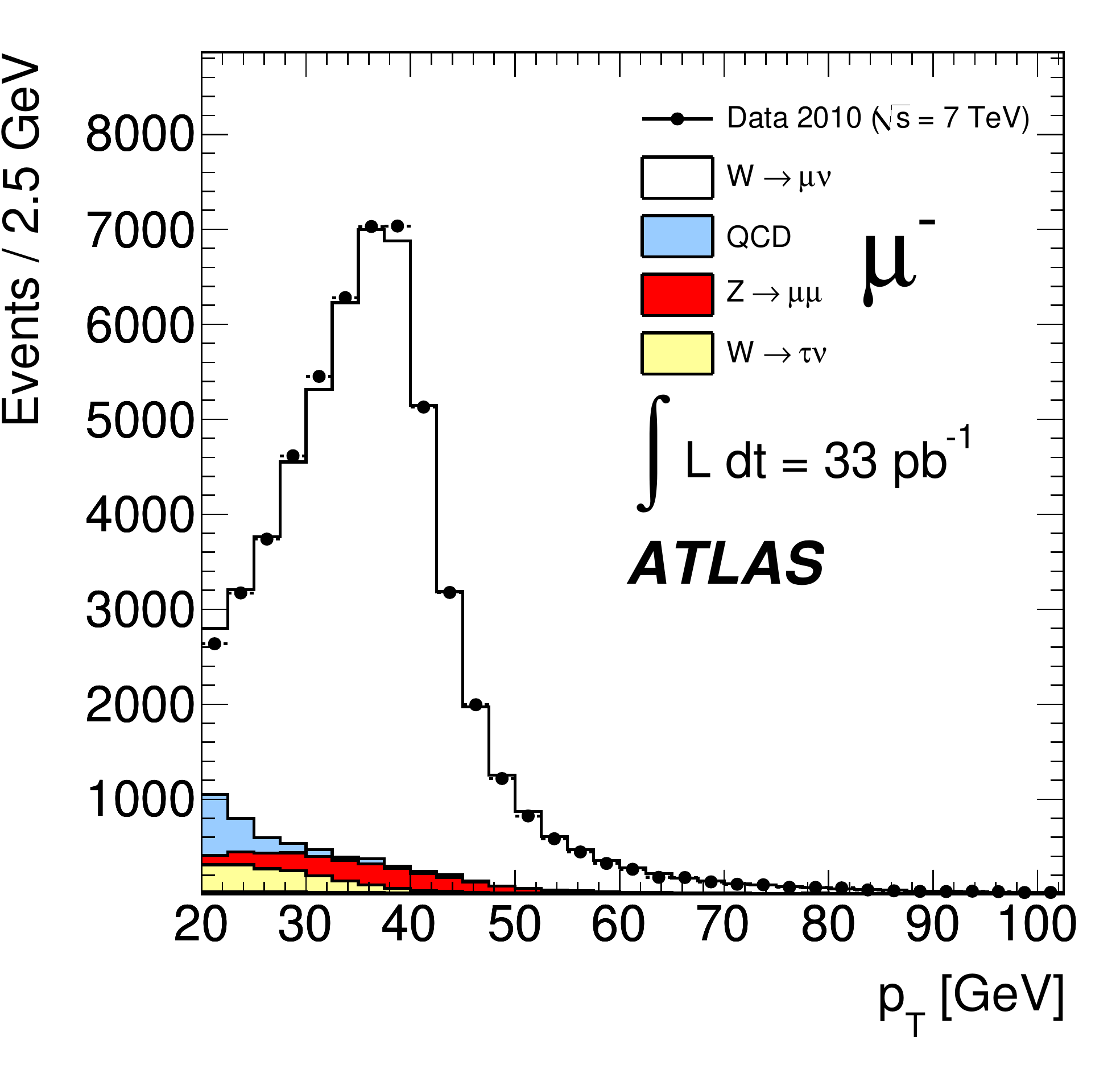}
    \caption{\it Muon transverse momentum distribution
      of candidate $W^+$ (top) and $W^-$ (bottom) events.
      The simulation is normalised to the data.
      The QCD background shape is taken from simulation and 
      normalised to the number of QCD events measured from data.
      \label{wmunu:fig:candWpt}}
  \end{center}
\end{figure}

\begin{figure}[b]
  \begin{center}
    \includegraphics[width=0.4\textwidth]{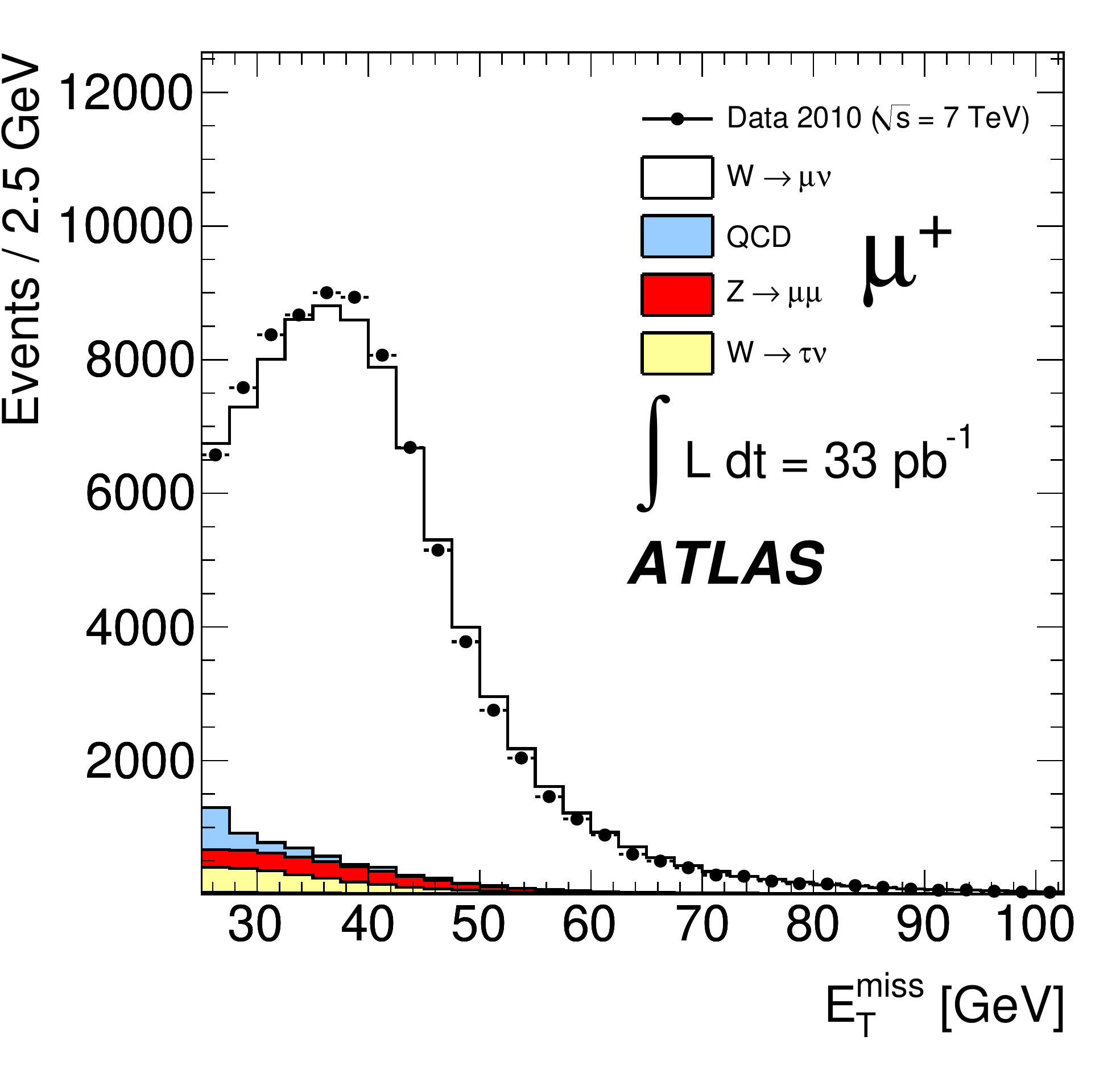}
    \includegraphics[width=0.4\textwidth]{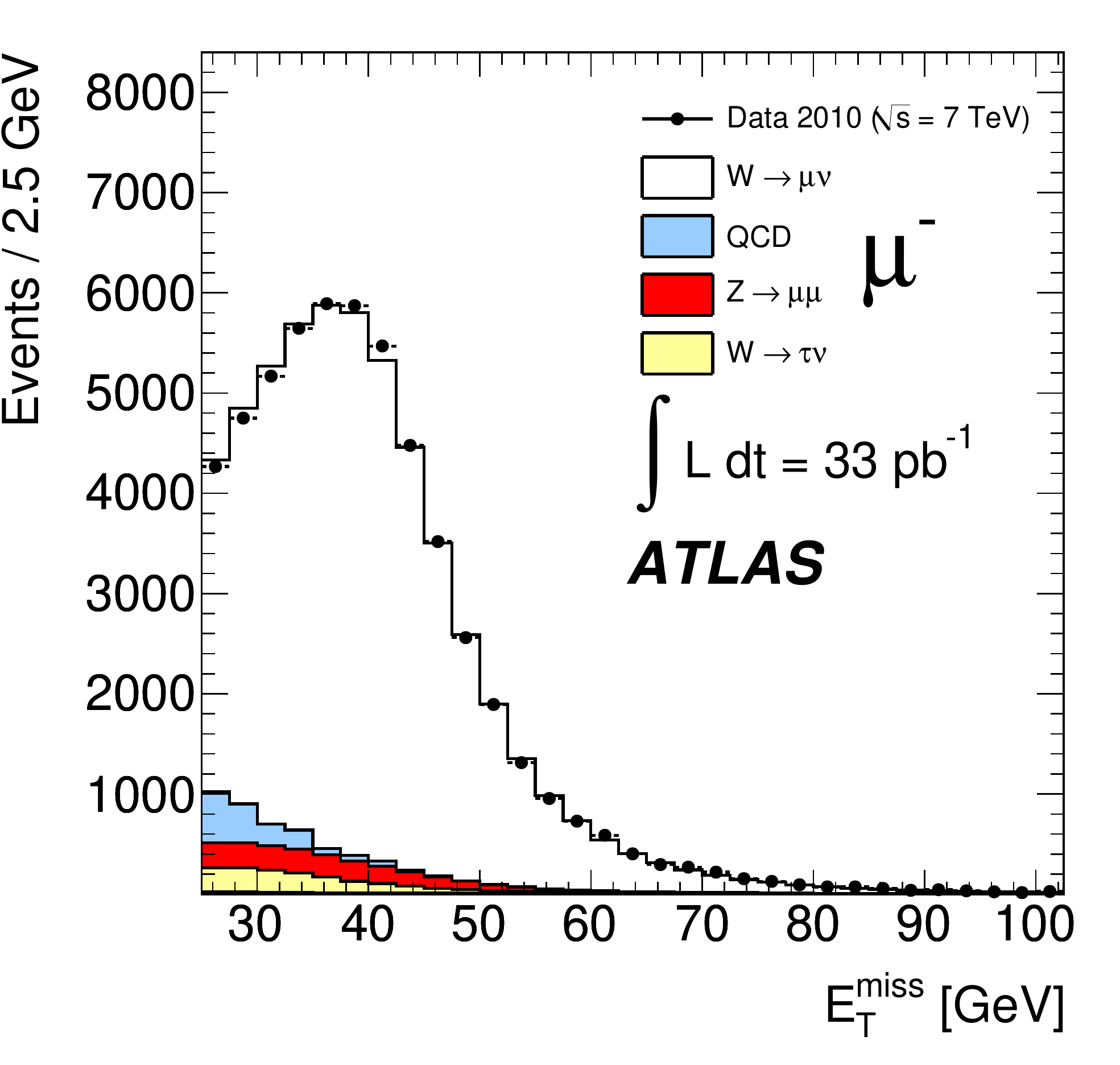}
    \caption{\it Missing transverse energy distribution
      of candidate $W^+$ (top) and $W^-$ (bottom) events.
      The simulation is normalised to the data.
      The QCD background shape is taken from simulation and 
      normalised to the number of QCD events measured from data.
      \label{wmunu:fig:candWmet}}
  \end{center}
\end{figure}

\begin{figure}[htbp]
  \begin{center}
    \includegraphics[width=0.4\textwidth]{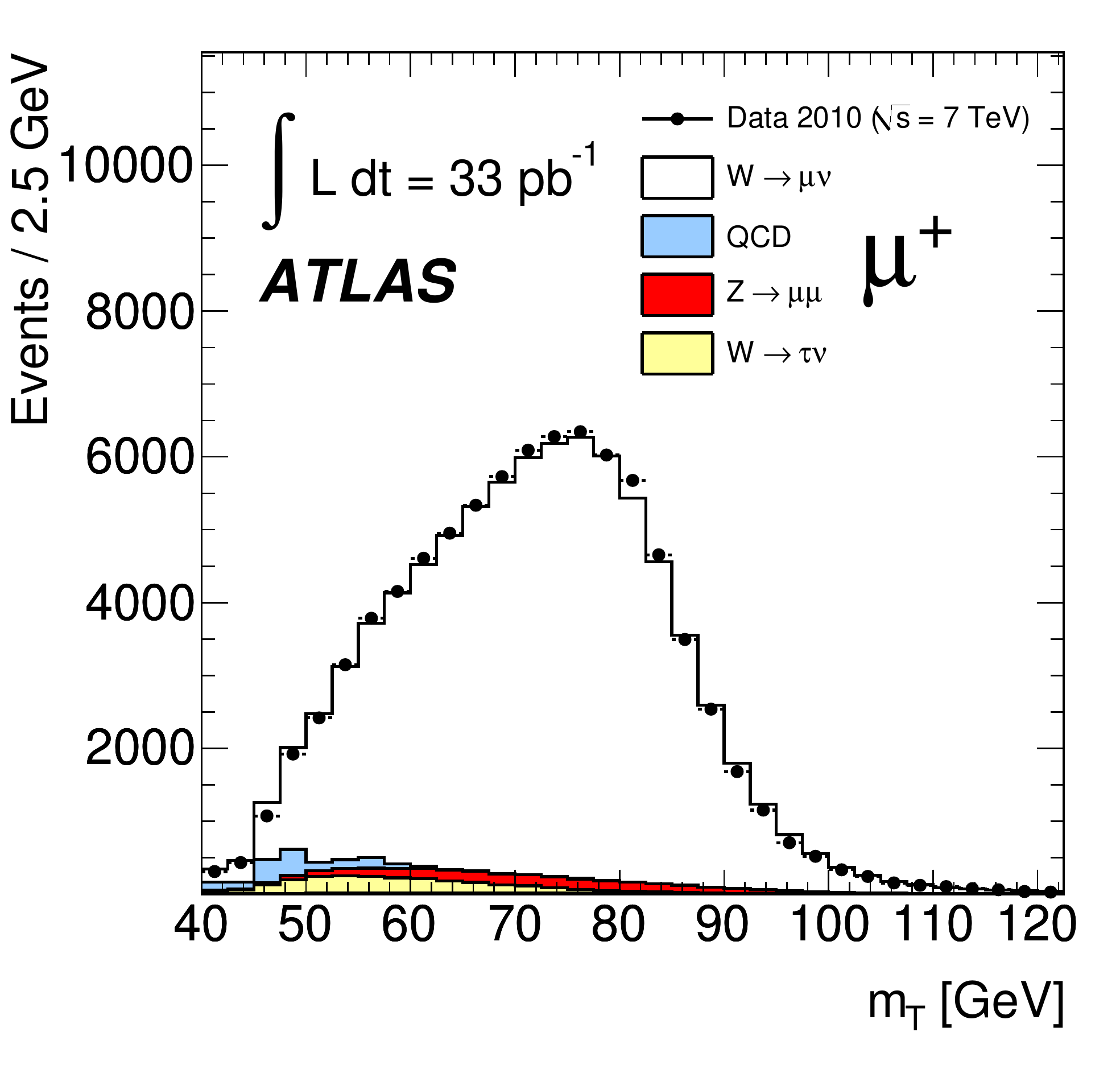}
    \includegraphics[width=0.4\textwidth]{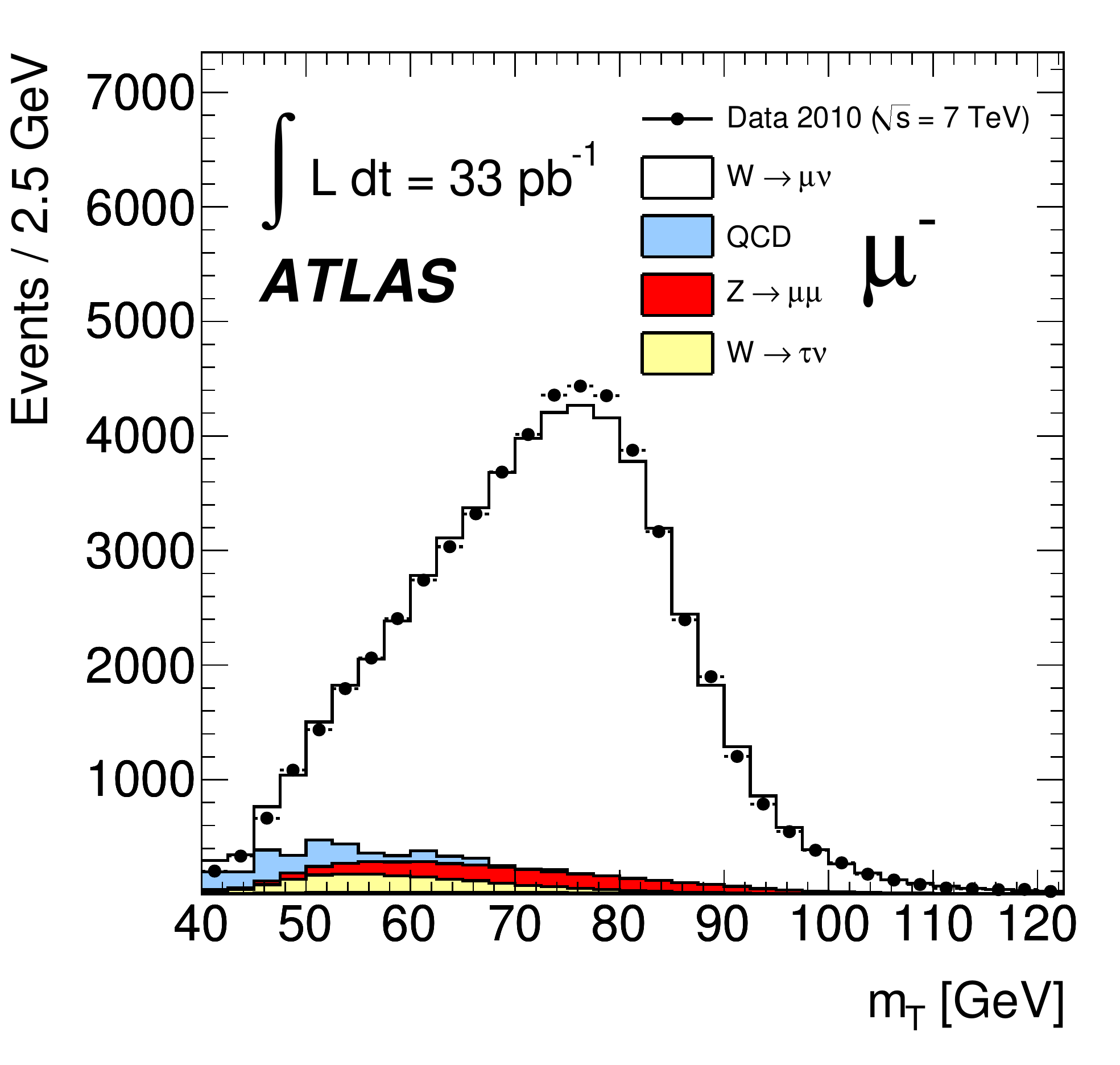}
    \caption{\it Transverse mass distribution of candidate $W^+$ (top) and $W^-$ (bottom) events.
      The simulation is normalised to the data.
      The QCD background shape is taken from simulation and 
      normalised to the number of QCD events measured from data.
      \label{wmunu:fig:candWmt}}
  \end{center}
\end{figure}

\begin{figure}[htbp]
  \begin{center}
    \includegraphics[width=0.4\textwidth]{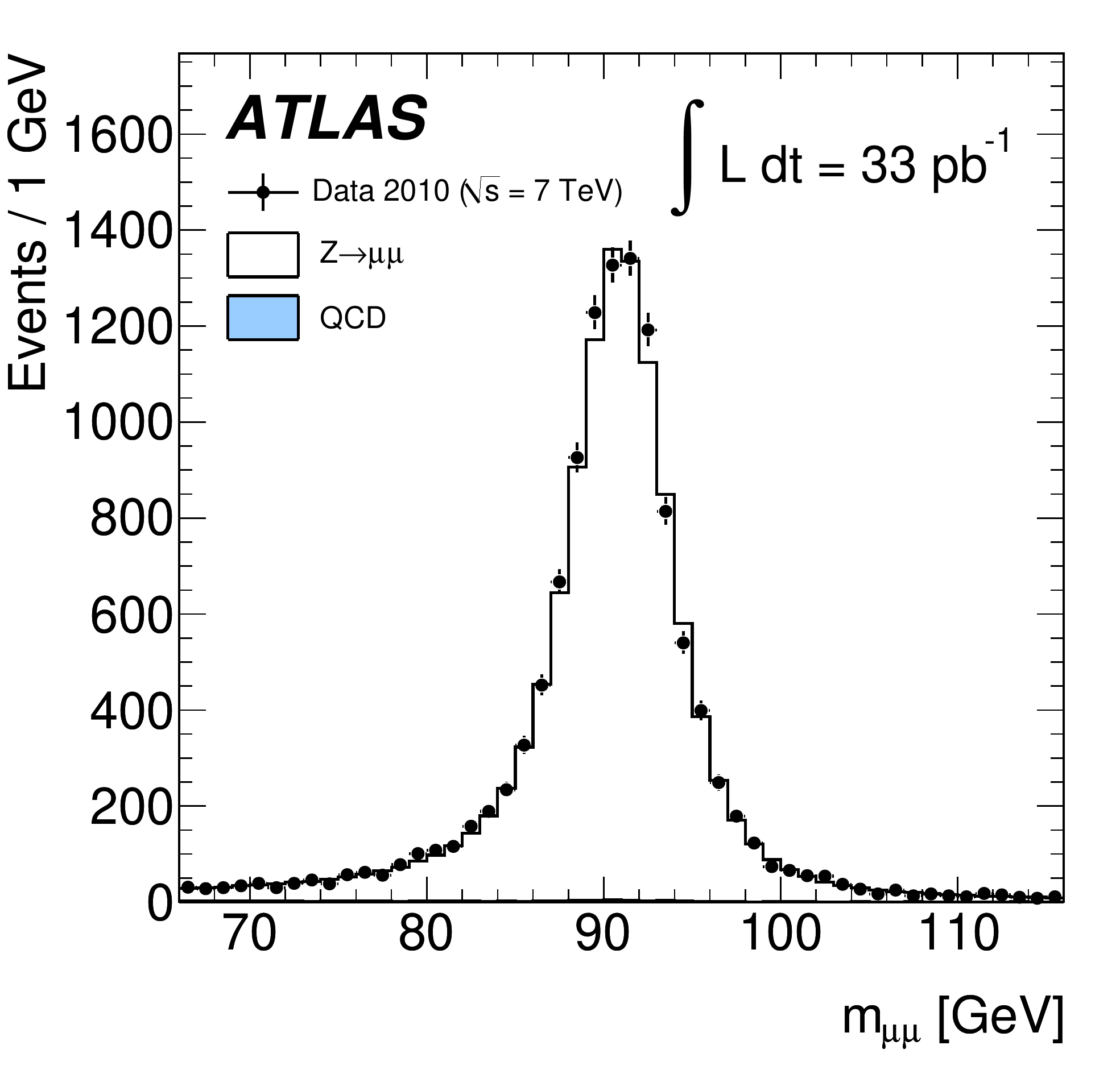}
    \includegraphics[width=0.4\textwidth]{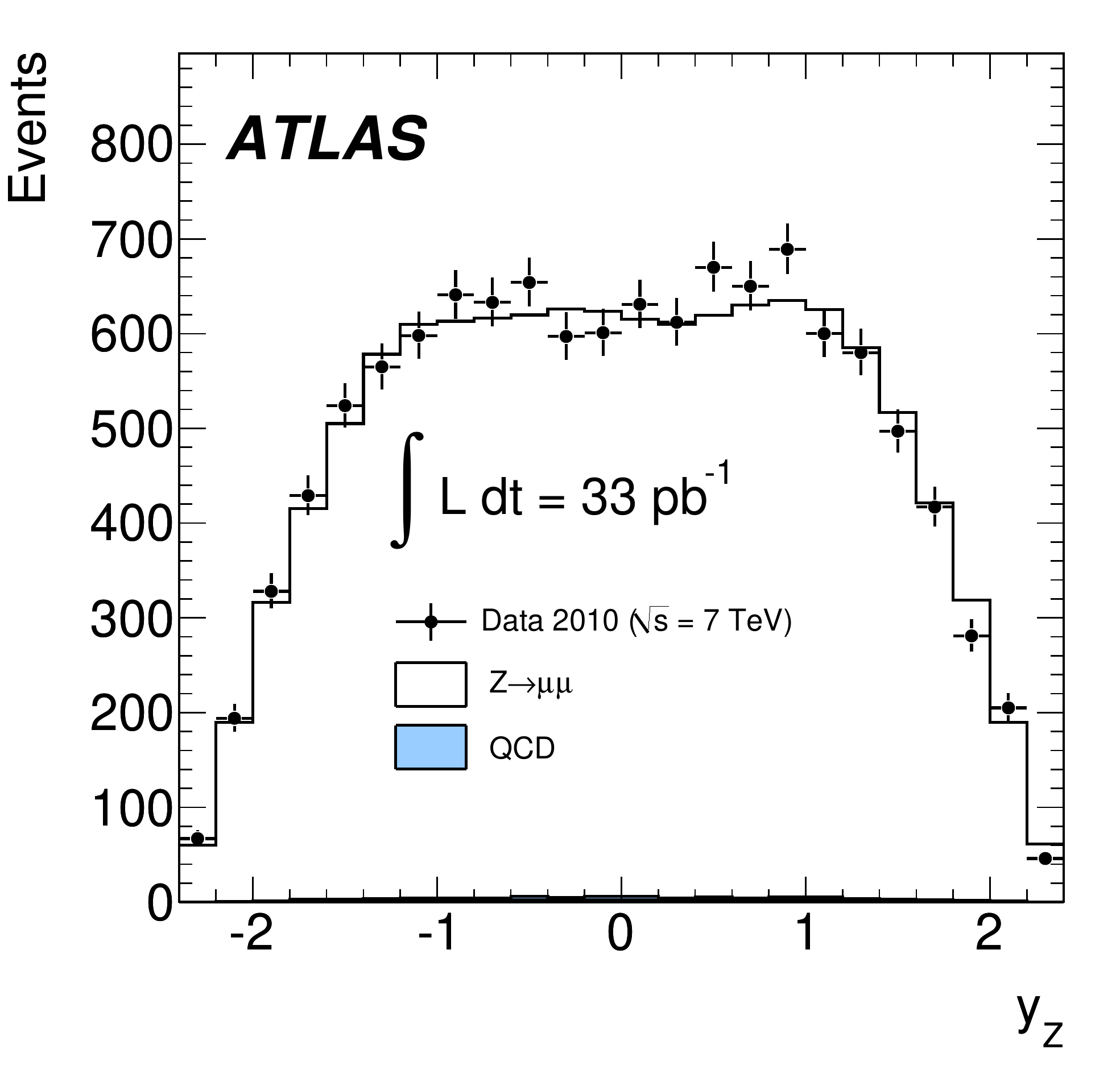}
    \caption{\it Invariant mass (top) and rapidity (bottom) distributions of candidate $Z$ bosons.
          The simulation is normalised to the data.
          The QCD background normalisation and shapes are taken 
          from control samples as described in the text.}
    \label{zmumu:fig:candZmPt}
  \end{center}
\end{figure}

A total of \NWmuplusCands\ $W^+$, \NWmuminusCands\ $W^{-}$ and
\NZmuCands\ $Z$ candidates are selected in the muon channels.
A few distributions of these candidate events are compared to the simulation for the signal and
the background contributions in the following.
\FFigs~\ref{wmunu:fig:candWpt} and \ref{wmunu:fig:candWmet} show the distributions of muon transverse momentum and the transverse missing energy of candidate $W$ events for positive and negative charges. 
The transverse mass distributions are shown in \Fig~\ref{wmunu:fig:candWmt}. The invariant mass distribution of
muon pairs, selected by the $Z$ analysis, and the boson rapidity distribution 
are shown in \Fig~\ref{zmumu:fig:candZmPt}. The agreement between data and Monte Carlo 
is good in all cases.

\paragraph{Results:}

\TTab~\ref{tab:WZaccMuon} reports the number of candidates, the estimated
background events and the \CWZ\ and \AWZ\ correction factors used for the different measurements.
The fiducial and total cross sections are reported in \Tab~\ref{tab:muonxsec} 
for all channels
with the uncertainties due to data statistics, luminosity, 
further experimental systematics and the acceptance extrapolation in case of
the total cross sections.

The breakdown of the systematic uncertainty in all channels is shown in \Tab~\ref{muonsyst}.
Apart from the luminosity contribution of \dlumi\,\%,
the \Wmn\ cross section is measured with an experimental uncertainty
of 1.6\%.
The largest contribution comes from the muon efficiencies (1.1\%),
followed by several contributions in the 0.3-0.8\%
range such as the QCD background, the transverse missing energy scale 
and resolution uncertainties and the uncertainty on the momentum scale
correction.

The \Zmm\ cross section is measured, apart from the luminosity contribution,  
with an experimental precision of 0.9\%. 
This is dominated by the uncertainty in the muon reconstruction efficiency (0.6\%),  
with about equal systematic and statistical components due to the 
limited  sample of \Zmm\ events.
The uncertainty of the momentum scale correction has an effect of 0.2\% while the uncertainty
from momentum resolution is again found to be negligible.
The impact of the QCD background uncertainty is at the level of 3 per mille.

The theoretical uncertainties on \CWZ\ are evaluated as in the electron channels
and found to be 0.7-0.8\% for \CW\ and 0.3\% for \CZ.

The uncertainty on the theoretical extrapolation from the fiducial
region to the total phase space for $W$ and $Z$ production is between 1.5\% and 2.1\%.

The cross sections measured as a function of the $W$ muon pseudorapidity,
for separated charges, and of the $Z$ rapidity are shown in
\Tabs~\ref{tab:shortmuwp}, \ref{tab:shortmuwm} and \ref{tab:shortzmu}.
The statistical, bin correlated and uncorrelated systematic and total
uncertainties are provided. The uncertainties on the extrapolation 
to the common fiducial volume, on electroweak and multijet
backgrounds, on the momentum scale and resolution are treated as fully
correlated between bins for both $W$ and $Z$ measurements. Other
uncertainties are considered as uncorrelated. 

The statistical uncertainties on the $W$ differential cross sections
are in the range 1-2\%, and the total uncertainties are in the
range of 2-3\%.

The differential $Z$ cross section is measured with a
statistical uncertainty of about 2\% up to $|y_Z|<1.6$, 2.6\% for
$1.6<|y_Z|<2.0$ and 4.4\% for $2.0<|y_Z|<2.4$. 
The available number of $Z$ events dominates the total uncertainty, with systematic sources
below 1.5\% in the whole rapidity range.

\begin{table}[htbp]
  \centering    
  \begin{tabular}{lcccc}
    \hline
    \hline
              &      $N$         &    $B$           & $\CWZ$               & $\AWZ$   \\
    \hline
    $W^+$     & \NWmuplusCands  & \nWmuplusBkg    & $0.796 \pm 0.016$    & $0.495 \pm 0.008$ \\
    $W^-$     & \NWmuminusCands & \nWmuminusBkg   & $0.779 \pm 0.015$    & $0.470 \pm 0.010$ \\
    $W^\pm$   & \NWmuCands      & \nWmuBkg        & $0.789 \pm 0.015$    & $0.485 \pm 0.007$ \\
    $Z$       & \NZmuCands      & \nZmuBkg        & $0.782 \pm 0.007$      & $0.487 \pm 0.010$ \\
    \hline 
    \hline
  \end{tabular}
  \caption{\it Number of observed candidates N and expected background
    events B, efficiency and acceptance correction factors for the $W$
    and $Z$ muon channels. Efficiency scale factors used to correct the simulation
    for differences between data and MC are included in the \CWZ\ factors. 
    The given uncertainties are the quadratic sum of
    statistical and systematic components.
    The statistical uncertainties on the $\CWZ$ and $\AWZ$ factors are negligible.}
  \label{tab:WZaccMuon}
\end{table}

\begin{table}[tbqh]
\small
\begin{center}
\begin{tabular}{lccccc}
\hline
\hline
  & \multicolumn{5}{c}{\bf $\sigma_W^{\rm fid} \cdot$ BR($W \to \mu \nu$) \ \ [nb]} \\
\hline
  &  & & \hspace{-0.45cm} sta & \hspace{-0.6cm} sys & \hspace{-0.9cm} lum  \\
\hline
${W^+} $      &   \multicolumn{5}{c}{$~\sigfidWmuplus$}  \\
${W^-} $      &   \multicolumn{5}{c}{$~\sigfidWmuminus$}  \\
$ W^\pm $         &   \multicolumn{5}{c}{$~\sigfidWmu$}       \\
\hline
  & \multicolumn{5}{c}{\bf $\sigma_W^{\rm tot} \cdot$ BR($W \to \mu \nu$) \ \ [nb]} \\
\hline
  & \hspace{1.26cm} & sta & \hspace{0.48cm} sys & \hspace{0.43cm} lum & \hspace{0.3cm} acc \\
\hline
${W^+} $      &   \multicolumn{5}{c}{$~\sigWmuplus$}  \\
${W^-} $      &   \multicolumn{5}{c}{$~\sigWmuminus$}  \\
$ W^\pm $         &   \multicolumn{5}{c}{$~\sigWmu$}       \\
\hline
\hline
 & \multicolumn{5}{c}{\bf $\sigma_{Z/\gamma^*}^{\rm fid} \cdot$ BR($Z/\gamma^* \to \mu\mu$)\ [nb]} \\
\hline
  &  & & \hspace{-0.45cm} sta & \hspace{-0.6cm} sys & \hspace{-0.9cm} lum  \\
\hline
$Z/\gamma^*$      &   \multicolumn{5}{c}{$~\sigfidZmu$}   \\
\hline
 & \multicolumn{5}{c}{\bf $\sigma_{Z/\gamma^*}^{\rm tot} \cdot$ BR($Z/\gamma^* \to \mu\mu$)\ [nb]} \\
\hline
  & \hspace{1.26cm} & sta & \hspace{0.48cm} sys & \hspace{0.43cm} lum & \hspace{0.3cm} acc \\
\hline
$Z/\gamma^*$      &   \multicolumn{5}{c}{$~\sigZmu$}   \\
\hline
\hline
\end{tabular}
\caption{\it Fiducial and total  cross sections times branching ratios 
   for $W^+$, $W^-$, $W^{\pm}$ and $\Zg$ production in the muon 
   decay channel.
   The muon fiducial regions are defined in \Sec~\ref{sec:sigdef}.  
   The uncertainties denote the statistical (sta), the experimental systematic (sys),
   the luminosity (lum), and the extrapolation (acc) uncertainties.}
\label{tab:muonxsec}
\end{center}
\end{table}

\begin{table}[tbqh]
\small
\begin{center}  
\begin{tabular}{lcccc}
\hline
\hline
 & $\delta \sigma_{W^\pm}$  &  $\delta \sigma_{W+}$    &  $\delta \sigma_{W-}$   & $\delta \sigma_Z$  \\
\hline
Trigger                          & $0.5$        &  $0.5$                  &  $0.5$                    & $0.1$   \\
Muon reconstruction              & $0.3$        &  $0.3$                  &  $0.3$                    & $0.6$   \\
Muon isolation                   & $0.2$        &  $0.2$                  &  $0.2$                    & $0.3$   \\
Muon $\pT$ resolution            & $0.04$        &  $0.03$                  &  $0.05$                    & $0.02$  \\
Muon $\pT$ scale                 & $0.4$         &  $0.6$                   &  $0.6$                     & $0.2$   \\
QCD background                   & $0.6$         &  $0.5$                   &  $0.8$                     & $0.3$   \\
Electroweak+$t\bar{t}$ background           & $0.4$         &  $0.3$                   &  $0.4$                     & $0.02$  \\
\met\                                           
resolution and scale             & $0.5$         &  $0.4$                   &  $0.6$                     &   -     \\
Pile-up modeling                & $0.3$         &  $0.3$                   &  $0.3$                     & $0.3$   \\
Vertex position                  & 0.1  & 0.1 & 0.1 & 0.1\\
\CWZ\ theoretical uncertainty    & $0.8$         &  $0.8$                   &  $0.7$                     & $0.3$    \\
\hline
Total experimental uncertainty   & $1.6$         &  $1.7$                   &  $1.7$                     & $0.9$   \\ 
\AWZ\ theoretical uncertainty    & $1.5$         &  $1.6$                   &  $2.1$                     & $2.0$   \\
\hline
\hline
Total excluding luminosity       & $2.1$         &  $2.3$                   &  $2.6$                     & $2.2$   \\
\hline
Luminosity                       &                \multicolumn{4}{c}{$\dlumi$}   \\
\hline
\hline
\end{tabular}
\caption{\it Summary of relative systematic uncertainties
  on the measured integrated cross sections in the muon channels 
  in per cent. 
  The efficiency systematic uncertainties are partially correlated between the trigger, reconstruction and
  isolation terms. This is taken into account in the computation of the total uncertainty
   quoted in the table. The theoretical uncertainty on \AWZ\ applies only
   to the total cross section.}
\label{muonsyst}       
\end{center}
\end{table}

\clearpage

\section{Combined Cross Sections and Comparison with Theory}
\label{sec:combicross}

\subsection{Data Combination}
\label{sec:combi}

Assuming lepton universality for the $W$ and $Z$ boson $e$ and $\mu$ decays,
the measured cross sections in both channels can be 
combined to decrease the statistical and systematic uncertainty. 
This combination cannot trivially be applied to the pure fiducial cross sections 
as somewhat different geometrical acceptances are used for the
electron and the muon measurements. 
This requires the introduction of the common kinematic regions, defined in Sec.~\ref{sec:sigdef},
where $W$ and $Z$ measurements can be combined.

The method of combination used here is an averaging procedure which
has been introduced and described in detail
in~\cite{Glazov:2005rn,Aaron:2009bp}.
It distinguishes different sources of systematic errors on the combination 
of the $W$ and $Z$ cross section measurements, in electron and muon channels.

The sources of uncertainty which are fully correlated between 
the electron and muon measurements are: the hadronic recoil uncertainty of the \met\ measurement (for $W$ measurements),
electroweak backgrounds, pile-up effects, uncertainties of 
the $z$-vertex position, the theoretical
uncertainties on the acceptance and extrapolation 
correction factors.

The sources of  uncertainty considered fully correlated bin-to-bin and across data sets are:
the extrapolation into non-covered phase space, normalisation
of the electroweak background, lepton energy or momentum scale and resolution,
and systematic effects on reconstruction efficiencies.

In addition, the QCD background systematics are bin-to-bin correlated but 
independent for the $e$ and $\mu$ data sets. The statistical
components of the lepton identification efficiencies
are largely bin-to-bin uncorrelated but correlated for the 
$W$ and $Z$ cross sections, whereas the statistical uncertainties
of the background and the electron isolation determinations are
fully uncorrelated sources.
Finally, some sources are considered as fully anti-correlated for 
$W^+$ and $W^-$ production, specifically the PDF uncertainty on $C_W$
and the charge misidentification.
The luminosity uncertainty is common to all data points and it is therefore
not used in the combination procedure.

In total there are $59$ differential cross section measurements 
entering the combination with
$30$ sources of correlated systematic uncertainties. 
The data are combined using the following
$\chi^2$ function~\cite{Aaron:2009bp} which is minimised in the averaging procedure
\begin{eqnarray*}
  \chi^2 &=& \sum_{k,i} w^i_k
  \frac{\left[m^i-\left({\mu^i_k} + \sum_j \gamma^i_{j,k} m^i b_j\right)\right]^2}
{(\delta^i_{\mathrm{sta}, k})^2 \mu^i_k (m^i-  \sum_j \gamma^i_{j,k} m^i b_j ) + (\delta^i_{\mathrm{unc},k} m^i)^2 }\\
&+& \sum_j b_j^2.
\label{eq:avebeta}
\end{eqnarray*}
The sums run over all measurement sets $k$ and points $i$ considered.
In case a specific set $k$ contributes a measurement $\mu^i_k$ to
point $i$ one has $w^i_k=1$, otherwise $w^i_k=0$. The deviations of
the combined measurements $m^i$ from the original measurements
$\mu^i_k$ are minimised. The correlated error sources $j$ can shift,
i.e. $b_j \neq 0$, where $b_j$ is expressed in units of standard
deviations, and such shifts incur a $\chi^2$ penalty of $b_j^2$. The
relative statistical and uncorrelated systematic uncertainties of a
specific measurement are labelled $\delta^i_{\mathrm{sta},k}$ and
$\delta^i_{\mathrm{unc},k}$, respectively. The relative correlated systematic uncertainties
are given by the matrix $\gamma^i_{j,k}$, which quantifies the
influence of the correlated systematic error source $j$ on the
measurement $i$ in the experimental data set $k$. In addition, total
correlated uncertainty $\delta^i_{\mathrm{corr},k}$ can be estimated as a
sum in quadrature of $\gamma^i_{j,k}$.

The combined $Z$, $W^-$ and $W^+$ differential cross sections are given
in \Tabs~\ref{tab:zfullcob}, \ref{tab:wmfullcob}, \ref{tab:wpfullcob}.
The data can be obtained electronically through the HepData repository~\cite{hepdata}.
The results are quoted with their statistical, uncorrelated and correlated
uncertainties per bin, where the influence of all correlated sources
is quantified individually with the matrix $\gamma^i_{j,k}$.

The data show good compatibility, with the total $\chi^2/\mathrm{dof} =33.9/29$.
A good level of agreement is also seen if combinations are 
performed separately for the $Z$
($\chi^2/\mathrm{dof}=15.5/9$), the $W^+$ ($\chi^2/\mathrm{dof} = 10.2/10$)
and the $W^-$ data ($\chi^2/\mathrm{dof} = 7.0/10$).

\subsection{Theoretical Calculations}
\label{sec:pqcd}
The precision of the current differential and integrated
cross section measurements has reached the per cent level.
Comparisons with QCD predictions therefore
are made at next-to-next-to-leading order 
in perturbation theory using recent NNLO sets
of PDFs. The dependence of the cross section predictions
on the renormalisation ($\mu_r$) and factorisation ($\mu_f$)
scales is reduced at NNLO. Varying $\mu_r$ and $\mu_f$
independently around their central values, taken to be $M_W$ or $M_Z$, 
with the constraint $0.5 < \mu_r /\mu_f < 2$,
a maximum effect of about $3$\,\% is observed on the NLO cross
sections, which is reduced to $0.6$\,\% at NNLO, using the
MSTW08 PDF sets.

The theoretical $Z/\gamma^*$ and
$W^{\pm}$ predictions, used in the following
for a comparison with the data, are obtained with
most recent versions of the programs
FEWZ~\cite{Anastasiou:2003ds,Gavin:2010az} 
and   DYNNLO~\cite{Catani:2007vq,Catani:2009sm},
which provide NNLO cross sections for vector boson
production and decays with full spin correlations and finite
width effects.   Calculations are performed using the
$G_{\mu}$ electroweak parameter scheme and 
those values of the strong coupling constant, $\alpha_s$, which belong
to the original determinations of the PDFs.
The predictions obtained with FEWZ and DYNNLO 
are found to agree to within $0.5$\,\% for
the total and to within $1$\,\% for the fiducial cross sections 
when using the same electroweak parameter settings and the Standard
Model predictions for the total and partial widths 
of the $W$ and $Z$ vector bosons,
which also account for higher order 
electroweak and QCD corrections~\cite{Nakamura:2010zzi}.

The NNLO QCD predictions do not include corrections due 
to pure weak and interference effects between initial
and final state radiation.
Both effects have been estimated using 
the SANC program~\cite{Andonov:2008ga}.
The interference effects are 
below $0.1$\,\% for all considered channels.
Pure weak effects may change the predicted cross
sections  by about $0.5$\,\%. Shape
modifications due to the pure weak corrections are calculated to be at
most $10$\,\% of the quoted correction values. 
Since the size of the pure weak
corrections is estimated to be
of the same order as the level of agreement of the NNLO QCD
predictions for the fiducial cross sections, they are not applied for the
subsequent comparison of the theory with the data.

For the following comparisons to data, 
all integrated cross section values, the $y_Z$ distributions 
and the normalisation of the $\eta_\ell$
distributions are taken from FEWZ.
The shapes of the pseudorapidity distributions are 
taken from DYNNLO which have a higher statistical
precision than the differential distributions
obtained with FEWZ.

\subsection{Differential Cross Sections}
\label{sec:sigdif}

\begin{figure}[b]
  \centerline{\includegraphics[width=0.45\textwidth]{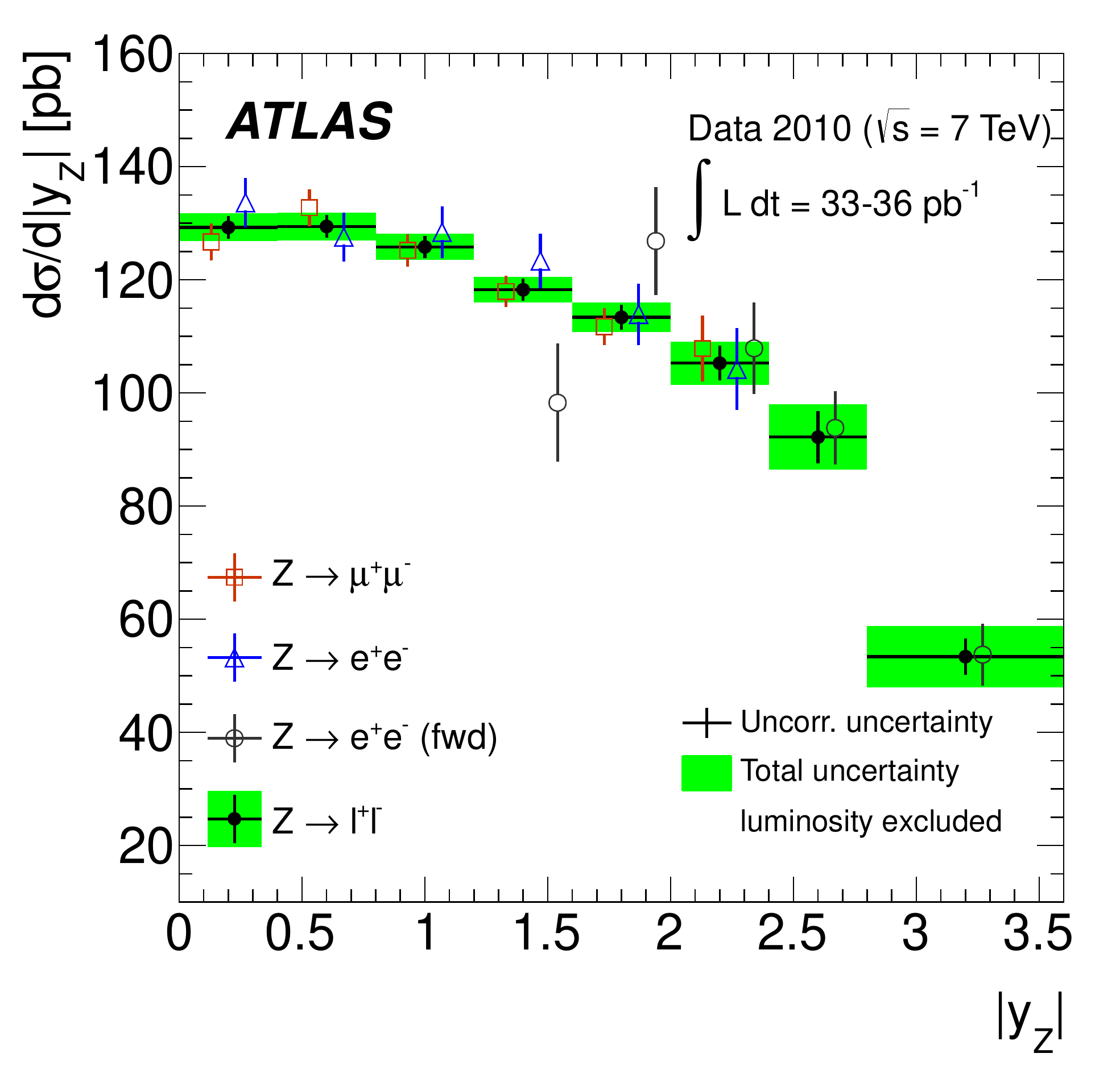}}
  \caption{\it \label{fig:combinedZ} The combined
    ${\rm d}\sigma/{\rm d}|y_Z|$ cross section, for $Z/\gamma^*\to\ell^+\ell^-$,
    compared to measurements obtained separately in the muon and electron (central
    and forward) channels. 
    The kinematic requirements are $66<m_{\ell\ell}
    <116\,\GeV$ and $p_{T,\ell} > 20\,\GeV$.
    For the combined result, the uncorrelated uncertainties 
    are shown as crosses and the total uncertainties as green boxes. 
    Only the total uncertainties are shown for uncombined measurements.
    The luminosity uncertainty is not included. 
    Points are displaced for clarity within each bin.}
\end{figure}

\begin{figure}[hb]
  \centering
  \includegraphics[width=0.45\textwidth]{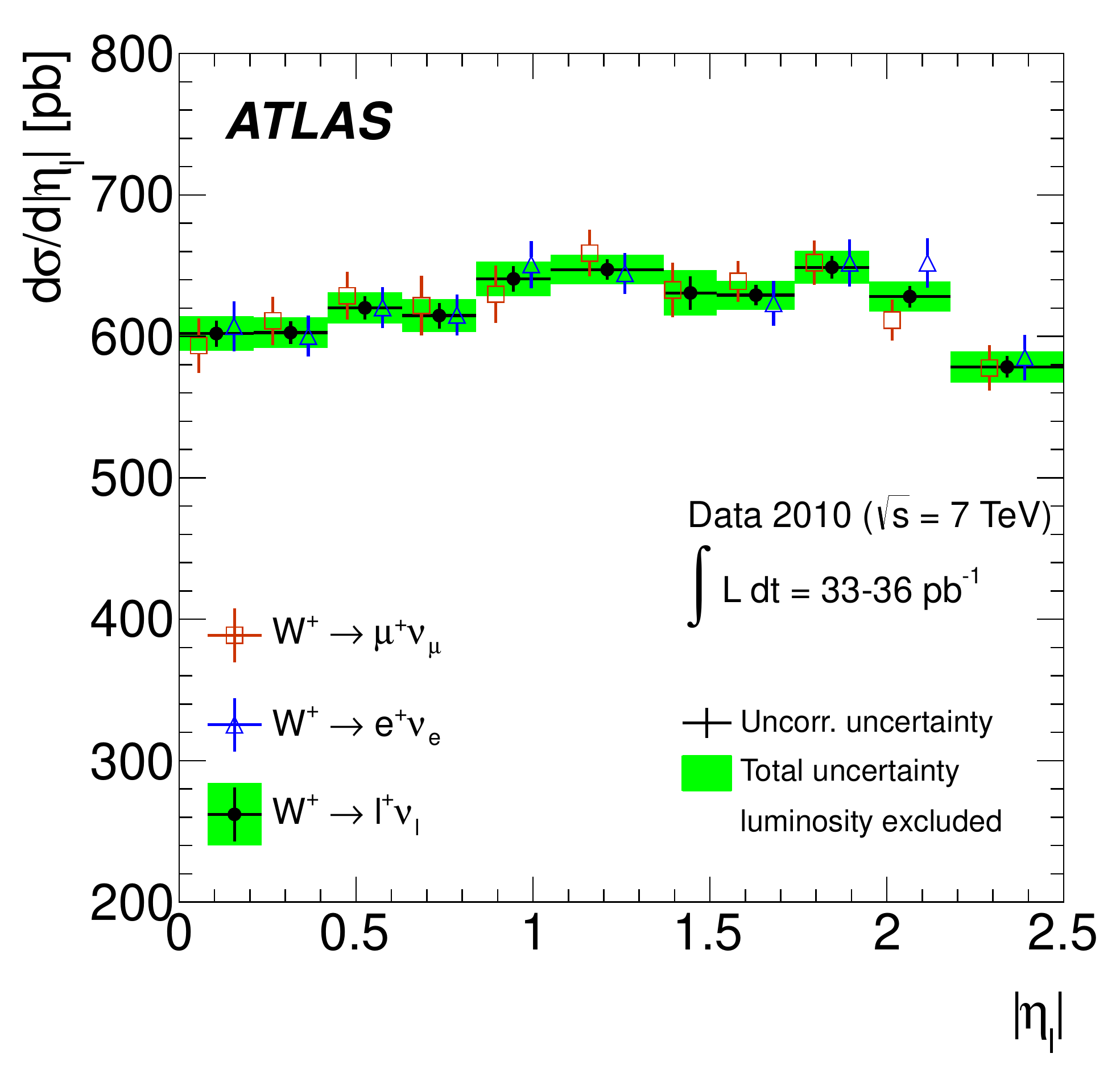} \\
  \includegraphics[width=0.45\textwidth]{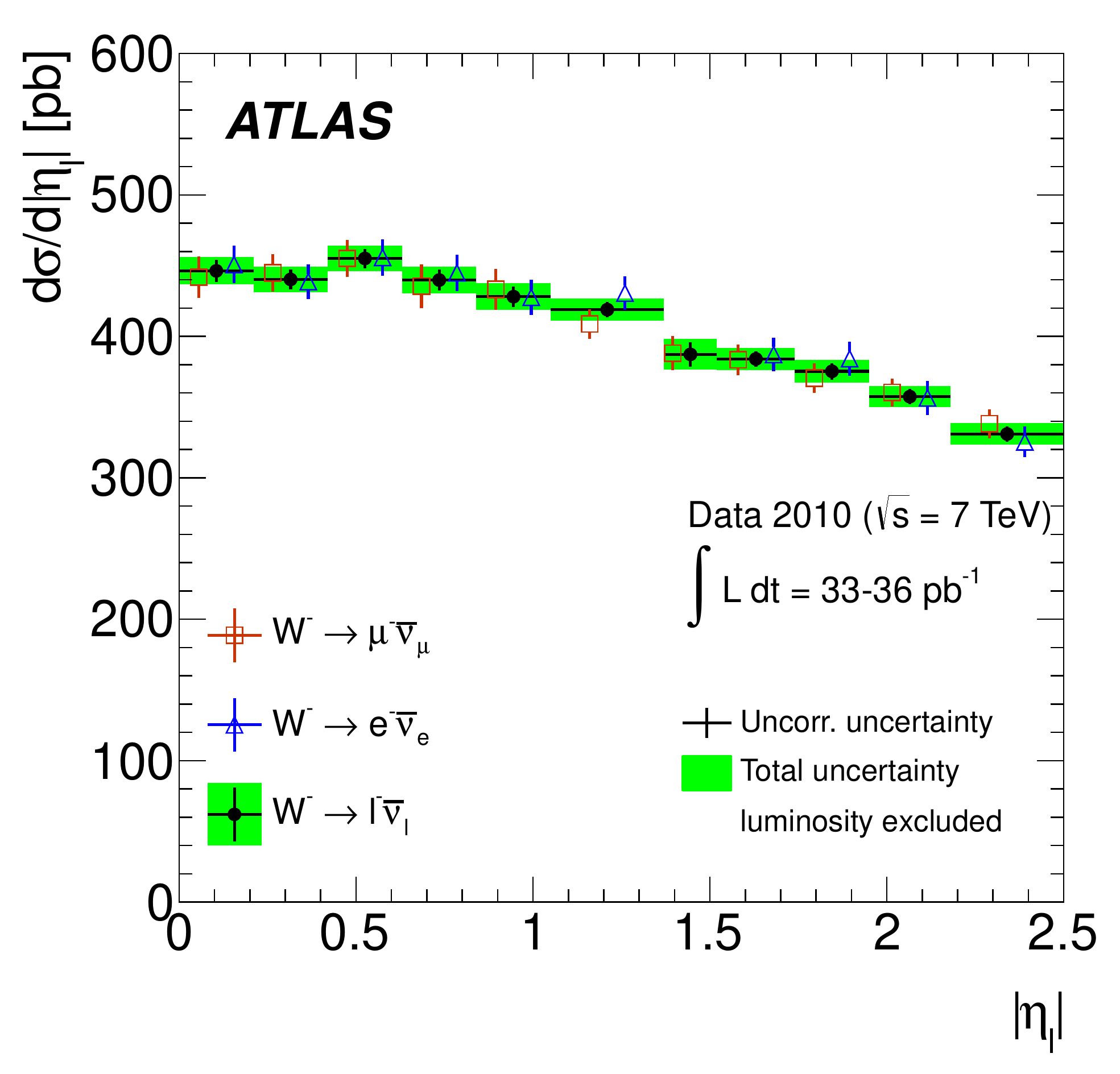}
  \caption{\it \label{fig:combinedWpm}The combined
    ${\rm d}\sigma/{\rm d}|\eta_\ell|$ cross sections, for $W^+$ (top) and
    $W^-$ (bottom), compared to measurements obtained separately in the
    electron and muon channels. 
    The kinematic requirements are $p_{T,\ell} > 20\,\GeV$, $p_{T,\nu} >
    25\,\GeV$ and $m_T > 40\,\GeV$.
    For the combined result, the uncorrelated uncertainties 
    are shown as crosses and the total uncertainties as green boxes.
    Only the total uncertainties
    are shown for uncombined measurements.
    The luminosity uncertainty is not included.
    Points are displaced for clarity within each bin.}
\end{figure}

The differential $Z$ and $W^{\pm}$ cross sections are shown in
\Figs~\ref{fig:combinedZ} and \ref{fig:combinedWpm}. The
measurements for different channels are seen to be in good agreement
with each other.
Excluding the overall luminosity normalisation uncertainty,
the data accuracy reaches about $2$\,\% in the central 
region of the $Z$ rapidity.
In the most forward region of the $Z$ cross section measurement, 
the accuracy is still limited to $6$~($10$)\,\% at $y_Z \simeq 2.6~(3.2)$.
For the $W$ cross section measurements, a precision of about $2$\,\% is 
obtained in each bin of $\eta_\ell$.

\begin{figure}[b]
  \centerline{\includegraphics[width=0.45\textwidth]{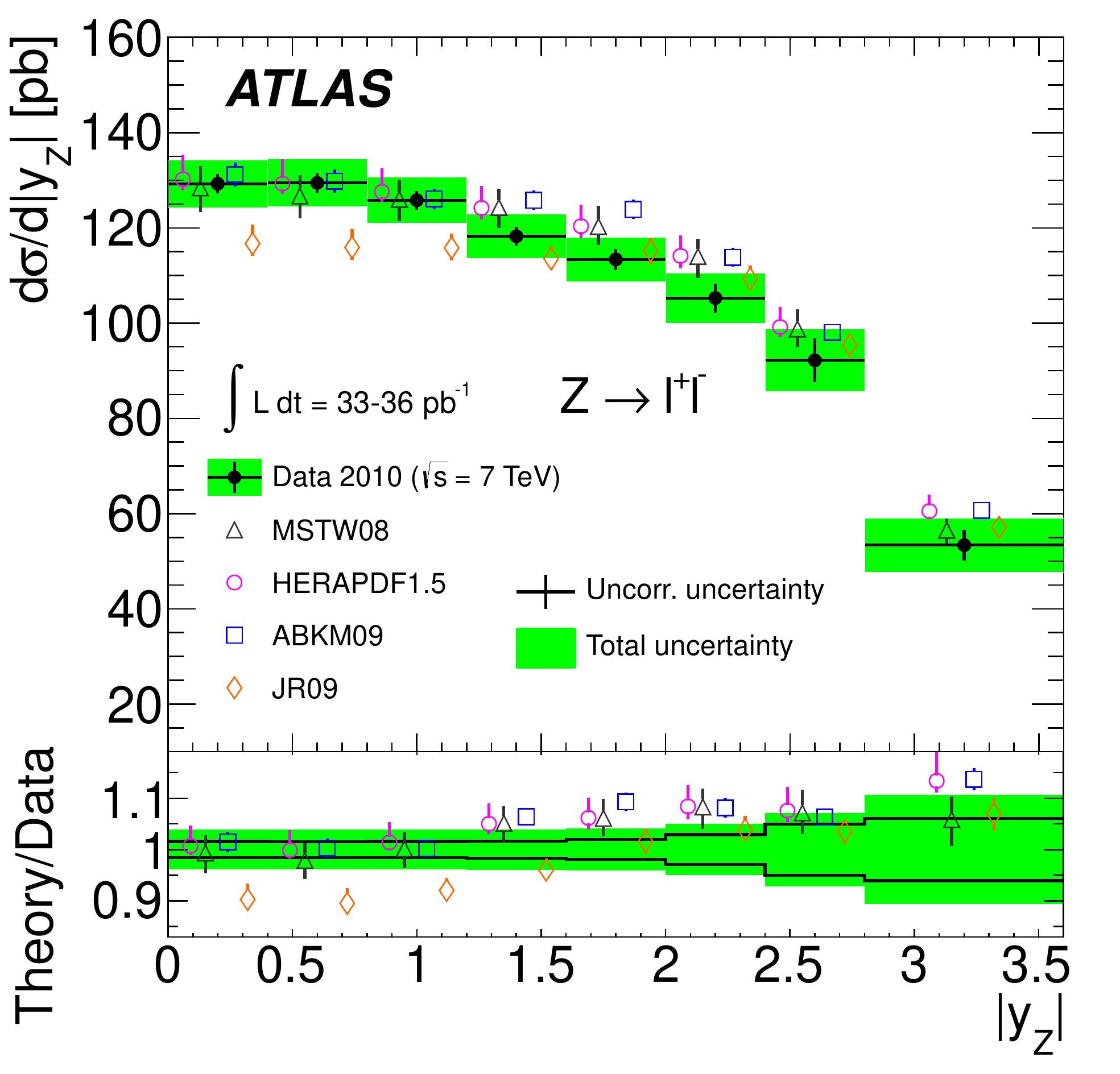}}
  \caption{\it \label{fig:ztheo}Differential ${\rm d}\sigma/{\rm d}|y_Z|$
    cross section measurement for \Zll\ compared to NNLO theory predictions
    using various PDF sets. The kinematic requirements are $66<m_{\ell\ell}
    <116\,\GeV$ and $p_{T,\ell} > 20\,\GeV$.
    The ratio of theoretical predictions to data is also shown.
    Theoretical points are displaced for clarity within each bin.}
\end{figure}

\begin{figure}[b]
  \begin{centering}
    \includegraphics[width=0.45\textwidth]{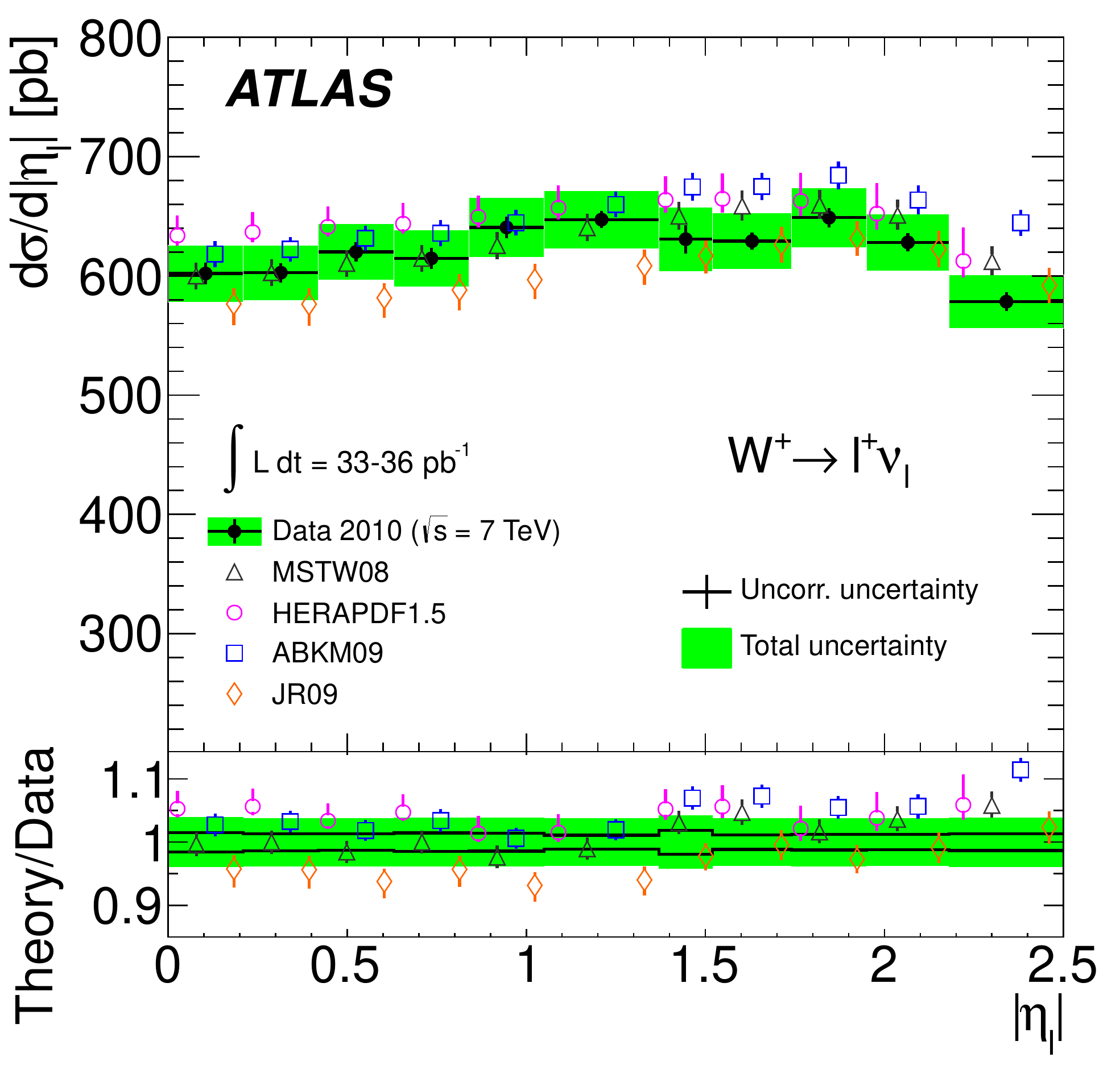}
    \includegraphics[width=0.45\textwidth]{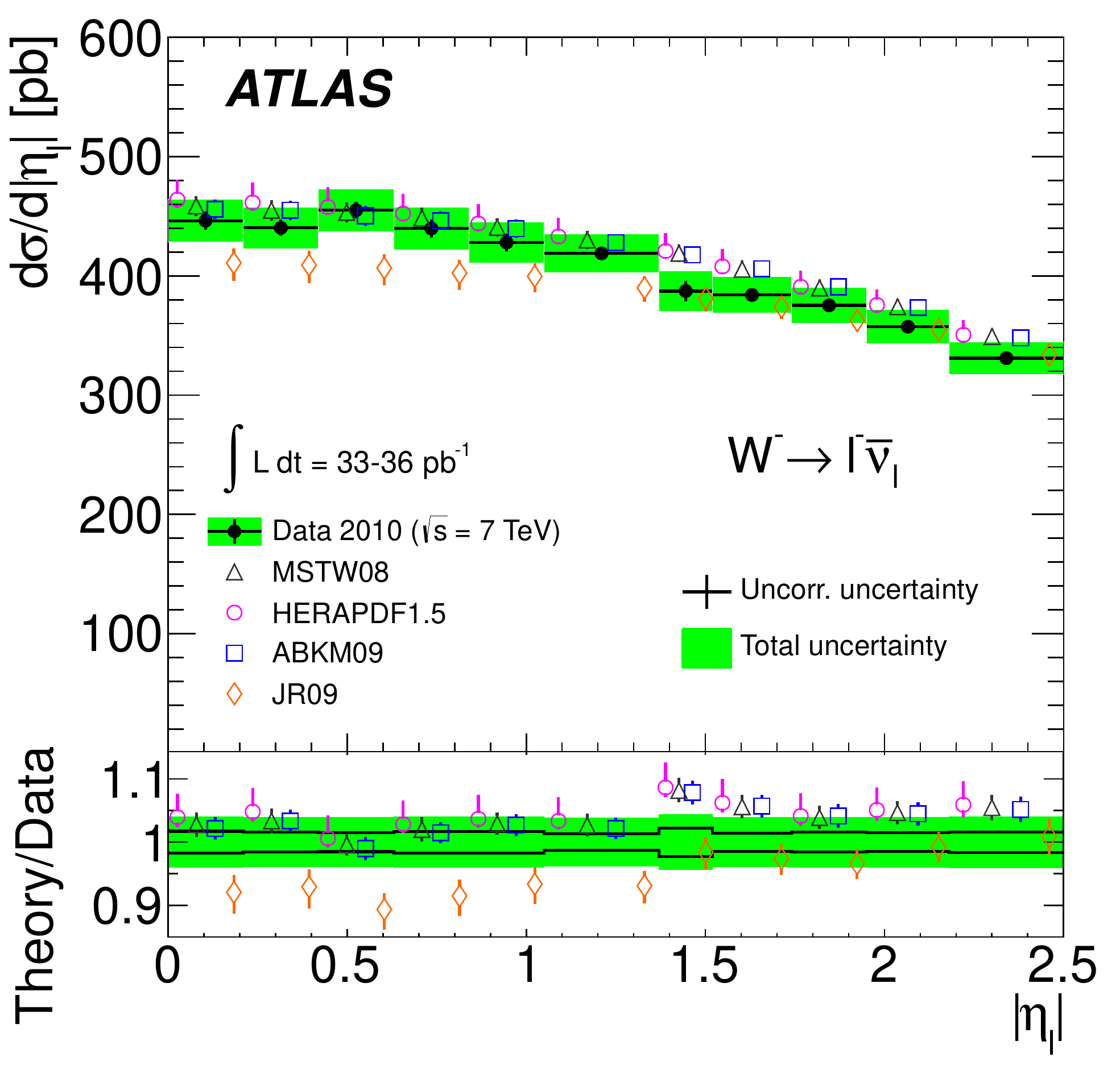}
    \caption{\it \label{fig:wptheo} Differential ${\rm d}\sigma/{\rm d}|\eta_{\ell^+}|$ (top)
      and ${\rm d}\sigma/{\rm d}|\eta_{\ell^-}|$ (bottom) cross section measurements
      for \Wln\ compared to the NNLO theory predictions using various PDF sets. 
      The kinematic requirements are $p_{T,\ell} > 20\,\GeV$, $p_{T,\nu} >
      25\,\GeV$ and $m_T > 40\,\GeV$.
      The ratio of theoretical predictions to data is also shown.
      Theoretical points are displaced for clarity within each bin.}
  \end{centering}
\end{figure}

The combined differential $Z$ and $W^{\pm}$
cross sections are compared in \Figs~\ref{fig:ztheo} and \ref{fig:wptheo}
with the calculated NNLO predictions using 
the JR09, ABKM09, HERAPDF1.5 and MSTW08
NNLO PDF sets.
The uncertainties of the bin-wise predictions are a convolution of the PDF uncertainties, considered
by the authors of the various PDF sets~\footnote{The HERAPDF analysis considers explicitly uncertainties
due to parameterisation and fit parameter choices. This leads to somewhat
enlarged and asymmetric errors as compared to the genuine 
experimental uncertainties, which in the HERAPDF analysis correspond
to a change of $\chi^2$ by one unit.}
 to correspond to $68$\,\% C.L., and a residual numerical uncertainty of below $0.5$\,\%.
One observes that the measured $y_Z$ and $\eta_\ell$
dependencies are
broadly described by the predictions of the PDF sets considered. Some deviations,
however, are visible, for example the lower $Z$ cross section at central rapidities
in the case of the JR09 PDF set, or the tendency of the ABKM09 prediction
to overshoot the $Z$ and the $W$ cross sections at larger $y_Z$ and
$\eta_\ell$, respectively. It thus can be expected that the differential
cross sections presented here will reduce the
uncertainties of PDF determinations and also influence the
central values.

The combined electron and muon data allow for an update of the 
measurement of the $W$ charge asymmetry
\begin{equation}
  A_\ell (\eta_\ell) = \frac{\mathrm{d}\sigma_{W^+}/\mathrm{d}\eta_{\ell} -
    \mathrm{d}\sigma_{W^-}/\mathrm{d}\eta_{\ell}}
  {\mathrm{d}\sigma_{W^+}/\mathrm{d}\eta_{\ell} +
    \mathrm{d}\sigma_{W^-}/\mathrm{d}\eta_{\ell}} \,,
\end{equation}
\noindent which 
\begin{figure}[htbp]
  \centering
  \includegraphics[width=0.45\textwidth]{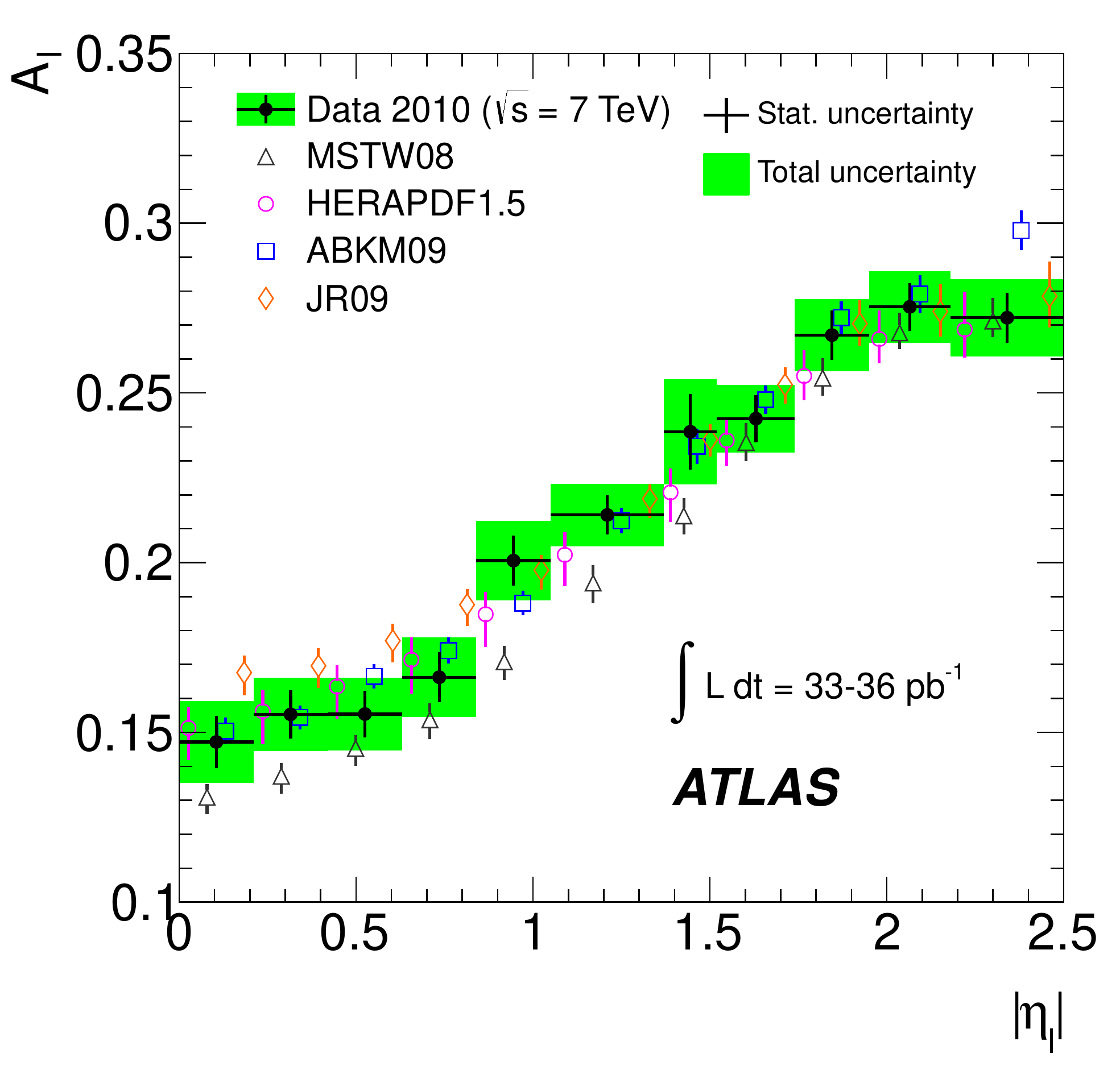}
  \caption{\it \label{fig:asy}
    Measured $W$ charge asymmetry as a function of lepton
    pseudorapidity $|\eta_\ell|$ compared with theoretical predictions
    calculated to NNLO.
    The kinematic requirements are $p_{T,\ell} > 20\,\GeV$, $p_{T,\nu} >
    25\,\GeV$ and $m_T > 40\,\GeV$.
    Theoretical points are displaced for clarity within each bin.}
\end{figure}
was previously published~\cite{Aad:2011yn} by ATLAS based
on initial muon measurements alone. The asymmetry values, obtained
in the $W$ fiducial region of this analysis, and their uncertainties are 
listed in \Tab~\ref{tab:leptasym}. The measurement accuracy ranges
between $4$ and $8$\,\%.  The previous and the new measurements
are consistent. Since the present measurement is more precise
and relies on the same data taking period, it supersedes the 
previous result.

\FFig~\ref{fig:asy} shows the measured $W$ charge asymmetry
together with the NNLO predictions obtained from the DYNNLO program. The ABKM09
and the HERAPDF 1.5 predictions give the best agreement with these results.
Some deviations from the measured $W^+$ cross section 
 of ABKM09 (HERAPDF 1.5) observed at larger (smaller) $|\eta_\ell|$, however,
illustrate that more sensitive information is inherent in
the separate $W^+$ and $W^-$ cross sections and their correlations
rather than in the asymmetry.

\subsection{Integrated Cross Sections}
\label{sec:crosec}

The combination procedure as outlined above is also used to
combine the integrated electron and muon
$Z$ and $W^{\pm}$ cross sections, separately for the
common fiducial and the total cross sections.

The integrated fiducial cross sections for the $W^+$, $W^-$, $W^{\pm}$ and 
$Z$ channels, listed in \Tab~\ref{tab:sigavfid} with their uncertainties,
are all measured to about $1$\,\% systematic uncertainty,
with significantly smaller uncertainties due to statistics and
essentially negligible uncertainties due to the extrapolation
to the common fiducial phase space. The luminosity uncertainty of
$3.4$\,\% is fully correlated between the measurements. 

\begin{table}[tbqh]
\small
\begin{center}
\begin{tabular}{lccccc}
\hline
\hline
  & \multicolumn{5}{c}{\bf $\sigma_W^{\rm fid} \cdot$ BR($W \to \ell \nu$) \ \ [nb]} \\
  & \multicolumn{5}{c}{$|\eta_\ell| < 2.5$, $p_{T,\ell} > 20\,\GeV$,} \\
  & \multicolumn{5}{c}{$p_{T,\nu} >  25\,\GeV$ and $m_T > 40\,\GeV$} \\
\hline
  & \hspace{1.26cm} & sta & \hspace{0.48cm} sys & \hspace{0.43cm} lum & \hspace{0.3cm} acc \\
\hline
${W^+} $      &   \multicolumn{5}{c}{$~\sigWplusfid$}  \\
${W^-} $      &   \multicolumn{5}{c}{$~\sigWminusfid$}  \\
$ W^{\pm} $         &   \multicolumn{5}{c}{$~\sigWfid$}       \\
\hline
\hline
 & \multicolumn{5}{c}{\bf $\sigma_{Z/\gamma^*}^{\rm fid} \cdot$ BR($Z/\gamma^* \to \ell \ell$)\ [nb]} \\
 & \multicolumn{5}{c}{$|\eta_\ell| < 2.5$, $p_{T,\ell} > 20\,\GeV$}\\
 & \multicolumn{5}{c}{and $66 < m_{\ell\ell} <116\,\GeV$}\\
\hline
  & \hspace{1.26cm} & sta & \hspace{0.48cm} sys & \hspace{0.43cm} lum & \hspace{0.3cm} acc \\
\hline
$Z/\gamma^*$      &   \multicolumn{5}{c}{$~\sigZfid$}   \\
\hline
\hline
\end{tabular}
\caption{\it  Combined cross sections times leptonic branching ratios
 for $W^+$, $W^-$, $W^{\pm}$ and $\Zg$ production within the corresponding
  fiducial regions of the measurements. 
   The uncertainties denote the statistical (sta), the experimental systematic (sys),
   the luminosity (lum), and the extrapolation (acc) uncertainties.}
\label{tab:sigavfid}
\end{center}
\end{table}
\begin{table}
  \begin{minipage}{0.2\textwidth}
    \begin{center}
      \begin{tabular}{lrrr}
        \hline
        \hline
  &  $Z$  &  $W^+$  &  $W^-$ \\ 
 \hline 
 $Z$ &       1.00 &       0.94 &       0.93 \\ 
 $W^+$ &       0.94 &       1.00 &       0.97 \\ 
 $W^-$ &       0.93 &       0.97 &       1.00 \\ 
        \hline
        \hline
      \end{tabular}
    \end{center}
  \end{minipage}
  \begin{minipage}{0.2\textwidth}
    \begin{center}
      \begin{tabular}{lrrr}
        \hline
        \hline
  &  $Z$  &  $W^+$  &  $W^-$ \\ 
 \hline 
 $Z$ &       1.00 &       0.48 &       0.44 \\ 
 $W^+$ &       0.48 &       1.00 &       0.79 \\ 
 $W^-$ &       0.44 &       0.79 &       1.00 \\ 
        \hline
        \hline
      \end{tabular}
    \end{center}
  \end{minipage}
  \caption{\it Correlation matrix for the measurements of the $Z$,
    $W^+$ and $W^-$ 
    cross sections in the fiducial volume, for the full uncertainty (left)
    and for all but the luminosity uncertainty (right).}
  \label{tab:wzfcor}
\end{table}

\begin{figure*}[htbp]
\includegraphics[width=0.45\textwidth,angle=0]{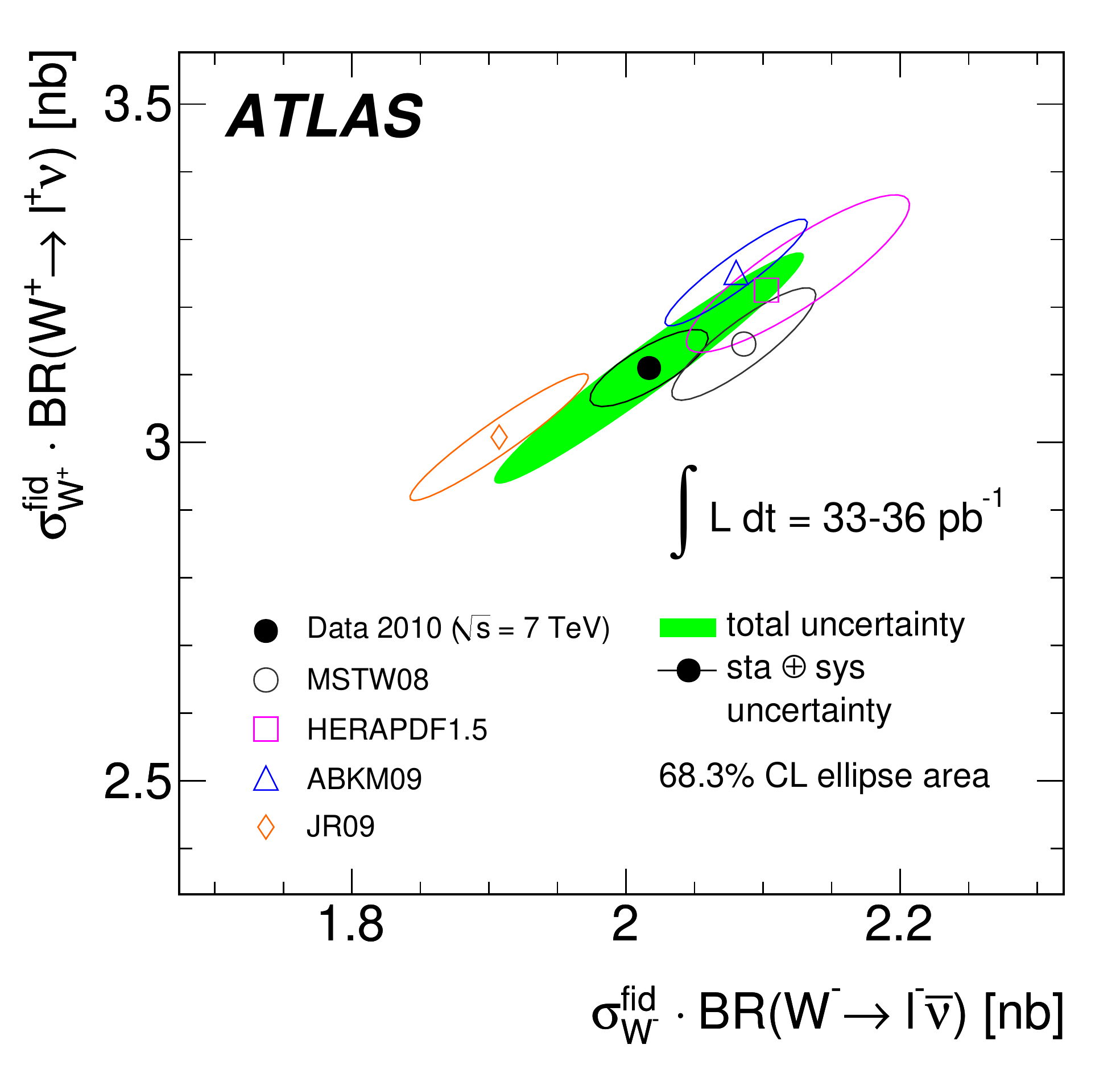}
\includegraphics[width=0.45\textwidth,angle=0]{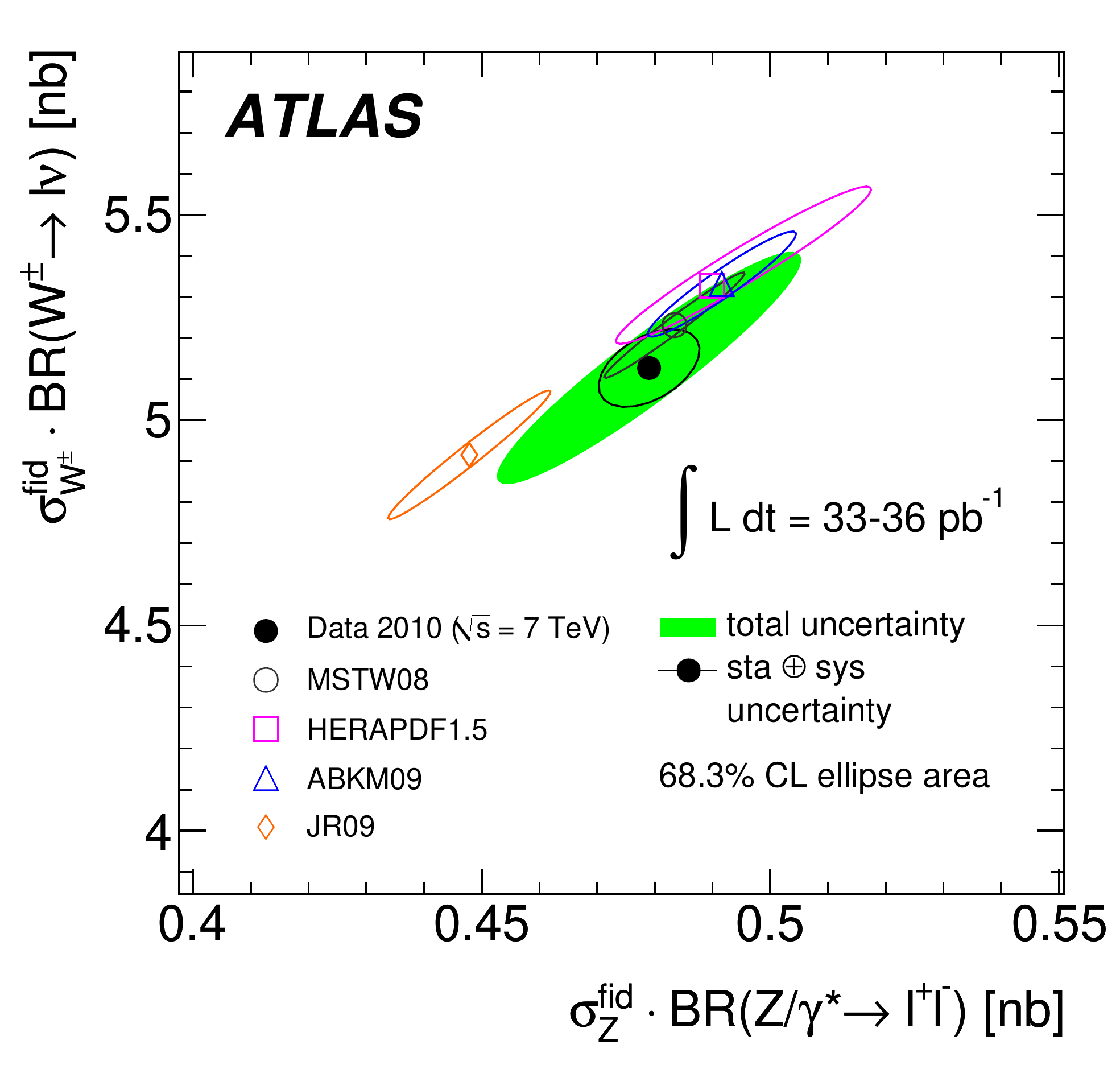}
\caption{\it Measured and predicted fiducial cross sections times leptonic branching ratios, 
$\sigma_{W^+}$ vs. $\sigma_{W^-}$ (left) and $(\sigma_{W^+} + \sigma_{W^-})$ vs. $\sigma_{\Zg}$ (right).
The ellipses illustrate the $68$\,\% CL coverage for total uncertainties (full green)
and excluding the luminosity uncertainty (open black).
The uncertainties of the theoretical predictions 
correspond to the PDF uncertainties only.
}
\label{fig:sigwpmzfid}
\end{figure*}

It is instructive to compare the measured integrated cross sections
with the theoretical predictions, evaluated in the fiducial
region of the measurement. The cross sections are calculated,
as described above, to NNLO using the FEWZ program and
the four NNLO PDF sets as used also for the differential comparisons.
\FFig~\ref{fig:sigwpmzfid} shows the $W^+$ and $W^-$
cross sections (left) and the  $(W^+ + W^-)$ and
$\Zg$ cross section (right). The outer ellipse 
is obtained using the correlation coefficients
for the total uncertainty, while the inner, much shorter
ellipse is obtained excluding the luminosity uncertainty.
The numerical values of these correlation coefficients
are given in \Tab~\ref{tab:wzfcor}. The theoretical ellipses result from the
PDF uncertainties, quoted to correspond to about $68$\,\% CL in their
two dimensional area~\footnote{All experimental and theoretical ellipses are defined such that their area
  corresponds to $68$\,\% CL. This implies that the projections onto the
  axes correspond to $1.52$ times the usual one-dimensional
  uncertainty. Note that this convention differs from the one chosen 
  in~\cite{Nadolsky:2008zw,Martin:2009iq,Watt:2011kp},
  in which the ellipses are narrower to reflect the 
  one-dimensional uncertainties.},
and the cross section correlations are
obtained from the different error eigenvector sets.
The measurement exhibits a sensitivity to 
differences in the predicted cross sections, which 
is hindered however by the luminosity uncertainty 
which dominates the error on the integrated cross section measurement.
\begin{figure*}[htbp]
\includegraphics[width=0.45\textwidth,angle=0]{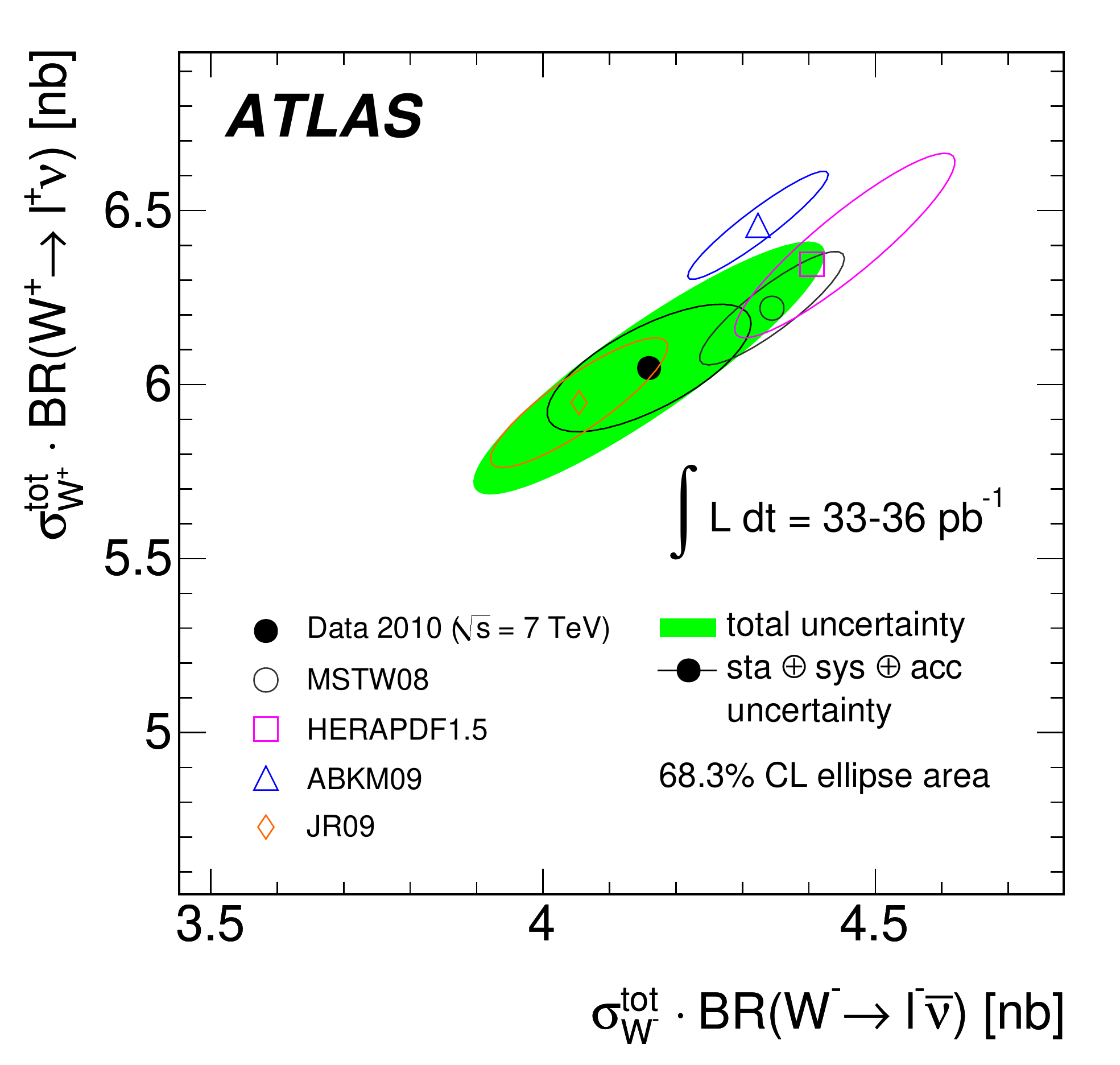}
\includegraphics[width=0.45\textwidth,angle=0]{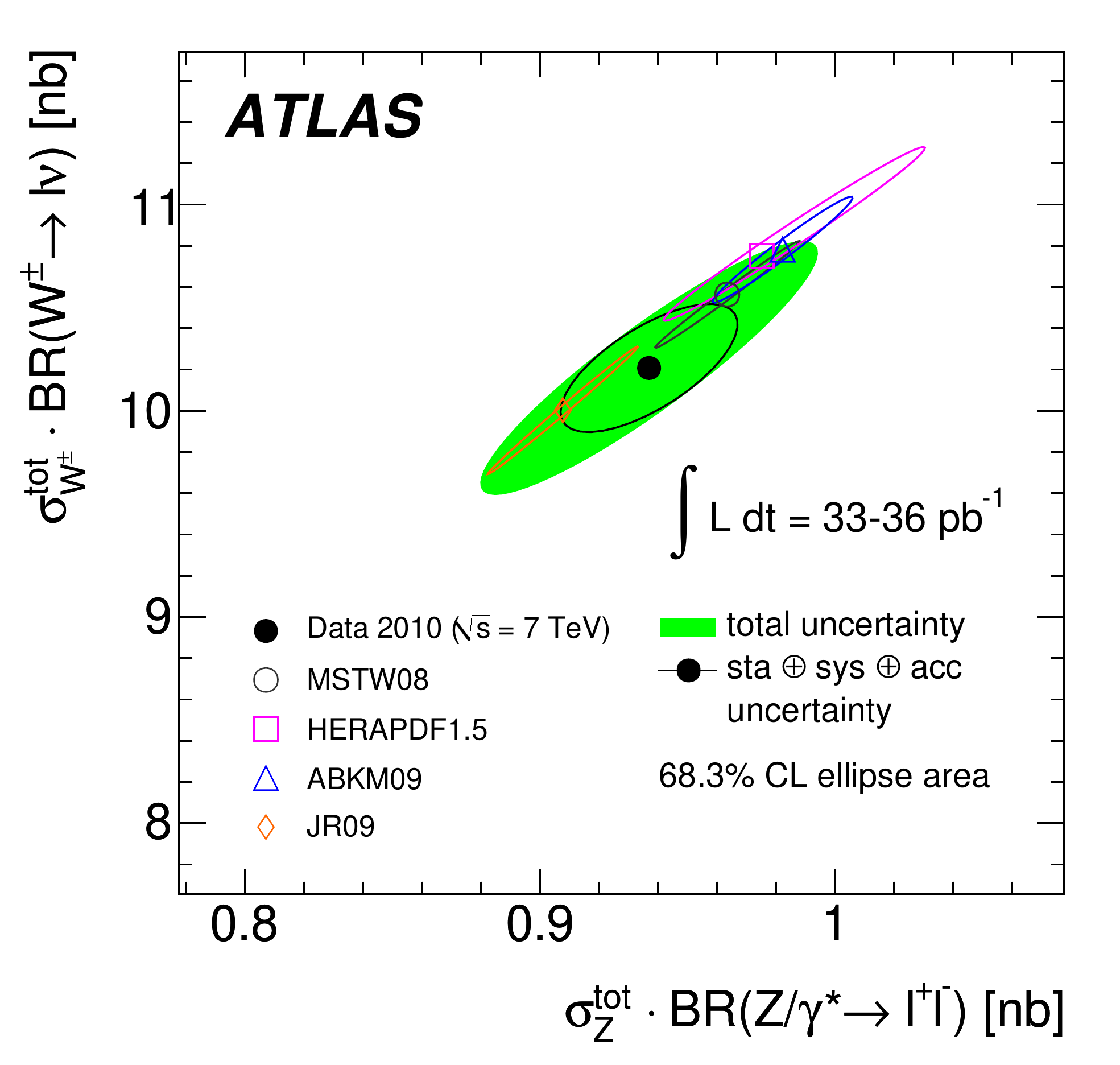}
\caption{\it Measured and predicted total cross sections times leptonic branching ratios: 
$\sigma_{W^+}$ vs. $\sigma_{W^-}$ (left) and $(\sigma_{W^+} + \sigma_{W^-})$ vs. $\sigma_{\Zg}$ (right). 
The ellipses illustrate the $68$\,\% CL coverage for total uncertainties (full green)
and excluding the luminosity uncertainty (open black).
The uncertainties of the theoretical predictions 
correspond to the PDF uncertainties only.
}
\label{fig:sigwpmz}
\end{figure*}

The predictions rely on the evolution
of the PDFs, determined mainly by deep inelastic scattering data
from HERA, into the region of the $W$ and $Z$ mass scales.
While possible deviations from the measured 
cross section values are of interest,
it is also remarkable, however, to note the overall agreement
between theory and experiment.
This is evidence that universality of the PDFs and 
perturbative QCD at high orders continue to work up to the
kinematic range probed in $W$ and $Z$ production at the LHC.

The combination and theory comparisons are also performed
with the total integrated cross sections, listed in 
\Tab~\ref{tab:sigavtot}.
The correlation coefficients are given in \Tab~\ref{tab:wztcor}.
 The pure experimental precision
of the total cross sections is as high as that of the fiducial
cross sections. However, the additional
extrapolation uncertainty, described in \Sec~\ref{sec:accext}, 
amounts to about $2$\,\%,
which is larger than  the experimental
systematic error.
The total cross section measurements
are thus less able to discriminate details of the PDFs, 
as may be deduced from comparing \Fig~\ref{fig:sigwpmz} with 
\Fig~\ref{fig:sigwpmzfid}. 
\begin{table}[tbqh]
\small
\begin{center}
\begin{tabular}{lccccc}
\hline
\hline
  & \multicolumn{5}{c}{\bf $\sigma_W^{\rm tot} \cdot$ BR($W \to \ell \nu$) \ \ [nb]} \\
\hline
  & \hspace{1.26cm} & sta & \hspace{0.48cm} sys & \hspace{0.43cm} lum & \hspace{0.3cm} acc \\
\hline
${W^+} $      &   \multicolumn{5}{c}{$~\sigWplus$}  \\
${W^-} $      &   \multicolumn{5}{c}{$~\sigWminus$}  \\
$ W^{\pm} $         &   \multicolumn{5}{c}{$~\sigW$}       \\
\hline
\hline
 & \multicolumn{5}{c}{\bf $\sigma_{Z/\gamma^*}^{\rm tot} \cdot$ BR($Z/\gamma^* \to \ell \ell$)\ [nb]} \\
 & \multicolumn{5}{c}{$66 < m_{\ell\ell} <116\,\GeV$}\\
\hline
  & \hspace{1.26cm} & sta & \hspace{0.48cm} sys & \hspace{0.43cm} lum & \hspace{0.3cm} acc \\
\hline
$Z/\gamma^*$      &   \multicolumn{5}{c}{$~\sigZ$}   \\
\hline
\hline
\end{tabular}
\caption{\it   Combined total cross sections times leptonic branching ratios
   for $W^+$, $W^-$, $W$ and $\Zg$ production. 
   The uncertainties denote the statistical (sta), the experimental systematic (sys),
   the luminosity (lum), and the extrapolation (acc) uncertainties.}
\label{tab:sigavtot}
\end{center}
\end{table}

\begin{table}
  \begin{minipage}{0.2\textwidth}
    \begin{center}
      \begin{tabular}{lrrr}
        \hline
        \hline
  &  $Z$  &  $W^+$  &  $W^-$ \\ 
 \hline 
 $Z$ &       1.00 &       0.91 &       0.91 \\ 
 $W^+$ &       0.91 &       1.00 &       0.91 \\ 
 $W^-$ &       0.91 &       0.91 &       1.00 \\ 
        \hline
        \hline
      \end{tabular}
    \end{center}
  \end{minipage}
  \begin{minipage}{0.2\textwidth}
    \begin{center}
      \begin{tabular}{lrrr}
        \hline
        \hline
  &  $Z$  &  $W^+$  &  $W^-$ \\ 
 \hline 
 $Z$ &       1.00 &       0.67 &       0.71 \\ 
 $W^+$ &       0.67 &       1.00 &       0.70 \\ 
 $W^-$ &       0.71 &       0.70 &       1.00 \\ 
        \hline
        \hline
      \end{tabular}
    \end{center}
  \end{minipage}
  \caption{\it Correlation matrix for the measurements 
    of the total Z, W$^+$ and W$^-$ 
    cross sections  for the full uncertainty (left)
    and  for all but the luminosity uncertainty (right).}
  \label{tab:wztcor}
\end{table}
Compared to the first total $W,~Z$ cross section
measurements by ATLAS~\cite{Aad:2010yt},
the statistical uncertainty  is improved by a factor of ten,
to 0.2\,(0.6)\,\% for $W$ ($Z$),  the systematic uncertainty by a factor of about five,
and the luminosity uncertainty by a factor of four, to \dlumi\,\%.

\subsection{Ratios of Cross Sections}
\label{sec:ratios}

\subsubsection{Electron-muon universality}

\begin{figure}[htbp]
  \begin{centering}
    \includegraphics[width=0.45\textwidth,angle=0]{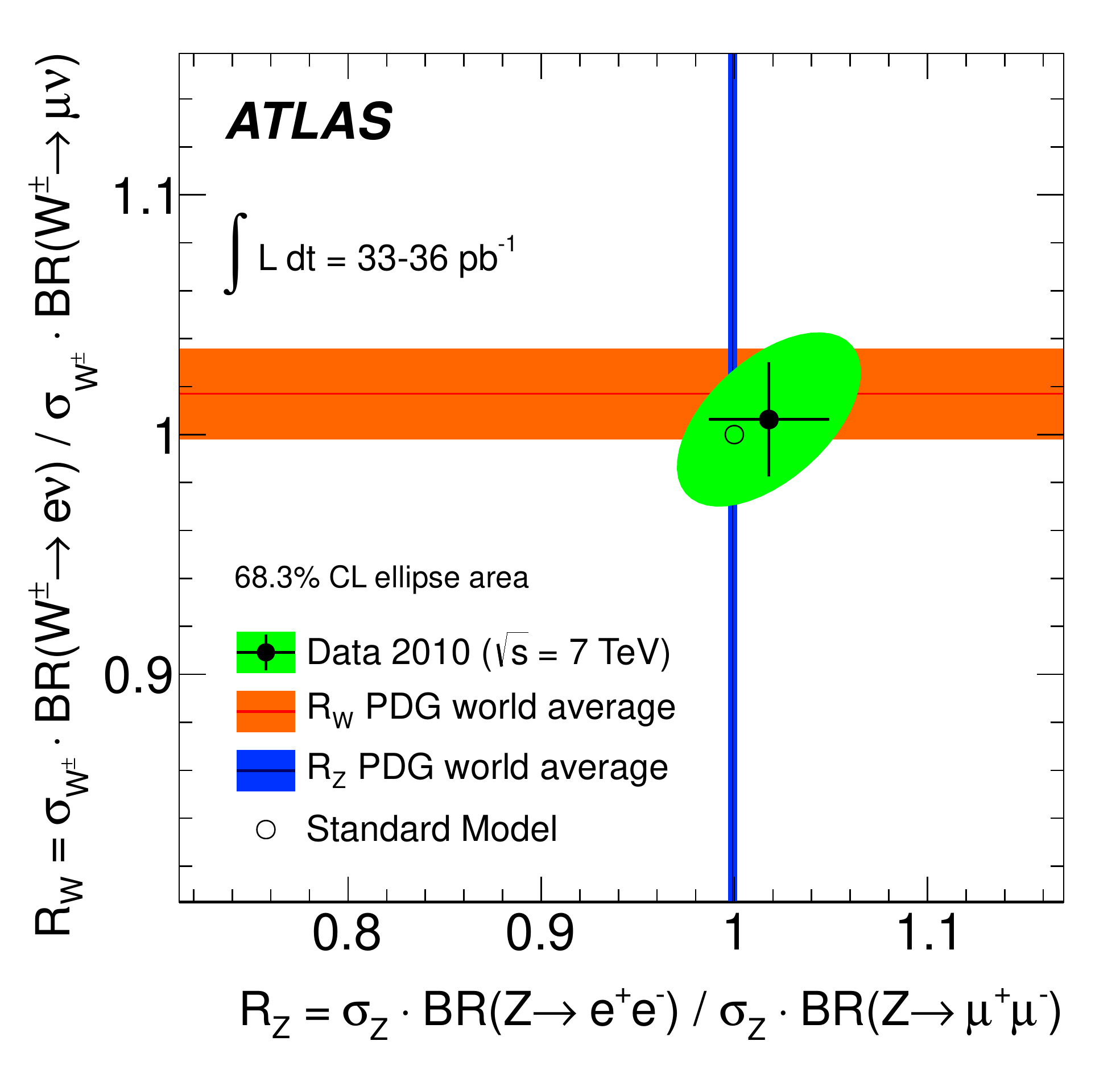}
    \caption{\it The correlated measurement of the electron-to-muon
      cross section ratios in the $W$ and the $Z$ channels.
      The vertical (horizontal) band represents the
      uncertainty of the corresponding $Z$ ($W$) branching fractions
      based on the current world average data. 
      The green ellipse illustrates the $68$\,\% CL
      for the correlated measurement of $R_W$ and $R_Z$, while the
      error bars correspond to the one-dimensional uncertainties of either
      $R_W$ or $R_Z$, respectively. }
    \label{fig:Remu}
  \end{centering}
\end{figure}

Ratios of electron and muon cross sections can be evaluated in the
common kinematic fiducial region. Since the production of the $W$ and
$Z$ bosons is
independent of the flavour of the decay lepton, the 
corresponding cross section ratios represent new measurements
of the ratios of the $e$ and $\mu$ branching fractions, i.e.
\begin{eqnarray*}
R_W &=& \frac{\sigma_W^e}{\sigma_W^{\mu}}=\frac{Br(W \rightarrow e \nu)}{Br(W \rightarrow \mu \nu)} \\
&=& 1.006 \pm 0.004\,\mathrm{(sta)} \pm 0.006\,\mathrm{(unc)} \pm 0.022\,\mathrm{(cor)}\\
 &=& 1.006 \pm 0.024. 
\end{eqnarray*}
This can be compared with the current world average of
$1.017 \pm 0.019$~\cite{Nakamura:2010zzi} and a similar measurement
performed by CDF giving $1.018 \pm 0.025$~\cite{Abulencia:2005ix}. 
Similarly one obtains for the $Z$ decays into electrons and muons a ratio
\begin{eqnarray*}
R_Z &=& \frac{\sigma_Z^e}{\sigma_Z^{\mu}}=\frac{Br(Z \rightarrow ee)}{Br(Z \rightarrow \mu \mu)} \\
&=& 1.018 \pm 0.014\,\mathrm{(sta)} \pm 0.016\,\mathrm{(unc)} \pm 0.028\,\mathrm{(cor)} \\
&=& 1.018 \pm 0.031. 
\end{eqnarray*}
This confirms $e$-$\mu$ universality in $Z$ decays as well, but the
result is much less accurate than the world average value of
$0.9991 \pm 0.0024$~\cite{Nakamura:2010zzi}.
If one uses this world average as a constraint on the analysis
presented here, the correlated systematic uncertainty on $R_W$ is
reduced, and an improved value $R_W = 0.999 \pm 0.020$ is obtained.
The correlation of $R_W$ and $R_Z$ and the comparison with the world average values
is illustrated in Fig.\,\ref{fig:Remu}.

\begin{figure}[thbp]
  \begin{center}
    \includegraphics[width=0.45\textwidth,angle=0]{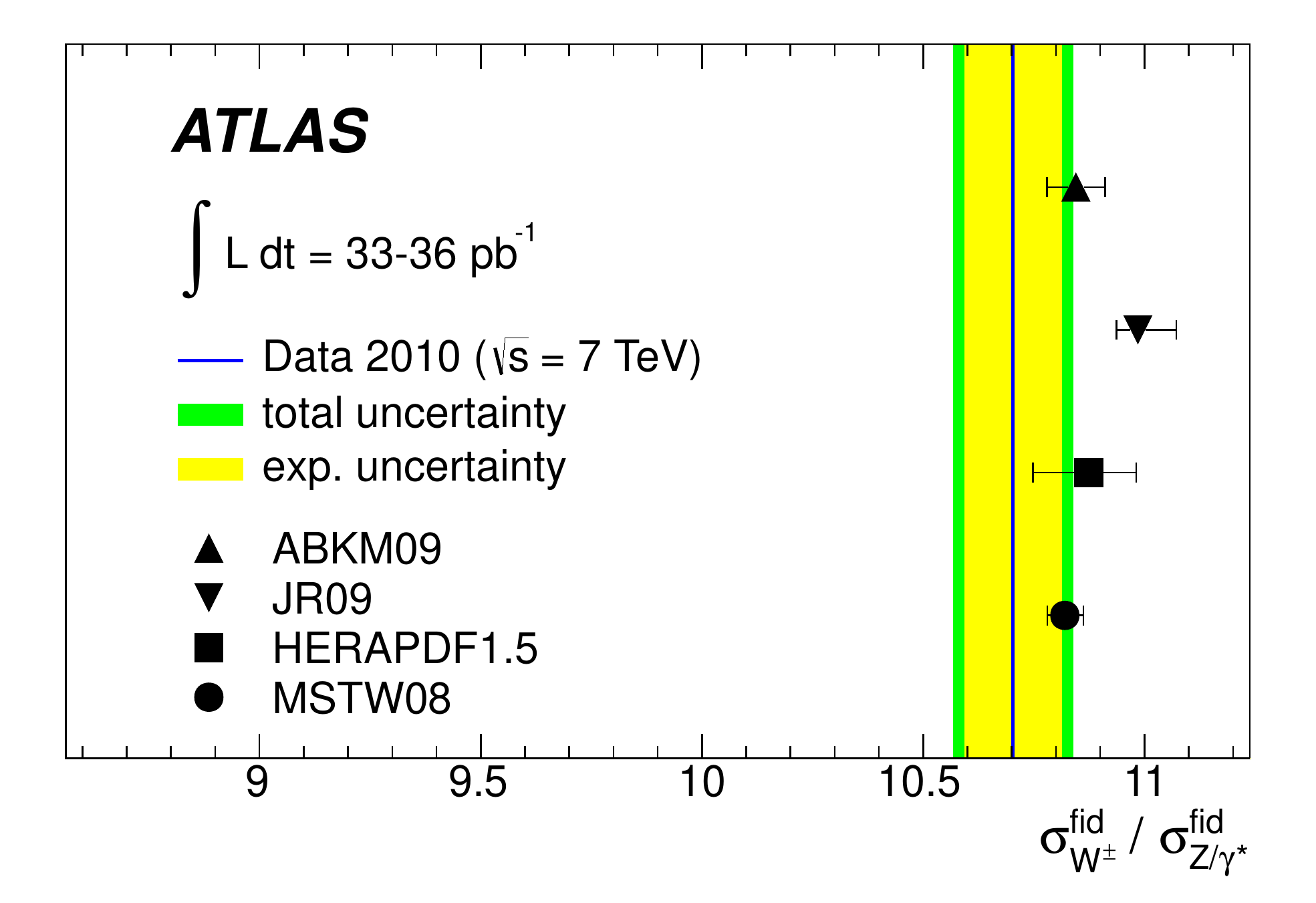}\\
    \includegraphics[width=0.45\textwidth,angle=0]{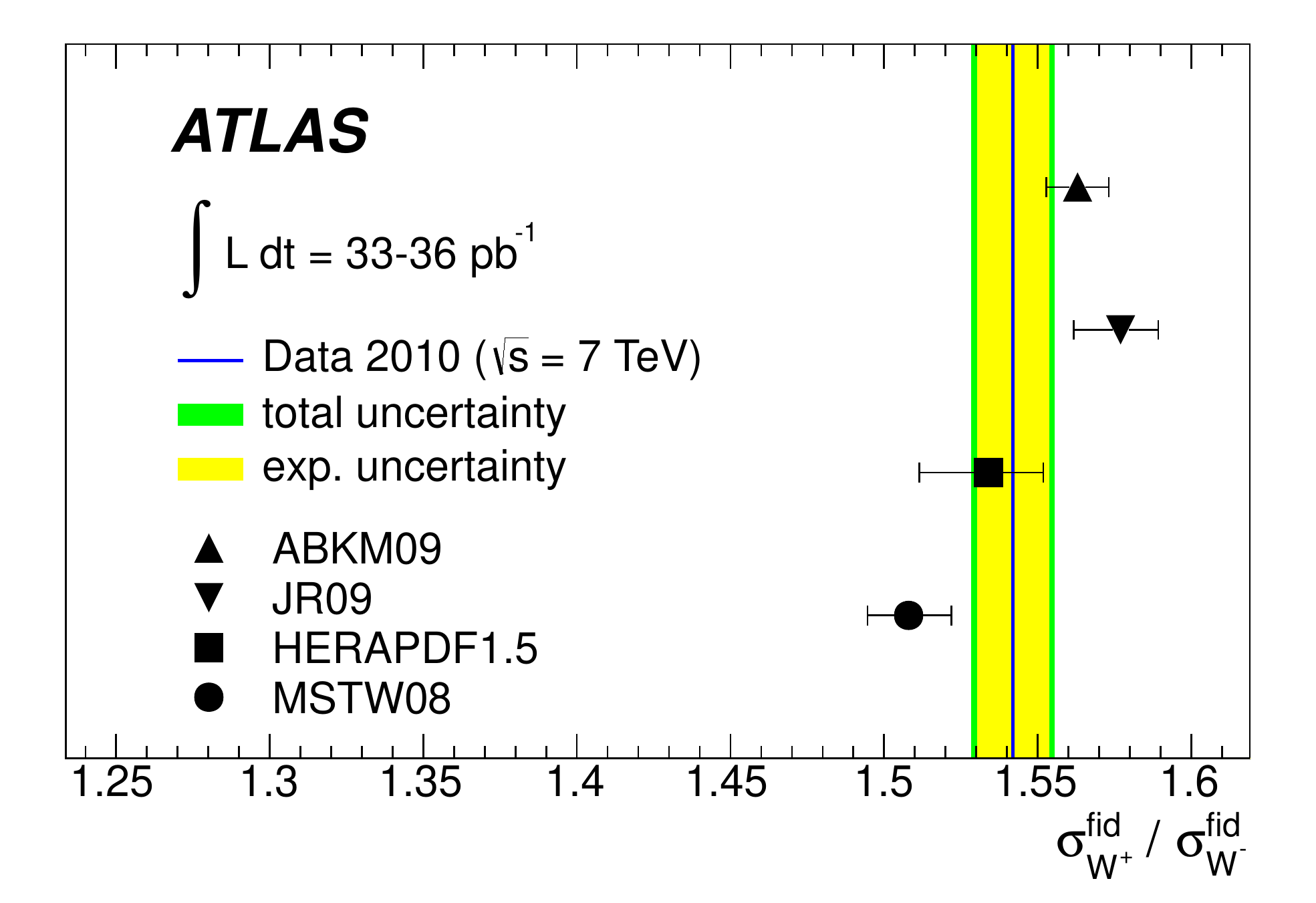}
  \end{center}
  \caption{\it  Measured and predicted fiducial cross section ratios,
    $(\sigma_{W^+} + \sigma_{W^-}) / \sigma_{\Zg}$ (top) 
    and $\sigma_{W^+} / \sigma_{W^-}$ (bottom).
    The experimental uncertainty (inner yellow band) 
    includes the experimental systematic errors. 
    The total uncertainty (outer green band) includes the
    statistical uncertainty and the small contribution from the acceptance correction.
    The uncertainties of the ABKM, JR and MSTW predictions
    are given by the PDF uncertainties considered to correspond
    to $68$\,\% CL and their correlations are derived from the
    eigenvector sets. The results for HERAPDF comprise
    all three sources of uncertainty of that set.}
  \label{fig:ratwzww_fid}
\end{figure}

\begin{figure*}[htbp]
  \includegraphics[width=0.45\textwidth,angle=0]{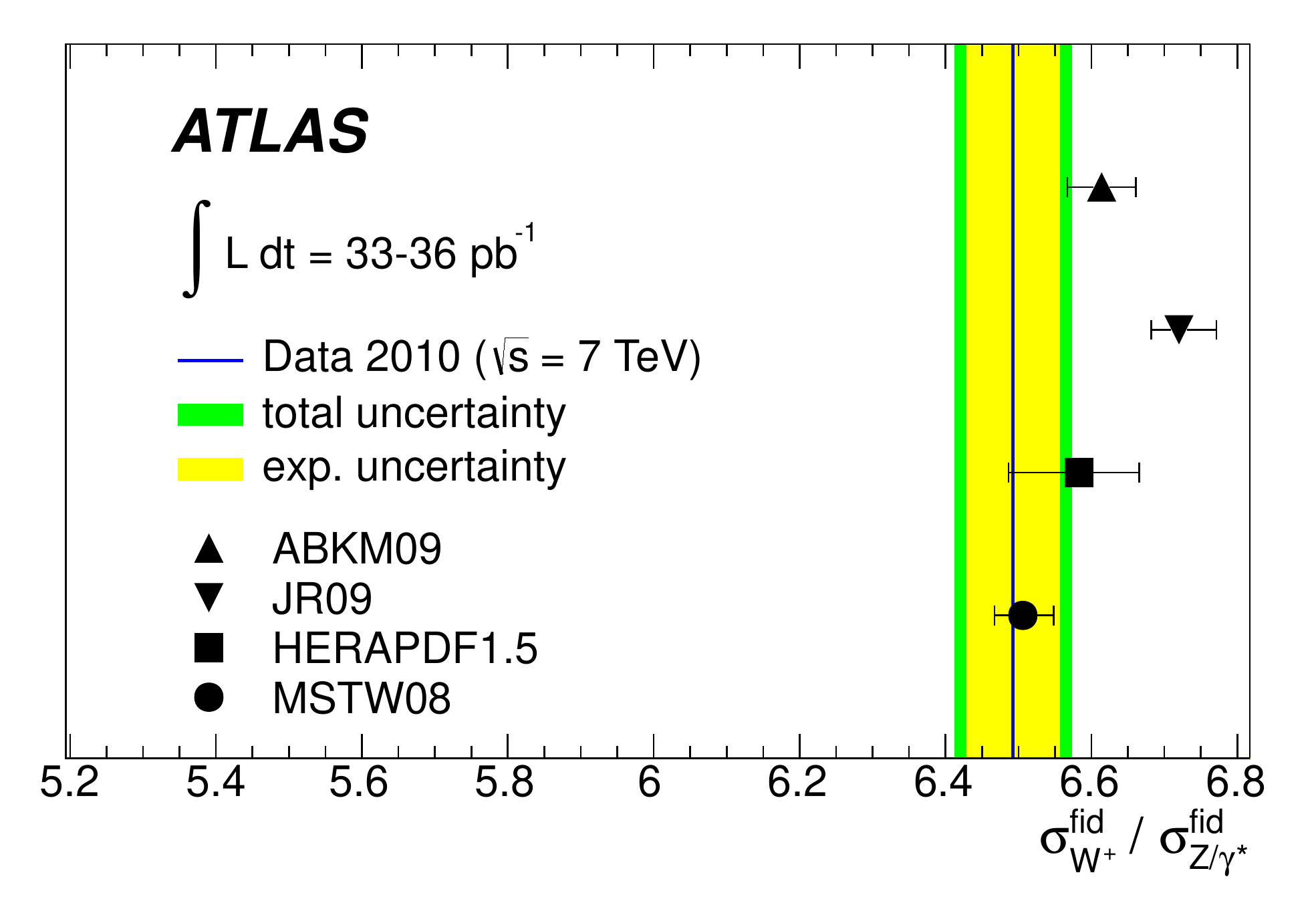}%
  \includegraphics[width=0.45\textwidth,angle=0]{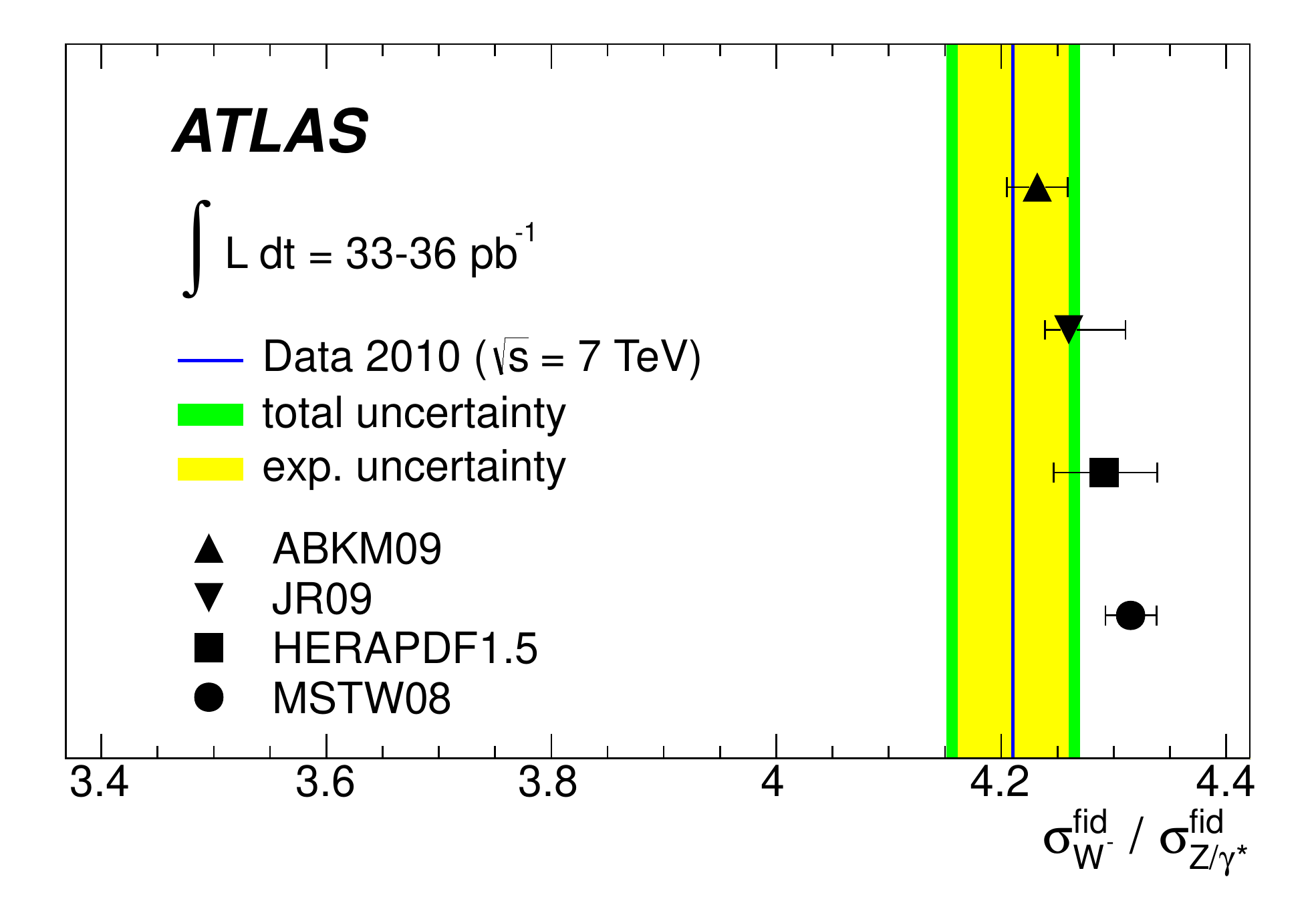}
  \caption{\it Measured and predicted fiducial cross section ratios,
    $\sigma_{W^+} / \sigma_{\Zg}$ (left) and 
    $\sigma_{W^-} / \sigma_{\Zg}$ (right).
    The experimental uncertainty (inner yellow band) 
    includes the experimental systematic errors. 
    The total uncertainty (outer green band) includes the
    statistical uncertainty and the small contribution from the acceptance correction.
    The uncertainties of the ABKM, JR and MSTW predictions
    are given by the PDF uncertainties considered to correspond
    to $68$\,\% CL and their correlations are derived from the
    eigenvector sets. The results for HERAPDF comprise
    all three sources of uncertainty of that set.}
  \label{fig:ratwpmz_fid}
\end{figure*}

\subsubsection{Combined cross section ratios}

Ratios of the $W^{\pm}$ and $Z$ cross sections
are calculated accounting for the correlations between uncertainties. 
The results obtained in the fiducial region are given in Tab.\,\ref{tab:ratiosfid}.

\begin{table}[tbqh]
  \small
  \begin{center}
    \begin{tabular}{lcccc}
      \hline
      \hline
      & \hspace{1.26cm} & sta & \hspace{0.48cm} sys & \hspace{0.3cm} acc \\
      \hline
      ${W^+}/{W^-}$ &   \multicolumn{4}{c}{$~\rWplusWminusfid$}   \\
      ${W^+/Z}$     &   \multicolumn{4}{c}{$~\rWplusZfid$} \\
      $ {W^-/Z}$    &   \multicolumn{4}{c}{$~\rWminusZfid$}       \\
      ${W^{\pm}/Z}$       &   \multicolumn{4}{c}{$~\rWZfid$}   \\
      \hline
      \hline
    \end{tabular}
    \caption{\it Measured ratios of the cross sections times leptonic
      branching ratios for $W^+/W^-$, $W^+/Z$, $W^-/Z$ and $(W^+ +
      W^-)/Z$, obtained in the fiducial regions and combining the electron
      and muon final states. The uncertainties denote the statistical
      (sta), the experimental systematic (sys), and the acceptance (acc)
      uncertainties.}
    \label{tab:ratiosfid}
  \end{center}
\end{table}

The precision of these measurements is very high, with
a total uncertainty of $0.9$\,\% for the $W^+/W^-$ ratio
and  of $1.3$\,\% for the $W^{\pm}/Z$ ratio.

\begin{table}[tbqh]
  \small
  \begin{center}
    \begin{tabular}{lcccc}
      \hline
      \hline
      & \hspace{1.26cm} & sta & \hspace{0.48cm} sys & \hspace{0.3cm} acc \\
      \hline
      ${W^+}/{W^-}$ &   \multicolumn{4}{c}{$~\rWplusWminus$}   \\
      ${W^+/Z}$     &   \multicolumn{4}{c}{$~\rWplusZ$} \\
      $ {W^-/Z}$    &   \multicolumn{4}{c}{$~\rWminusZ$}       \\
      ${W^{\pm}/Z}$       &   \multicolumn{4}{c}{$~\rWZ$}   \\
      \hline
      \hline
    \end{tabular}
    \caption{\it 
      Measured ratios of the total
      cross sections times leptonic branching ratios
      for  $W^+/W^-$, $W^+/Z$, $W^-/Z$ and $(W^+ + W^-)/Z$,
      combining the  electron and muon final states. The uncertainties
      denote the statistical (sta), the experimental systematic (sys),
      and the acceptance (acc) uncertainties.}
    \label{tab:ratios}
  \end{center}
\end{table}

Ratios for the total cross sections are given in Tab.\,\ref{tab:ratios}.
The uncertainties of the total cross section ratios are 
enlarged significantly by the additional acceptance
contribution.
Compared to the fiducial cross section ratios, 
the  uncertainties are almost doubled,
with a value of $1.8$\,\% for the $W^+/W^-$ ratio
and  of $1.6$\,\% for the $W^{\pm}/Z$ ratio.

The cross section ratios, determined in the fiducial regions
of the $W$ and $Z$  measurements, are compared in 
\Figs~\ref{fig:ratwzww_fid} and \ref{fig:ratwpmz_fid} with the
theoretical predictions accounting for the correlations 
inherent in the PDF determinations.

The mean boson rapidity for the data presented 
here is about zero, and  thus  on average the Bjorken $x$
values of the incoming partons are equal, $x_1=x_2 \simeq 0.01$.
In a rough leading order calculation,
neglecting the heavy quark and Cabibbo disfavoured parts of the 
cross sections and the $\gamma^\ast$ contribution to
the $Z$ cross section, and also assuming the light sea and anti-quark 
distributions to be all the same, $xs$,
the $(W^+ + W^-)/Z$ ratio is found to be proportional to
$(u_v + d_v + 2s)/[(v_u^2 +a_u^2) (u_v+s) +(v_d^2 +a_d^2) (d_v+s)]$.
Here $xu_v~(xd_v)$ is the up~(down) valence-quark
momentum distribution and
$v_{u,d}$ and $a_{u,d}$ are the vector and axial-vector weak neutral current
couplings of the light quarks. As the numerical values for the Z
coupling to the up and down quarks, $v_{u,d}^2 +a_{u,d}^2$, are of
similar size, the $W^{\pm}/Z$ ratio measures a rather PDF insensitive quantity,
provided that the sea is flavour symmetric. Since this 
symmetry assumption, with a small deviation to account for some light sea
quark asymmetry near Bjorken $x \simeq 0.1$, is inherent in all
major PDF fit determinations,
there is indeed not  much difference
observed between the various $W^{\pm}/Z$ ratio predictions, see 
Fig.\,\ref{fig:ratwzww_fid} (top).
The agreement with the present measurement therefore supports 
the assumption of a flavour
independent light quark sea at high scales, and Bjorken $x$ near to $0.01$.
The predictions for the charge dependent $W^+/W^-$, $W^+/Z$
and $W^-/Z$ ratios, shown in
\Figs~\ref{fig:ratwzww_fid} (bottom) and \ref{fig:ratwpmz_fid},
exhibit more significant deviations as they are more sensitive to
up-down quark distribution differences.

\section{Summary}

New measurements are presented of the
inclusive cross sections of Drell-Yan $W^{\pm}$ and $\Zg$
production in the electron and muon decay channels.
They are based
on the full data sample collected
by the ATLAS experiment at 
the LHC in 2010 at a centre-of-mass energy of $7$\,TeV.
With an integrated luminosity of about $35$\,pb$^{-1}$,
a total of about $270,000$ $W$ boson decays into
an electron or muon and the associated neutrino
and a total of about $24,000$
$\Zg$ decays into electron or muon pairs
have been observed. 

The cross sections are measured 
in a well defined
kinematic range within the detector acceptance,
defined by charged lepton pseudorapidity 
and charged lepton and neutrino
transverse momentum cuts.
Integrated cross sections are determined in these
fiducial regions and are also extrapolated
to the full kinematic range to obtain
the total integrated $W$ and $\Zg$  cross sections.
 
The $W^{\pm}$ cross sections  are
measured  differentially as a function of
the lepton pseudorapidity, extending to
$|\eta_\ell| \leq 2.5$. The $Z/\gamma^*$ cross
section is measured as a function of the boson rapidity
$|y_Z|$ up to a value of $2.4$.
An extension to $|y_Z| \leq 3.6$ is obtained through 
the electron channel measurements,
which include the forward detector region and
$|\eta_e|$ as large as $4.9$.

The electron and muon measurements are found to be 
consistent in the three channels, $W^+$, $W^-$ and $\Zg$. The
data sets are therefore combined
using a  method which accounts for the
different systematic error correlations.

This combination provides the most
accurate integrated inclusive $W$ and $\Zg$
cross sections so far obtained by the ATLAS Collaboration
and the first measurements of rapidity dependent
cross sections. An update is also presented
of the $W$ charge asymmetry as a function
of $|\eta_\ell|$.

The precision of the integrated $W$ and $\Zg$
cross sections in the fiducial region is $\sim\,1.2$\,\%
with an additional uncertainty of $3.4$\,\%
resulting from the luminosity error.
The uncertainties on the total integrated cross
sections are about twice as large because of
the extrapolation uncertainties in the
determination of the acceptance correction.
The differential cross sections are determined
in the fiducial region
with a typical precision of $2$\,\%, apart
from the most forward part of $y_Z$.

The results are compared with QCD predictions
calculated to NNLO in the fiducial regions of
the measurements which allows for maximum sensitivity
to details of the parton distributions used in 
these calculations. 

The broad agreement of the theory predictions
at the few per cent level
with the data supports the validity of the
QCD evolution equations, as the results
rely on lower scale parton distribution
functions evolved to the $W$ and $Z$ kinematic region,
at the average value of Bjorken $x$ of about $0.01$.

Interesting differences between sets
of parton distributions are observed, both in the
integrated and the differential fiducial cross sections.
The results presented in this paper
therefore provide a further basis 
for  sensitive tests of perturbative
QCD and determinations of the partonic content of the proton. 

\section{Acknowledgements}

We thank CERN for the very successful operation of the LHC, as well as the
support staff from our institutions without whom ATLAS could not be
operated efficiently.

We acknowledge the support of ANPCyT, Argentina; YerPhI, Armenia; ARC,
Australia; BMWF, Austria; ANAS, Azerbaijan; SSTC, Belarus; CNPq and FAPESP,
Brazil; NSERC, NRC and CFI, Canada; CERN; CONICYT, Chile; CAS, MOST and
NSFC, China; COLCIENCIAS, Colombia; MSMT CR, MPO CR and VSC CR, Czech
Republic; DNRF, DNSRC and Lundbeck Foundation, Denmark; ARTEMIS, European
Union; IN2P3-CNRS, CEA-DSM/IRFU, France; GNAS, Georgia; BMBF, DFG, HGF, MPG
and AvH Foundation, Germany; GSRT, Greece; ISF, MINERVA, GIF, DIP and
Benoziyo Center, Israel; INFN, Italy; MEXT and JSPS, Japan; CNRST, Morocco;
FOM and NWO, Netherlands; RCN, Norway; MNiSW, Poland; GRICES and FCT,
Portugal; MERYS (MECTS), Romania; MES of Russia and ROSATOM, Russian
Federation; JINR; MSTD, Serbia; MSSR, Slovakia; ARRS and MVZT, Slovenia;
DST/NRF, South Africa; MICINN, Spain; SRC and Wallenberg Foundation,
Sweden; SER, SNSF and Cantons of Bern and Geneva, Switzerland; NSC, Taiwan;
TAEK, Turkey; STFC, the Royal Society and Leverhulme Trust, United Kingdom;
DOE and NSF, United States of America.

The crucial computing support from all WLCG partners is acknowledged
gratefully, in particular from CERN and the ATLAS Tier-1 facilities at
TRIUMF (Canada), NDGF (Denmark, Norway, Sweden), CC-IN2P3 (France),
KIT/GridKA (Germany), INFN-CNAF (Italy), NL-T1 (Netherlands), PIC (Spain),
ASGC (Taiwan), RAL (UK) and BNL (USA) and in the Tier-2 facilities
worldwide.

\bibliography{WZInclusiveCrossSec}

\begin{table*}
\begin{center}
\begin{tabular}{ccccccc}
\hline\hline
$\eta_{min}$ & $\eta_{max}$ & ${\rm d}\sigma/{\rm d}\eta$ & $\delta_{\rm sta}$ & $\delta_{\rm unc}$ & $\delta_{\rm cor}$ & $\delta_{\rm tot}$ \\
           &             &           pb          &     \%            &   \%                   &     \%                &     \%            \\
\hline
    0.00 &     0.21 &    607.1 &    1.29 &     1.32 &     2.26 &     2.92 \\ 
    0.21 &     0.42 &    600.2 &    1.26 &     1.16 &     1.71 &     2.42 \\ 
    0.42 &     0.63 &    620.3 &    1.19 &     1.15 &     1.62 &     2.31 \\ 
    0.63 &     0.84 &    615.1 &    1.21 &     1.25 &     1.56 &     2.34 \\ 
    0.84 &     1.05 &    650.8 &    1.18 &     1.19 &     1.96 &     2.58 \\ 
    1.05 &     1.37 &    644.5 &    0.99 &     1.00 &     1.76 &     2.26 \\ 
    1.52 &     1.74 &    623.3 &    1.26 &     1.15 &     1.86 &     2.52 \\ 
    1.74 &     1.95 &    652.0 &    1.31 &     1.21 &     1.85 &     2.57 \\ 
    1.95 &     2.18 &    651.9 &    1.24 &     1.23 &     2.02 &     2.67 \\ 
    2.18 &     2.50 &    585.1 &    1.31 &     1.35 &     2.01 &     2.75 \\ 
\hline
\hline
\end{tabular}
\end{center}
\caption{\it \label{tab:shortelwp}Differential cross section for the $W^+\to e^+\nu$ process,
extrapolated to the common fiducial region.
The statistical ($\delta_{\rm sta}$), uncorrelated systematic ($\delta_{\rm unc}$),
correlated systematic ($\delta_{\rm cor}$), and total ($\delta_{\rm tot}$) uncertainties 
 are given in percent of the cross section values. 
The overall $3.4\%$ luminosity uncertainty is not included.
}
\end{table*}
\begin{table*}
\begin{center}
\begin{tabular}{ccccccc}
\hline\hline
$\eta_{min}$ & $\eta_{max}$ & ${\rm d}\sigma/{\rm d}\eta$ & $\delta_{\rm sta}$ & $\delta_{\rm unc}$ & $\delta_{\rm cor}$ & $\delta_{\rm tot}$ \\
           &             &           pb          &     \%            &   \%                   &     \%                &     \%            \\
\hline
    0.00 &     0.21 &    450.7 &    1.51 &     1.52 &     2.01 &     2.94 \\ 
    0.21 &     0.42 &    438.7 &    1.48 &     1.42 &     1.94 &     2.83 \\ 
    0.42 &     0.63 &    455.8 &    1.40 &     1.41 &     2.03 &     2.84 \\ 
    0.63 &     0.84 &    444.9 &    1.46 &     1.53 &     1.99 &     2.90 \\ 
    0.84 &     1.05 &    427.6 &    1.47 &     1.55 &     1.93 &     2.88 \\ 
    1.05 &     1.37 &    430.5 &    1.21 &     1.25 &     2.10 &     2.73 \\ 
    1.52 &     1.74 &    387.2 &    1.62 &     1.62 &     1.97 &     3.02 \\ 
    1.74 &     1.95 &    384.2 &    1.70 &     1.64 &     2.04 &     3.13 \\ 
    1.95 &     2.18 &    356.5 &    1.68 &     1.53 &     2.47 &     3.35 \\ 
    2.18 &     2.50  &    325.4 &    1.73 &     1.67 &     2.26 &     3.30 \\ 
\hline
\hline
\end{tabular}
\end{center}
\caption{\it \label{tab:shortelwm}Differential cross section for the $W^-\to e^-\bar{\nu}$ process,
extrapolated to the common fiducial region.
The statistical ($\delta_{\rm sta}$), uncorrelated systematic ($\delta_{\rm unc}$),
correlated systematic  ($\delta_{\rm cor}$), and total ($\delta_{\rm tot}$) uncertainties 
 are given in percent of the cross section values.  
The overall $3.4\%$ luminosity uncertainty is not included.
}
\end{table*}

\begin{table*}
\begin{center}
\begin{tabular}{ccccccc}
\hline\hline
$y_{min}$ & $y_{max}$ & ${\rm d}\sigma/{\rm d}y$ & $\delta_{\rm sta}$ & $\delta_{\rm unc}$ & $\delta_{\rm cor}$ & $\delta_{\rm tot}$ \\
           &             &           pb          &     \%            &   \%                   &     \%                &     \%            \\
\hline
     0.0 &      0.4 &    133.6 &    2.06 &     0.68 &     2.41 &     3.25 \\ 
     0.4 &      0.8 &    127.6 &    2.17 &     0.67 &     2.49 &     3.37 \\ 
     0.8 &      1.2 &    128.4 &    2.26 &     0.64 &     2.66 &     3.55 \\ 
     1.2 &      1.6 &    123.3 &    2.52 &     0.65 &     2.92 &     3.91 \\ 
     1.6 &      2.0 &    113.9 &    3.30 &     0.73 &     3.38 &     4.78 \\ 
     2.0 &      2.4 &    104.2 &    5.07 &     0.90 &     4.65 &     6.94 \\ 
\hline
\hline
\end{tabular}
\end{center}
\caption{\it \label{tab:shortzcc} Differential cross section for the
central $Z\to e^+e^-$ selection, extrapolated to the common fiducial region.
The statistical ($\delta_{\rm sta}$), uncorrelated systematic ($\delta_{\rm unc}$),
correlated systematic  ($\delta_{\rm cor}$), and total ($\delta_{\rm tot}$) uncertainties 
 are given in percent of the cross section values. 
The overall $3.4\%$ luminosity uncertainty is not included.
}
\end{table*}

\begin{table*}
\begin{center}
\begin{tabular}{ccccccc}
\hline\hline
$y_{min}$ & $y_{max}$ & ${\rm d}\sigma/{\rm d}y$ & $\delta_{\rm sta}$ & $\delta_{\rm unc}$ & $\delta_{\rm cor}$ & $\delta_{\rm tot}$ \\
           &             &           pb          &     \%            &   \%                   &     \%                &     \%            \\
\hline
     1.2 &      1.6 &     98.3 &    7.31 &     4.94 &     5.94 &    10.64 \\ 
     1.6 &      2.0 &    126.9 &    3.74 &     3.16 &     5.74 &     7.54 \\ 
     2.0 &      2.4 &    107.9 &    3.28 &     4.30 &     5.21 &     7.51 \\ 
     2.4 &      2.8 &     93.8 &    3.21 &     3.81 &     4.80 &     6.92 \\ 
     2.8 &      3.6 &     53.7 &    4.20 &     4.37 &     8.22 &    10.21 \\ 
\hline
\hline
\end{tabular}
\end{center}
\caption{\it \label{tab:shortzcf} Differential cross section for the
forward $Z\to e^+e^-$ selection, extrapolated to the common fiducial region.
The statistical ($\delta_{\rm sta}$), uncorrelated systematic ($\delta_{\rm unc}$),
correlated systematic  ($\delta_{\rm cor}$), and total ($\delta_{\rm tot}$) uncertainties 
 are given in percent of the cross section values. 
The overall $3.4\%$ luminosity uncertainty is not included.
}
\end{table*}

\begin{table*}
\begin{center}
\begin{tabular}{ccccccc}
\hline\hline
$\eta_{min}$ & $\eta_{max}$ & ${\rm d}\sigma/{\rm d}\eta$ & $\delta_{\rm sta}$ & $\delta_{\rm unc}$ & $\delta_{\rm cor}$ & $\delta_{\rm tot}$ \\
           &             &           pb          &     \%            &   \%                   &     \%                &     \%            \\
\hline
    0.00 &     0.21 &    593.5 &    1.48 &     2.32 &     1.76 &     3.26 \\ 
    0.21 &     0.42 &    611.0 &    1.31 &     1.79 &     1.69 &     2.79 \\ 
    0.42 &     0.63 &    628.7 &    1.27 &     1.72 &     1.62 &     2.68 \\ 
    0.63 &     0.84 &    621.7 &    1.38 &     2.34 &     2.04 &     3.40 \\ 
    0.84 &     1.05 &    629.8 &    1.37 &     2.32 &     1.81 &     3.24 \\ 
    1.05 &     1.37 &    658.8 &    1.01 &     1.43 &     1.78 &     2.50 \\ 
    1.37 &     1.52 &    632.8 &    1.37 &     1.30 &     2.38 &     3.04 \\ 
    1.52 &     1.74 &    638.9 &    1.13 &     1.07 &     1.67 &     2.28 \\ 
    1.74 &     1.95 &    652.1 &    1.17 &     1.26 &     1.70 &     2.42 \\ 
    1.95 &     2.18 &    611.5 &    1.15 &     1.22 &     1.68 &     2.37 \\ 
    2.18 &     2.50 &    577.6 &    1.21 &     1.43 &     2.05 &     2.78 \\ 
\hline
\hline
\end{tabular}
\end{center}
\caption{\it \label{tab:shortmuwp}Differential cross section for the $W^+\to \mu^+\nu$ process,
extrapolated to the common fiducial region.
The statistical ($\delta_{\rm sta}$), uncorrelated systematic ($\delta_{\rm unc}$),
correlated systematic  ($\delta_{\rm cor}$), and total ($\delta_{\rm tot}$) uncertainties 
 are given in percent of the cross section values. 
The overall $3.4\%$ luminosity uncertainty is not included.
}
\end{table*}
\begin{table*}
\begin{center}
\begin{tabular}{ccccccc}
\hline\hline
$\eta_{min}$ & $\eta_{max}$ & ${\rm d}\sigma/{\rm d}\eta$ & $\delta_{\rm sta}$ & $\delta_{\rm unc}$ & $\delta_{\rm cor}$ & $\delta_{\rm tot}$ \\
           &             &           pb         &     \%            &   \%                   &     \%                &     \%            \\
\hline
    0.00 &     0.21 &    441.9 &    1.73 &     2.34 &     1.63 &     3.33 \\ 
    0.21 &     0.42 &    444.9 &    1.56 &     1.82 &     1.79 &     2.99 \\ 
    0.42 &     0.63 &    455.1 &    1.52 &     1.75 &     1.75 &     2.91 \\ 
    0.63 &     0.84 &    435.5 &    1.68 &     2.39 &     2.07 &     3.57 \\ 
    0.84 &     1.05 &    433.2 &    1.67 &     2.36 &     1.68 &     3.34 \\ 
    1.05 &     1.37 &    408.8 &    1.32 &     1.47 &     1.66 &     2.58 \\ 
    1.37 &     1.52 &    388.1 &    1.79 &     1.35 &     2.14 &     3.10 \\ 
    1.52 &     1.74 &    383.5 &    1.50 &     1.11 &     2.15 &     2.85 \\ 
    1.74 &     1.95 &    370.5 &    1.59 &     1.32 &     1.94 &     2.83 \\ 
    1.95 &     2.18 &    360.3 &    1.53 &     1.26 &     1.88 &     2.73 \\ 
    2.18 &     2.50 &    338.3 &    1.60 &     1.47 &     2.11 &     3.03 \\ 
\hline
\hline
\end{tabular}
\end{center}
\caption{\it \label{tab:shortmuwm}Differential cross section for the $W^-\to \mu^-\nu$ process,
extrapolated to the common fiducial region.
The statistical ($\delta_{\rm sta}$), uncorrelated systematic ($\delta_{\rm unc}$),
correlated systematic  ($\delta_{\rm cor}$), and total ($\delta_{\rm tot}$) uncertainties 
 are given in percent of the cross section values.  
The overall $3.4\%$ luminosity uncertainty is not included.
}
\end{table*}

\begin{table*}
\begin{center}
\begin{tabular}{ccccccc}
\hline\hline
$y_{min}$ & $y_{max}$ & ${\rm d}\sigma/{\rm d}y$ & $\delta_{\rm sta}$ & $\delta_{\rm unc}$ & $\delta_{\rm cor}$ & $\delta_{\rm tot}$ \\
           &             &           pb          &     \%            &   \%                   &     \%                &     \%            \\
\hline
     0.0 &      0.4 &    126.7 &    2.04 &     0.97 &     1.22 &     2.57 \\ 
     0.4 &      0.8 &    132.7 &    1.97 &     0.73 &     1.23 &     2.44 \\ 
     0.8 &      1.2 &    125.2 &    2.01 &     0.68 &     0.82 &     2.27 \\ 
     1.2 &      1.6 &    117.9 &    2.16 &     0.55 &     0.82 &     2.38 \\ 
     1.6 &      2.0 &    111.7 &    2.63 &     0.65 &     1.08 &     2.92 \\ 
     2.0 &      2.4 &    107.8 &    4.43 &     1.32 &     2.88 &     5.45 \\ 
\hline
\hline
\end{tabular}
\end{center}
\caption{\it \label{tab:shortzmu} Differential cross section for the $Z\to\mu^+\mu^-$ process,
extrapolated to the common fiducial region.
The statistical ($\delta_{\rm sta}$), uncorrelated systematic ($\delta_{\rm unc}$),
correlated systematic  ($\delta_{\rm cor}$), and total ($\delta_{\rm tot}$) uncertainties 
 are given in percent of the cross section values. 
The overall $3.4\%$ luminosity uncertainty is not included.
}
\end{table*}

\begin{table*}[htbp]
\begin{center}
\begin{scriptsize}
\begin{tabular}{lrrrrrrrr}
\hline
\hline
$y_{min}-y_{max}$ & 0.0-0.4 & 0.4-0.8 & 0.8-1.2 & 1.2-1.6 & 1.6-2.0 & 2.0-2.4 & 2.4-2.8 & 2.8-3.6 \\
\hline
   ${\rm d} \sigma/ {\rm d} y$ [pb]&    129.27&    129.44&    125.81&    118.23&    113.37&    105.26&     92.18&     53.38\\
              $\delta_{\rm sta},\%$&      1.46&      1.47&      1.50&      1.61&      1.84&      2.57&      3.24&      4.21\\
              $\delta_{\rm unc},\%$&      0.59&      0.50&      0.47&      0.45&      0.63&      1.37&      3.81&      4.37\\
              $\delta_{\rm cor},\%$&      1.07&      1.08&      0.93&      0.97&      1.26&      2.19&      3.77&      8.06\\
              $\delta_{\rm tot},\%$&      1.90&      1.89&      1.83&      1.94&      2.32&      3.65&      6.26&     10.09\\
 \hline
   $\gamma_{  1},\%$&      0.29&      0.29&      0.29&      0.29&      0.29&      0.29&      0.29&      0.29\\
   $\gamma_{  2},\%$&      0.09&      0.09&      0.09&      0.09&      0.09&      0.09&      0.09&      0.09\\
   $\gamma_{  3},\%$&      0.04&      0.04&      0.04&      0.04&      0.04&      0.04&      0.04&      0.04\\
   $\gamma_{  4},\%$&      0.19&      0.19&      0.19&      0.19&      0.19&      0.19&      0.19&      0.18\\
   $\gamma_{  5},\%$&      0.07&      0.07&      0.05&      0.04&      0.01&      0.00&      0.06&      0.18\\
   $\gamma_{  6},\%$&     -0.13&     -0.10&     -0.08&     -0.05&     -0.04&     -0.07&     -0.06&     -0.03\\
   $\gamma_{  7},\%$&      0.05&      0.04&      0.05&      0.04&      0.05&      0.09&      0.58&      1.76\\
   $\gamma_{  8},\%$&     -0.07&     -0.09&     -0.07&     -0.09&     -0.08&     -0.19&     -0.42&     -1.16\\
   $\gamma_{  9},\%$&     -0.03&     -0.02&     -0.05&      0.01&      0.05&      0.18&      0.61&      1.28\\
   $\gamma_{ 10},\%$&      0.12&      0.13&      0.11&      0.08&      0.03&     -0.05&     -0.40&     -0.93\\
   $\gamma_{ 11},\%$&     -0.10&     -0.10&     -0.10&     -0.05&      0.01&      0.13&      0.63&      1.87\\
   $\gamma_{ 12},\%$&      0.06&      0.06&      0.06&      0.15&      0.33&      0.76&      2.26&      4.97\\
   $\gamma_{ 13},\%$&     -0.28&     -0.29&     -0.17&     -0.15&      0.15&      0.18&      0.11&     -0.39\\
   $\gamma_{ 14},\%$&     -0.02&      0.01&     -0.03&      0.05&     -0.01&      0.23&      1.16&      3.19\\
   $\gamma_{ 15},\%$&      0.07&      0.06&      0.01&      0.03&      0.02&      0.23&      1.18&      2.70\\
   $\gamma_{ 16},\%$&     -0.10&     -0.08&     -0.08&     -0.03&     -0.09&      0.04&      0.23&      1.64\\
   $\gamma_{ 17},\%$&     -0.53&     -0.55&     -0.43&     -0.37&     -0.37&     -0.58&     -0.82&     -1.95\\
   $\gamma_{ 18},\%$&      0.07&      0.02&      0.03&      0.07&      0.17&      0.17&      0.45&      0.56\\
   $\gamma_{ 19},\%$&     -0.16&     -0.16&     -0.13&     -0.06&     -0.07&     -0.06&      0.03&      0.37\\
   $\gamma_{ 20},\%$&      0.34&      0.32&      0.22&      0.30&      0.41&      0.66&     -0.03&     -0.83\\
   $\gamma_{ 21},\%$&     -0.15&     -0.17&     -0.15&     -0.09&      0.04&      0.13&      0.04&     -0.03\\
   $\gamma_{ 22},\%$&     -0.10&     -0.15&      0.00&     -0.25&     -0.45&     -1.15&     -0.28&      1.39\\
   $\gamma_{ 23},\%$&      0.05&      0.02&      0.00&     -0.23&     -0.49&     -0.85&     -0.09&      0.78\\
   $\gamma_{ 24},\%$&      0.22&      0.23&      0.23&      0.16&      0.00&      0.15&      0.49&      0.28\\
   $\gamma_{ 25},\%$&      0.17&      0.16&      0.12&      0.14&      0.08&      0.01&      0.26&      0.26\\
   $\gamma_{ 26},\%$&      0.18&      0.25&      0.28&      0.18&      0.24&      0.69&      0.03&     -1.13\\
   $\gamma_{ 27},\%$&      0.00&     -0.01&     -0.04&     -0.04&     -0.06&     -0.20&     -0.19&     -0.04\\
   $\gamma_{ 28},\%$&      0.50&      0.47&      0.45&      0.52&      0.66&      0.62&      0.70&      0.26\\
   $\gamma_{ 29},\%$&      0.17&      0.18&      0.16&      0.13&     -0.06&     -0.14&     -1.68&     -0.46\\
   $\gamma_{ 30},\%$&     -0.12&     -0.11&     -0.14&     -0.12&     -0.11&     -0.20&     -0.21&     -0.21\\
\hline
\hline
\end{tabular}
\end{scriptsize}
\end{center}
\caption{\it Combined differential cross section
  ${\rm d}\sigma/{\rm d}y_Z$ for the $Z\to\ell^+\ell^-$ process
  measured for $66<m_{\ell\ell}<116$~GeV and
$p_{T,\ell}>20$~GeV. All uncertainties are quoted in $\%$ with
respect to the cross section values. $\delta_{\rm sta}$, $\delta_{\rm unc}$,
$\delta_{\rm cor}$, and $\delta_{\rm tot}$ represent statistical, uncorrelated systematic, 
correlated systematic, and total uncertainties. 
$\gamma_{1}-\gamma_{30}$ represent diagonalised correlated systematic uncertainties, which
are correlated bin-to-bin and across the $W^+$, $W^-$ and $Z$ measurements.
The overall $3.4\%$ luminosity uncertainty is not included.
\label{tab:zfullcob}
}
\end{table*}

\begin{table*}[htbp]
\begin{center}
\begin{scriptsize}
\begin{tabular}{lrrrrrrrrrrr}
\hline
\hline
$\eta_{min}-\eta_{max}$ & 0.00-0.21 & 0.21-0.42 & 0.42-0.63& 0.63-0.84& 0.84-1.05& 1.05-1.37& 1.37-1.52 & 1.52-1.74& 1.74-1.95& 1.95-2.18& 2.18-2.50 \\
\hline
${\rm d} \sigma/ {\rm d} \eta$ [pb]&    446.32&    440.26&    455.06&    439.81&    428.07&    418.89&    387.27&    384.03&    375.29&    357.39&    330.99\\
              $\delta_{\rm sta},\%$&      1.16&      1.08&      1.04&      1.12&      1.12&      0.90&      1.79&      1.11&      1.17&      1.13&      1.18\\
              $\delta_{\rm unc},\%$&      1.29&      1.13&      1.10&      1.30&      1.30&      0.95&      1.35&      0.93&      1.03&      0.98&      1.10\\
              $\delta_{\rm cor},\%$&      1.30&      1.29&      1.31&      1.35&      1.36&      1.37&      1.67&      1.47&      1.48&      1.50&      1.64\\
              $\delta_{\rm tot},\%$&      2.16&      2.02&      2.00&      2.18&      2.19&      1.90&      2.80&      2.06&      2.15&      2.12&      2.30\\
 \hline
   $\gamma_{  1},\%$&      0.10&      0.10&      0.10&      0.10&      0.10&      0.10&      0.10&      0.10&      0.10&      0.10&      0.10\\
   $\gamma_{  2},\%$&      0.18&      0.18&      0.18&      0.18&      0.18&      0.18&      0.18&      0.18&      0.18&      0.18&      0.18\\
   $\gamma_{  3},\%$&      0.23&      0.23&      0.23&      0.23&      0.23&      0.23&      0.23&      0.23&      0.23&      0.23&      0.23\\
   $\gamma_{  4},\%$&      0.19&      0.19&      0.19&      0.19&      0.19&      0.19&      0.19&      0.19&      0.19&      0.19&      0.19\\
   $\gamma_{  5},\%$&      0.03&      0.03&      0.04&      0.04&      0.03&      0.04&      0.05&      0.04&      0.05&      0.04&      0.06\\
   $\gamma_{  6},\%$&     -0.01&      0.00&     -0.01&     -0.01&     -0.01&      0.00&     -0.02&      0.00&      0.00&     -0.01&      0.00\\
   $\gamma_{  7},\%$&      0.04&      0.05&      0.06&      0.05&      0.04&      0.05&      0.04&      0.06&      0.04&      0.04&      0.14\\
   $\gamma_{  8},\%$&      0.05&      0.06&      0.03&      0.06&      0.06&      0.03&      0.06&      0.04&      0.03&      0.05&      0.01\\
   $\gamma_{  9},\%$&      0.07&      0.07&      0.07&      0.08&      0.05&      0.11&      0.12&      0.11&      0.10&      0.10&      0.02\\
   $\gamma_{ 10},\%$&      0.11&      0.11&      0.09&      0.10&      0.11&      0.13&      0.19&      0.15&      0.11&      0.10&      0.14\\
   $\gamma_{ 11},\%$&      0.17&      0.18&      0.16&      0.18&      0.21&      0.18&      0.20&      0.20&      0.19&      0.20&      0.35\\
   $\gamma_{ 12},\%$&      0.06&      0.09&      0.06&      0.09&      0.07&      0.13&      0.09&      0.13&      0.13&      0.14&      0.12\\
   $\gamma_{ 13},\%$&     -0.42&     -0.42&     -0.45&     -0.49&     -0.46&     -0.49&     -0.62&     -0.53&     -0.54&     -0.52&     -0.49\\
   $\gamma_{ 14},\%$&      0.08&      0.05&      0.05&      0.08&      0.09&      0.11&      0.14&      0.11&      0.13&      0.15&      0.16\\
   $\gamma_{ 15},\%$&     -0.18&     -0.20&     -0.20&     -0.22&     -0.23&     -0.32&     -0.25&     -0.36&     -0.35&     -0.36&     -0.36\\
   $\gamma_{ 16},\%$&     -0.29&     -0.29&     -0.31&     -0.31&     -0.27&     -0.34&     -0.38&     -0.36&     -0.37&     -0.32&     -0.31\\
   $\gamma_{ 17},\%$&     -0.57&     -0.48&     -0.52&     -0.49&     -0.61&     -0.60&     -0.81&     -0.74&     -0.61&     -0.64&     -0.84\\
   $\gamma_{ 18},\%$&      0.39&      0.44&      0.50&      0.56&      0.53&      0.37&      0.52&      0.35&      0.47&      0.36&      0.40\\
   $\gamma_{ 19},\%$&      0.30&      0.37&      0.32&      0.38&      0.26&      0.33&      0.25&      0.23&      0.33&      0.15&      0.08\\
   $\gamma_{ 20},\%$&      0.34&      0.34&      0.32&      0.44&      0.29&      0.45&      0.58&      0.40&      0.48&      0.53&      0.58\\
   $\gamma_{ 21},\%$&     -0.41&     -0.38&     -0.28&     -0.36&     -0.47&     -0.47&     -0.53&     -0.43&     -0.44&     -0.56&     -0.55\\
   $\gamma_{ 22},\%$&     -0.11&     -0.08&     -0.08&     -0.03&     -0.09&     -0.02&     -0.05&     -0.01&     -0.01&     -0.02&     -0.10\\
   $\gamma_{ 23},\%$&      0.12&      0.17&      0.19&      0.14&      0.07&      0.03&      0.01&      0.01&      0.11&      0.06&     -0.01\\
   $\gamma_{ 24},\%$&     -0.11&     -0.18&     -0.14&     -0.09&     -0.22&     -0.15&      0.00&     -0.29&     -0.23&     -0.32&     -0.45\\
   $\gamma_{ 25},\%$&     -0.02&     -0.16&     -0.14&     -0.07&     -0.12&     -0.11&      0.13&     -0.22&     -0.10&     -0.04&     -0.04\\
   $\gamma_{ 26},\%$&      0.51&      0.41&      0.50&      0.32&      0.36&      0.25&      0.19&      0.26&      0.36&      0.42&      0.24\\
   $\gamma_{ 27},\%$&     -0.08&     -0.15&     -0.18&     -0.08&     -0.07&     -0.02&     -0.25&     -0.09&     -0.08&      0.00&      0.09\\
   $\gamma_{ 28},\%$&      0.11&      0.21&      0.12&      0.21&      0.21&      0.27&      0.34&      0.32&      0.25&      0.25&      0.29\\
   $\gamma_{ 29},\%$&      0.07&      0.07&      0.05&      0.07&      0.06&      0.09&      0.12&      0.08&      0.08&      0.09&      0.09\\
   $\gamma_{ 30},\%$&     -0.16&     -0.10&     -0.10&     -0.13&     -0.14&     -0.10&      0.31&      0.03&     -0.01&     -0.04&      0.00\\
\hline
\hline
\end{tabular}
\end{scriptsize}
\end{center}
\caption{\it Combined differential cross section
  ${\rm d}\sigma/{\rm d}\eta_{\ell^-}$ for the $W^-\to\ell^-\bar{\nu}$ process
  measured for $p_{T,\ell}>20$~GeV,
  $p_{T,\nu}>25$~GeV and $m_{T}>40$~GeV. All uncertainties are quoted in $\%$ with
  respect to the cross section values. $\delta_{\rm sta}$, $\delta_{\rm unc}$,
  $\delta_{\rm cor}$, and $\delta_{\rm tot}$ represent statistical,
  uncorrelated systematic,  correlated systematic, 
  and total uncertainties. $\gamma_{1}-\gamma_{30}$
  represent diagonalised correlated systematic uncertainties, which 
  are correlated bin-to-bin and across the $W^+$, $W^-$ and $Z$ measurements. 
  The overall $3.4\%$ luminosity uncertainty is not included.
  \label{tab:wmfullcob}
}
\end{table*}

\begin{table*}[htbp]
\begin{center}
\begin{scriptsize}
\begin{tabular}{lrrrrrrrrrrr}
\hline
\hline
$\eta_{min}-\eta_{max}$ & 0.00-0.21 & 0.21-0.42 & 0.42-0.63& 0.63-0.84& 0.84-1.05& 1.05-1.37& 1.37-1.52 & 1.52-1.74& 1.74-1.95& 1.95-2.18& 2.18-2.50 \\
\hline
${\rm d} \sigma/ {\rm d} \eta$ [pb]&    602.00&    602.67&    620.15&    614.69&    640.65&    647.21&    630.74&    629.17&    648.85&    628.13&    578.39\\
              $\delta_{\rm sta},\%$&      1.00&      0.93&      0.89&      0.95&      0.94&      0.72&      1.37&      0.84&      0.88&      0.85&      0.89\\
              $\delta_{\rm unc},\%$&      1.16&      0.99&      0.97&      1.12&      1.07&      0.83&      1.30&      0.78&      0.88&      0.87&      0.98\\
              $\delta_{\rm cor},\%$&      1.33&      1.17&      1.17&      1.20&      1.26&      1.19&      1.73&      1.15&      1.29&      1.21&      1.39\\
              $\delta_{\rm tot},\%$&      2.03&      1.79&      1.76&      1.89&      1.90&      1.62&      2.56&      1.63&      1.79&      1.71&      1.92\\
 \hline
   $\gamma_{  1},\%$&      0.23&      0.23&      0.23&      0.23&      0.23&      0.23&      0.23&      0.23&      0.23&      0.23&      0.23\\
   $\gamma_{  2},\%$&     -0.18&     -0.18&     -0.18&     -0.18&     -0.18&     -0.18&     -0.18&     -0.18&     -0.18&     -0.18&     -0.18\\
   $\gamma_{  3},\%$&      0.09&      0.09&      0.09&      0.09&      0.09&      0.09&      0.09&      0.09&      0.09&      0.09&      0.09\\
   $\gamma_{  4},\%$&      0.20&      0.20&      0.20&      0.20&      0.20&      0.20&      0.20&      0.20&      0.20&      0.20&      0.20\\
   $\gamma_{  5},\%$&      0.03&      0.03&      0.03&      0.04&      0.04&      0.04&      0.06&      0.04&      0.05&      0.03&      0.02\\
   $\gamma_{  6},\%$&     -0.01&     -0.01&     -0.01&     -0.01&      0.00&     -0.01&     -0.02&     -0.01&     -0.01&      0.00&     -0.01\\
   $\gamma_{  7},\%$&      0.03&      0.03&      0.02&      0.04&      0.04&      0.02&      0.02&      0.02&      0.02&      0.03&      0.08\\
   $\gamma_{  8},\%$&      0.03&      0.03&      0.04&      0.02&      0.04&      0.00&     -0.03&      0.00&      0.00&      0.03&     -0.03\\
   $\gamma_{  9},\%$&      0.08&      0.08&      0.06&      0.09&      0.07&      0.11&      0.17&      0.11&      0.12&      0.11&      0.05\\
   $\gamma_{ 10},\%$&      0.10&      0.11&      0.14&      0.12&      0.11&      0.15&      0.18&      0.11&      0.12&      0.12&      0.11\\
   $\gamma_{ 11},\%$&      0.13&      0.12&      0.13&      0.12&      0.19&      0.11&      0.12&      0.12&      0.13&      0.13&      0.24\\
   $\gamma_{ 12},\%$&      0.08&      0.08&      0.07&      0.07&      0.10&      0.10&      0.07&      0.11&      0.11&      0.11&      0.01\\
   $\gamma_{ 13},\%$&     -0.43&     -0.39&     -0.40&     -0.47&     -0.48&     -0.46&     -0.61&     -0.49&     -0.51&     -0.44&     -0.48\\
   $\gamma_{ 14},\%$&      0.03&      0.01&      0.04&      0.01&      0.05&      0.08&      0.08&      0.06&      0.09&      0.07&      0.05\\
   $\gamma_{ 15},\%$&     -0.17&     -0.15&     -0.16&     -0.16&     -0.22&     -0.23&     -0.18&     -0.26&     -0.26&     -0.26&     -0.29\\
   $\gamma_{ 16},\%$&     -0.22&     -0.19&     -0.17&     -0.23&     -0.23&     -0.22&     -0.26&     -0.21&     -0.24&     -0.17&     -0.24\\
   $\gamma_{ 17},\%$&     -0.59&     -0.60&     -0.55&     -0.63&     -0.46&     -0.66&     -1.00&     -0.64&     -0.62&     -0.60&     -0.82\\
   $\gamma_{ 18},\%$&      0.37&      0.34&      0.47&      0.47&      0.64&      0.36&      0.40&      0.37&      0.46&      0.36&      0.46\\
   $\gamma_{ 19},\%$&      0.21&      0.31&      0.44&      0.35&      0.28&      0.29&      0.10&      0.17&      0.28&      0.21&      0.23\\
   $\gamma_{ 20},\%$&      0.30&      0.20&      0.17&      0.23&      0.28&      0.29&      0.48&      0.27&      0.37&      0.25&      0.19\\
   $\gamma_{ 21},\%$&     -0.18&     -0.19&     -0.28&     -0.10&     -0.31&     -0.17&     -0.30&     -0.11&     -0.16&     -0.25&     -0.33\\
   $\gamma_{ 22},\%$&     -0.05&     -0.08&     -0.12&     -0.01&     -0.08&      0.00&      0.03&      0.01&      0.11&     -0.01&      0.08\\
   $\gamma_{ 23},\%$&      0.09&      0.08&      0.06&      0.05&      0.01&     -0.05&     -0.19&     -0.11&     -0.10&     -0.10&     -0.19\\
   $\gamma_{ 24},\%$&     -0.08&     -0.14&     -0.05&     -0.13&      0.26&     -0.04&     -0.07&      0.14&     -0.12&     -0.03&     -0.04\\
   $\gamma_{ 25},\%$&     -0.24&     -0.13&      0.06&     -0.09&     -0.03&      0.02&      0.02&      0.03&     -0.02&     -0.07&     -0.01\\
   $\gamma_{ 26},\%$&      0.74&      0.45&      0.05&      0.18&      0.04&      0.22&      0.49&      0.13&      0.27&      0.44&      0.27\\
   $\gamma_{ 27},\%$&      0.09&      0.16&      0.23&      0.14&      0.13&      0.18&      0.41&      0.18&      0.22&      0.18&      0.19\\
   $\gamma_{ 28},\%$&      0.12&      0.21&      0.25&      0.26&      0.18&      0.21&      0.46&      0.18&      0.29&      0.24&      0.31\\
   $\gamma_{ 29},\%$&      0.07&      0.06&      0.03&      0.03&      0.05&      0.05&      0.15&      0.05&      0.08&      0.09&      0.06\\
   $\gamma_{ 30},\%$&     -0.18&     -0.10&     -0.07&     -0.09&     -0.19&     -0.09&      0.36&     -0.03&     -0.02&     -0.06&     -0.02\\
\hline
\hline
\end{tabular}
\end{scriptsize}
\end{center}
\caption{\it Combined differential cross section
  ${\rm d}\sigma/{\rm d}\eta_{\ell^+}$ for the $W^+\to\ell^+\nu$ process
  measured for $p_{T,\ell}>20$~GeV,
  $p_{T,\nu}>25$~GeV and $m_{T}>40$~GeV. All uncertainties are quoted in $\%$ with
  respect to the cross section values. $\delta_{\rm sta}$, $\delta_{\rm unc}$,
  $\delta_{\rm cor}$, and $\delta_{\rm tot}$ represent statistical,
  uncorrelated systematic,  correlated systematic, 
  and total uncertainties. $\gamma_{1}-\gamma_{30}$
  represent diagonalised correlated systematic uncertainties, which 
  are correlated bin-to-bin and across the $W^+$, $W^-$ and $Z$ measurements. 
  The overall $3.4\%$ luminosity uncertainty is not included.
\label{tab:wpfullcob}
}
\end{table*}

\begin{table*}
\begin{center}
\begin{tabular}{ccccccc}
\hline
\hline
  $\eta_{min}$ & $\eta_{max}$ & $A_{\ell}$ & $\Delta_{\rm sta}$ & $\Delta_{\rm unc}$ & $\Delta_{\rm cor}$ & $\Delta_{\rm tot}$ \\
\hline
      0.00 &       0.21 &      0.149 &      0.008 &      0.009 &      0.003 &      0.012\\
      0.21 &       0.42 &      0.156 &      0.007 &      0.008 &      0.003 &      0.011\\
      0.42 &       0.63 &      0.154 &      0.007 &      0.007 &      0.004 &      0.011\\
      0.63 &       0.84 &      0.166 &      0.007 &      0.009 &      0.003 &      0.012\\
      0.84 &       1.05 &      0.199 &      0.007 &      0.008 &      0.004 &      0.012\\
      1.05 &       1.37 &      0.214 &      0.006 &      0.006 &      0.003 &      0.009\\
      1.37 &       1.52 &      0.239 &      0.011 &      0.010 &      0.005 &      0.016\\
      1.52 &       1.74 &      0.242 &      0.007 &      0.006 &      0.004 &      0.010\\
      1.74 &       1.95 &      0.267 &      0.007 &      0.007 &      0.003 &      0.011\\
      1.95 &       2.18 &      0.275 &      0.007 &      0.007 &      0.003 &      0.010\\
      2.18 &       2.50 &      0.272 &      0.007 &      0.008 &      0.004 &      0.011\\
\hline
\hline
\end{tabular}
\end{center}
\caption{\it The combined lepton charge asymmetry $A_{\ell}$ from W boson decays in 
  bins of absolute lepton pseudorapidity measured for $p_{T,\ell}>20$~GeV,
  $p_{T,\nu}>25$~GeV, and $m_{T}>40$~GeV.
  $\Delta_{\rm sta}$, $\Delta_{\rm unc}$, $\Delta_{\rm cor}$, and $\Delta_{\rm tot}$ represent statistical, uncorrelated systematic,
  correlated systematic, and total uncertainty. \label{tab:leptasym}
}
\end{table*}

\clearpage

\begin{flushleft}
{\Large The ATLAS Collaboration}

\bigskip

G.~Aad$^{\rm 48}$,
B.~Abbott$^{\rm 111}$,
J.~Abdallah$^{\rm 11}$,
A.A.~Abdelalim$^{\rm 49}$,
A.~Abdesselam$^{\rm 118}$,
O.~Abdinov$^{\rm 10}$,
B.~Abi$^{\rm 112}$,
M.~Abolins$^{\rm 88}$,
H.~Abramowicz$^{\rm 153}$,
H.~Abreu$^{\rm 115}$
% ...
\bigskip

% ...
$^{10}$ Institute of Physics, Azerbaijan Academy of Sciences, Baku, Azerbaijan\\
$^{11}$ Institut de F\'isica d'Altes Energies and Departament de F\'isica de la Universitat Aut\`onoma  de Barcelona and ICREA, Barcelona, Spain\\
% ...
$^{48}$ Fakult\"{a}t f\"{u}r Mathematik und Physik, Albert-Ludwigs-Universit\"{a}t, Freiburg i.Br., Germany\\
$^{49}$ Section de Physique, Universit\'e de Gen\`eve, Geneva, Switzerland\\
% ...
$^{88}$ Department of Physics and Astronomy, Michigan State University, East Lansing MI, United States of America\\
% ...
$^{111}$ Homer L. Dodge Department of Physics and Astronomy, University of Oklahoma, Norman OK, United States of America\\
$^{112}$ Department of Physics, Oklahoma State University, Stillwater OK, United States of America\\
% ...
$^{115}$ LAL, Univ. Paris-Sud and CNRS/IN2P3, Orsay, France\\
% ...
$^{118}$ Department of Physics, Oxford University, Oxford, United Kingdom\\
% ...
$^{153}$ Raymond and Beverly Sackler School of Physics and Astronomy, Tel Aviv University, Tel Aviv, Israel\\
% ...
$^{*}$ Deceased\end{flushleft}


\end{document}